\documentclass[10pt]{iopart}
\usepackage{iopams}  
\usepackage{graphicx}
\usepackage[breaklinks=true,colorlinks=true,linkcolor=blue,urlcolor=blue,citecolor=blue]{hyperref}

\newcommand{\DL}{\textrm{\scriptsize DL}}
\newcommand{\E}{\textrm{\scriptsize E}}
\newcommand{\w}{\textrm{\scriptsize w}}
\newcommand{\ths}{\textrm{\scriptsize th}}
\newcommand{\bb}{\textrm{\scriptsize B}}
\newcommand{\F}{\textrm{\scriptsize F}}
\newcommand{\s}{\textrm{\scriptsize s}}

\newcommand{\D}{\textrm{\scriptsize D}}

\newcommand{\vc}{\mathbf}
\usepackage{graphicx}
\usepackage{color}

\begin{document}


\title[Sheaths, double layers and fireballs]{Interaction of Biased Electrodes and Plasmas: Sheaths, Double Layers and Fireballs}


\author{Scott D.~Baalrud, Brett Scheiner}
\address{Department of Physics and Astronomy, University of Iowa, Iowa City, Iowa 52242, USA}


\author{Benjamin Yee, Matthew M.~Hopkins and Edward Barnat}
\address{Sandia National Laboratories, Albuquerque, New Mexico 87185, USA}


\date{\today}

\begin{abstract}

Biased electrodes are common components of plasma sources and diagnostics. 
The plasma-electrode interaction is mediated by an intervening sheath structure that influences properties of the electrons and ions contacting the electrode surface, as well as how the electrode influences properties of the bulk plasma. 
A rich variety of sheath structures have been observed, including ion sheaths, electron sheaths, double sheaths, double layers, anode glow, and fireballs. 
These represent complex self-organized responses of the plasma that depend not only on the local influence of the electrode, but also on the global properties of the plasma and the other boundaries that it is in contact with. 
This review summarizes recent advances in understanding the conditions under which each type of sheath forms, what the basic stability criteria and steady-state properties of each are, and the ways in which each can influence plasma-boundary interactions and bulk plasma properties. 
These results may be of interest to a number of application areas where biased electrodes are used, including diagnostics, plasma modification of materials, plasma sources, electric propulsion, and the interaction of plasmas with objects in space.

\end{abstract}

\pacs{52.40.Kh,52.40.Hf}


\vspace{2pc}
\noindent{\it Keywords}: Sheath, double layer, fireball, anode spot, electrostatic instability\\
\submitto{\PSST}





\section{Introduction\label{sec:intro}}

Sheaths are fascinating examples of plasma self-organization. 
They are thin regions of strong electric field separating a quasineutral plasma from a material boundary that naturally form due to the surface charge generated as ions and electrons diffuse from the plasma at different rates~\cite{lang:23,lang:29,riem:91,fran:03a,lieb:05,robe:13}. 
Sheaths act to balance the electron and ion losses at steady-state~\cite{hers:05}. 
An accurate description of sheaths is essential for many plasma-based applications and experiments. 
For example, sheaths provide the directed energy necessary to etch semiconductors or alter surface properties of materials~\cite{cobu:79,wu:10}. 
They influence the particle and energy exhaust, wall erosion, and recycling in fusion energy experiments~\cite{stan:84}. 
Interpretation of diagnostics such as Langmuir probes rely on an accurate description of their properties~\cite{lieb:05,merl:07}. 
Sheaths are a critical feature of the interaction between objects (such as the moon) and space plasmas (such as the solar wind)~\cite{hale:11}, as well as spacecraft charging~\cite{garr:81} and interpretation of their onboard diagnostics~\cite{jaco:10}. 
Understanding sheaths is important. 

Sheaths have been studied since the beginning of plasma physics research~\cite{lang:23,lang:29}. 
Most studies have focused on ion sheaths, which are thin (several Debye length long) ion-rich regions where the electric field points from the plasma to the boundary with monotonically increasing magnitude~\cite{riem:91}. 
Ion sheaths are the most common type of sheath because electrons are typically much more mobile than ions in a plasma. 
This leads to a balance between negative charge on boundary surfaces and positive sheath charge in the plasma. 
The sheath acts to reduce the electron flux so that it balances the ion flux reaching the boundary. 
The basic properties of ion sheaths are well understood. 
However, a rich variety of different types of sheaths can be generated near biased electrodes~\cite{hers:05}. 
Not all of these are well understood. 

This review summarizes recent progress in understanding sheaths and related space-charge structures near biased electrodes in low-temperature, low-pressure plasmas; plasmas with electron temperature of a few eV, ion temperatures near room temperature, and pressures of approximately $10^{-2}-10^2$ mTorr. 
These include ion sheaths, electron sheaths, double sheaths, double layers, anode glow, and fireballs. 
Whereas the typical description of ion sheaths is based on a local analysis of a boundary interacting with an infinite plasma, the type of sheath that forms near a biased electrode often depends on global properties of the plasma and confinement chamber. 
Descriptions of these structures thus depend on the non-local physics of global plasma self-organization. 
This review discusses experimental conditions where each type of sheath may be expected to form, the basic properties of each type of sheath, ways in which the sheath influences bulk plasma properties, and how the different types of sheaths have been used to create advantageous outcomes in a variety of applications. 

Section~\ref{sec:structures} uses an example experimental configuration to illustrate how global conditions influence the type of sheath structure that will form near a biased electrode. 
The example geometry consists of a single electrode of surface area $A_\E$ biased at a potential $\phi_\E$ with respect to a grounded chamber wall of area $A_\w$. 
Conditions of global current balance in steady-state are shown to distinguish between the variety of possible sheath types, which can be categorized as ion sheaths, electron sheaths, double sheaths, anode glow (a type of double layer), or fireballs. 
Applications associated with this configuration are discussed. 
The remainder of the review focuses on recent advances in fundamental physics and applications associated with each of these sheath types. 

Section~\ref{sec:is} discusses ion sheaths. 
Although the basics of ion sheaths are generally well understood, a few recent advances are highlighted. 
These focus on features particular to biased electrodes, such as kinetic effects that arise as the electrode bias approaches the plasma potential~\cite{baal:11c,sche:16}. 
It also includes a review of recent theory, simulations, and experiments that have established the importance of ion-flow-driven instabilities in the presheath~\cite{baal:16}. 
These include ion-acoustic instabilities in plasmas with one ion species, and ion-ion two-stream instabilities in plasmas with multiple ion species. 
These instabilities have been observed to influence plasma properties, such as the ion velocity distribution function, velocity and density profiles, near the sheath in low-temperature low-pressure plasmas. 

Section~\ref{sec:es} discusses electron sheaths. 
Electron sheaths are thin regions of negative space charge in which the electric field points from the electrode to the plasma, and which monotonically increases in magnitude from the plasma toward the electrode. 
These are observed near small electrodes biased positive with respect to the plasma. 
A number of interesting features of electron sheaths have been discovered recently. 
These include the presence of an electron presheath~\cite{yee:17}, which is a long region with an electron pressure gradient that acts to accelerate electrons toward the boundary. 
Electrons are observed to gain a drift approaching the electron thermal speed as they near the electron sheath~\cite{sche:15}. 
The differential streaming between electrons and ions excites electron-ion two-stream instabilities near the ion plasma frequency~\cite{sche:15}. 
In addition, high frequency instabilities near the electron plasma frequency have been observed~\cite{sten:88}. 
The use of electron sheaths in applications such as electron source design and in the control of the electron energy distribution function (EEDF) are also discussed. 

Section~\ref{sec:ds} discusses double sheaths. 
Double sheaths (also known as virtual cathodes) are regions of alternating positive and negative space charge near the electrode~\cite{hers:94}. 
These can form due to current balance considerations associated with the global confinement geometry~\cite{baal:07}, due to local geometric effects of other surfaces near the biased electrode~\cite{hood:16}, or due to electron emission from the electrode~\cite{intr:88,hers:87,hers:95}. 
Recent advances have deepened our understanding of the multitude of different mechanisms responsible for double sheath formation, as well as the role of ion pumping mechanisms required to remove ions from the potential well that forms in a steady-state double sheath~\cite{fore:86}. 
These include ion-acoustic instabilities that cause the potential well to oscillate, as well as steady-state potential structures that can form to allow ions to leak out of the well to surrounding surfaces. 

Section~\ref{sec:fb} discusses fireballs~\cite{sten:08}. 
Fireballs are a secondary discharge near the electrode that is separated from the bulk plasma by a double layer.
They form from a thin region of positive space charge that develops within an electron sheath due to a localized increase in the ionization rate generated by sheath-accelerated electrons. 
When the positive space charge builds to a sufficiently high level, a secondary quasineutral discharge rapidly forms near the electrode~\cite{baal:09}. 
Recent advances include a more detailed understanding of fireball onset, steady-state properties, stability, and hysteresis that is observed in the electrode bias required for onset and disappearance of the fireball. 
This understanding has recently been advanced by new laser collision-induced fluorescence (LCIF) diagnostics and the first 2D particle-in-cell (PIC) simulations of fireball formation~\cite{sche:17}. 
Fireballs have been proposed as a means to control flows in plasmas~\cite{sten:11e}, as well as to generate thrust for plasma-based propulsion systems~\cite{makr:14}. 

Section~\ref{sec:con} concludes the review with a brief discussion of connections with related topics and open questions. 
These include measurements that are not yet understood, as well as how these phenomena may behave in related systems such as high pressure plasmas, magnetized plasmas, rf capacitively coupled plasmas, and electronegative plasmas. 
Answers to these questions will lead to a deeper understanding of these phenomena and are likely to enable new applications.

\section{Observed Sheath Structures\label{sec:structures}} 

\subsection{Geometric Considerations\label{sec:gc}} 

\begin{figure}
\begin{center}
\includegraphics[width=6.5cm]{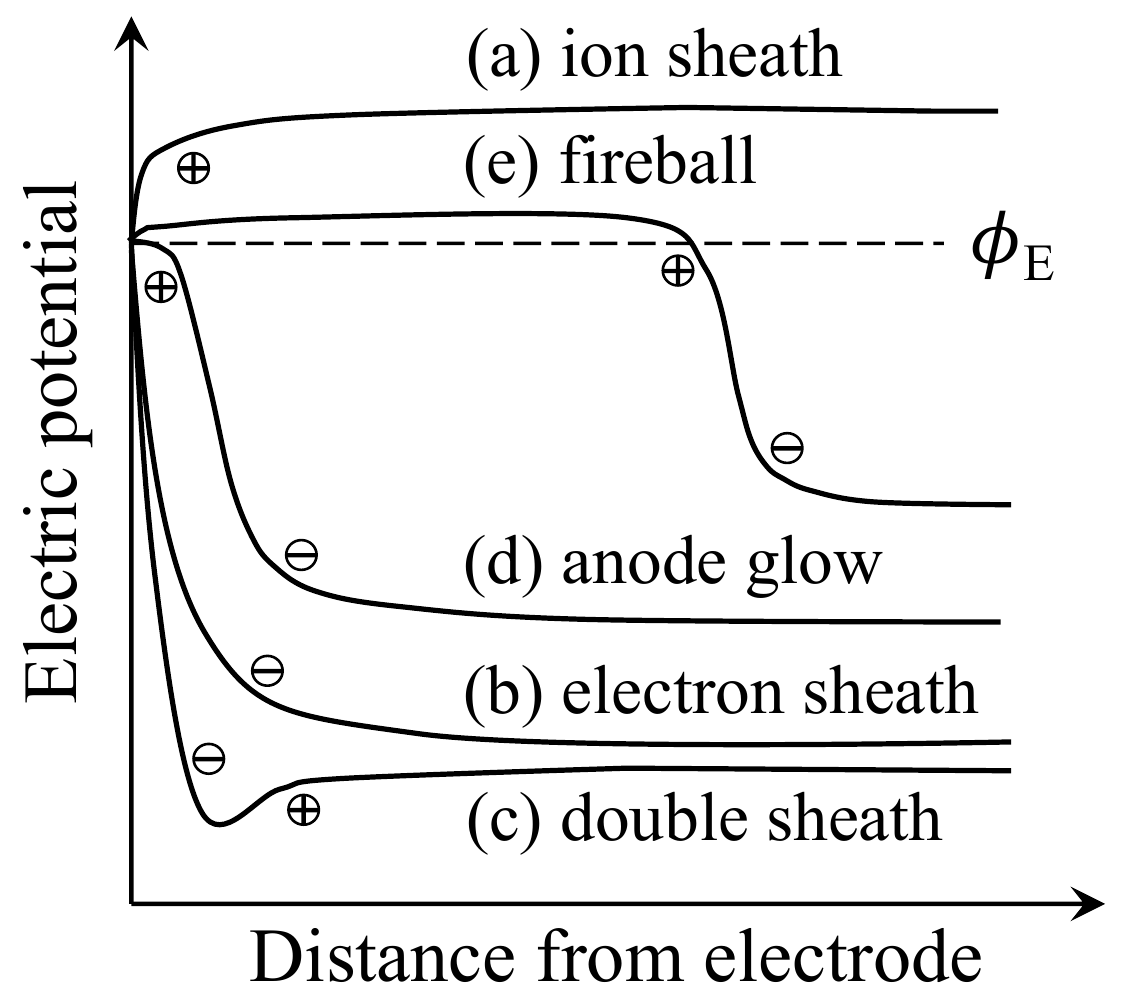}
\caption{Sketch of the electrostatic potential profile of various types of sheaths that can form near a biased electrode: (a) ion sheath, (b) electron sheath, (c) double sheath, (d) anode glow and (e) fireball. Plus signs denote regions of positive space charge and minus signs denote regions of negative space charge. }
\label{fg:type}
\end{center}
\end{figure}

\begin{figure}
\begin{center}
\includegraphics[width=6.5cm]{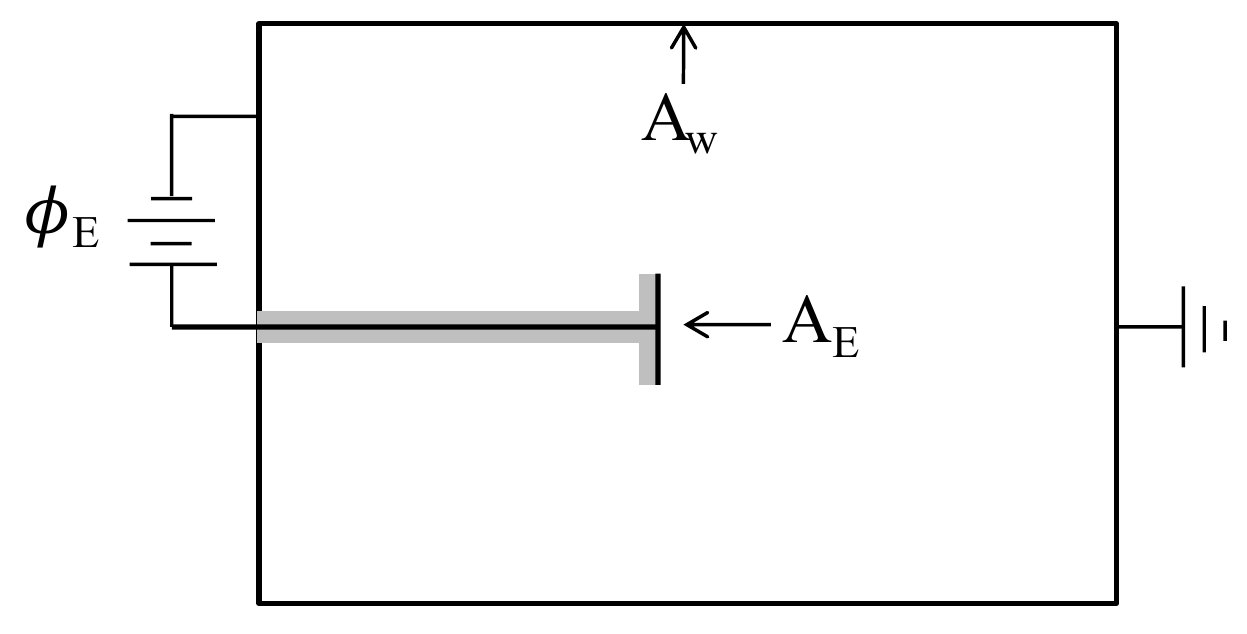}
\caption{Illustration of a hypothetical experimental setup with a biased electrode of area $A_\E$ in a confinement chamber of area $A_\w$. The gray area denotes an insulator, such that the only the conducting face of the electrode is exposed to the plasma. }
\label{fg:electrode}
\end{center}
\end{figure}

Figure~\ref{fg:type} illustrates the potential profile associated with a variety of sheath structures that have been observed near electrodes biased positive with respect to the confinement chamber walls. 
To understand when each might form, consider the simple geometry of a planar conducting electrode placed in a plasma confined by a conducting chamber wall, as depicted in figure~\ref{fg:electrode}. 
The chamber wall potential will be considered the reference potential (ground) $\phi_\w=0$.
Even if the electrode is biased much more positive than the chamber wall, it may or may not be positive with respect to the plasma potential. 
The plasma is assumed to be quasineutral with a uniform density and potential except in the sheaths. 
At steady-state, the plasma potential is determined by balancing the total current of electrons and ions lost from the plasma. 
As such, the resulting sheath structure depends on the effective area of the electrode for collecting plasma, $A_\E$, as well as the area of the chamber wall, $A_\w$~\cite{baal:07}. 
Here, $A_\E$ and $A_\w$ denote effective areas, which may differ from the geometrical surface areas. 
For instance, sheath expansion increases $A_\E$ compared to the geometrical area~\cite{sher:00}, while obstructions, such as confining cusp magnetic fields, decrease $A_\w$~\cite{hers:92,hubb:14,arth:18} in comparison to the geometric wall area. 
Despite their importance, such factors are particular to specific experimental arrangements. 
For simplicity, the following discussion focuses on the hypothetical geometry of figure~\ref{fg:electrode} where the effective areas can either be equated with geometric areas, or there is sufficient information available to determine $A_\E$ and $A_\w$ from the geometric areas. 

The electrode must be sufficiently small to be biased above the plasma potential. 
Otherwise, it would collect more electron current than the ion current lost to $A_\w$. 
Consider current balance. 
The electron current lost to the chamber wall is $I_{e,\w} = e \Gamma_{e,\ths} \exp(-e\phi_p/T_e) A_\w$, where $\Gamma_{e,\ths} = \frac{1}{4} \bar{v}_e n_o$ is the random electron flux incident on the ion sheath, $\bar{v}_e = \sqrt{8k_\bb T_e/\pi m_e}$ is the mean electron speed, and $\exp(-e\phi_p/T_e)$ is the Boltzmann factor associated with the electron density drop from the plasma potential $\phi_p$ to the grounded wall. 
Assume that the sheath near the electrode is an electron sheath that monotonically decreases from the electrode to the plasma potential. 
In this case, the electron current lost to the electrode is conventionally thought to be the random thermal flux incident on the electrode $e \Gamma_{e,\ths} A_\E$, representing a half-Maxwellian electron velocity distribution function at the electron sheath edge. 
However, recent work has shown the existence of an electron presheath, which establishes a flow shift of the electron distribution function by the sheath edge that satisfies an electron sheath analog of the Bohm criterion: $V_e = \sqrt{T_e/m_e} \equiv v_{e,\bb}$~\cite{yee:17,sche:15}. 
Further 2D-3V PIC simulations revealed that both a combination of flow-shift and loss cone distribution contribute to the electron flux~\cite{sche:16}. 
To account for these, we take $I_{e,\E} = \alpha_e e \Gamma_{e,\ths} A_\E$, where $\alpha_e =1$ represents the random flux limit and $\alpha_e = \sqrt{2\pi} \exp(-1/2) \approx 1.5$ represents the electron Bohm flux limit. 

\begin{figure}
\begin{center}
\includegraphics[width=7cm]{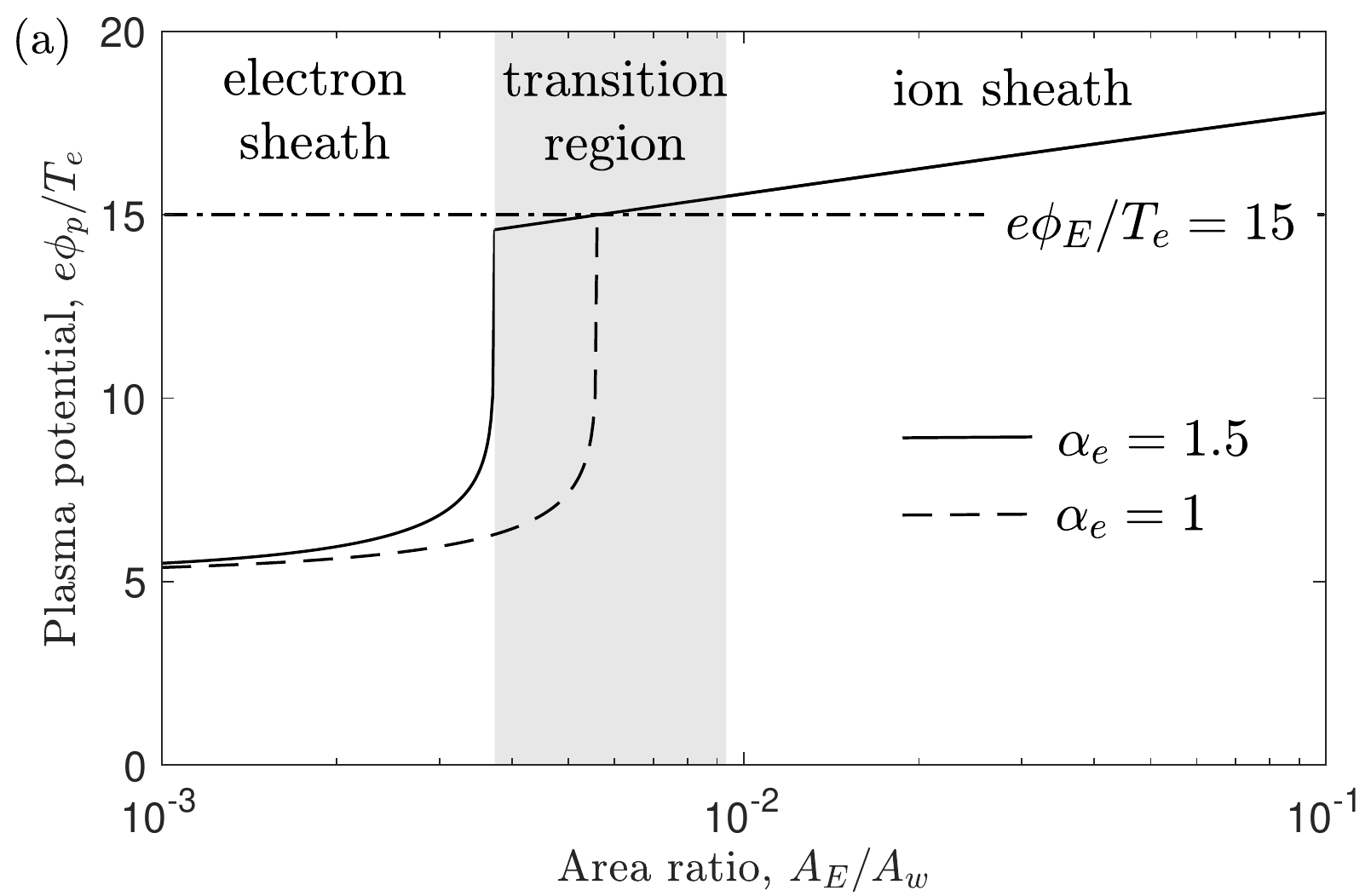}\\
\includegraphics[width=7cm]{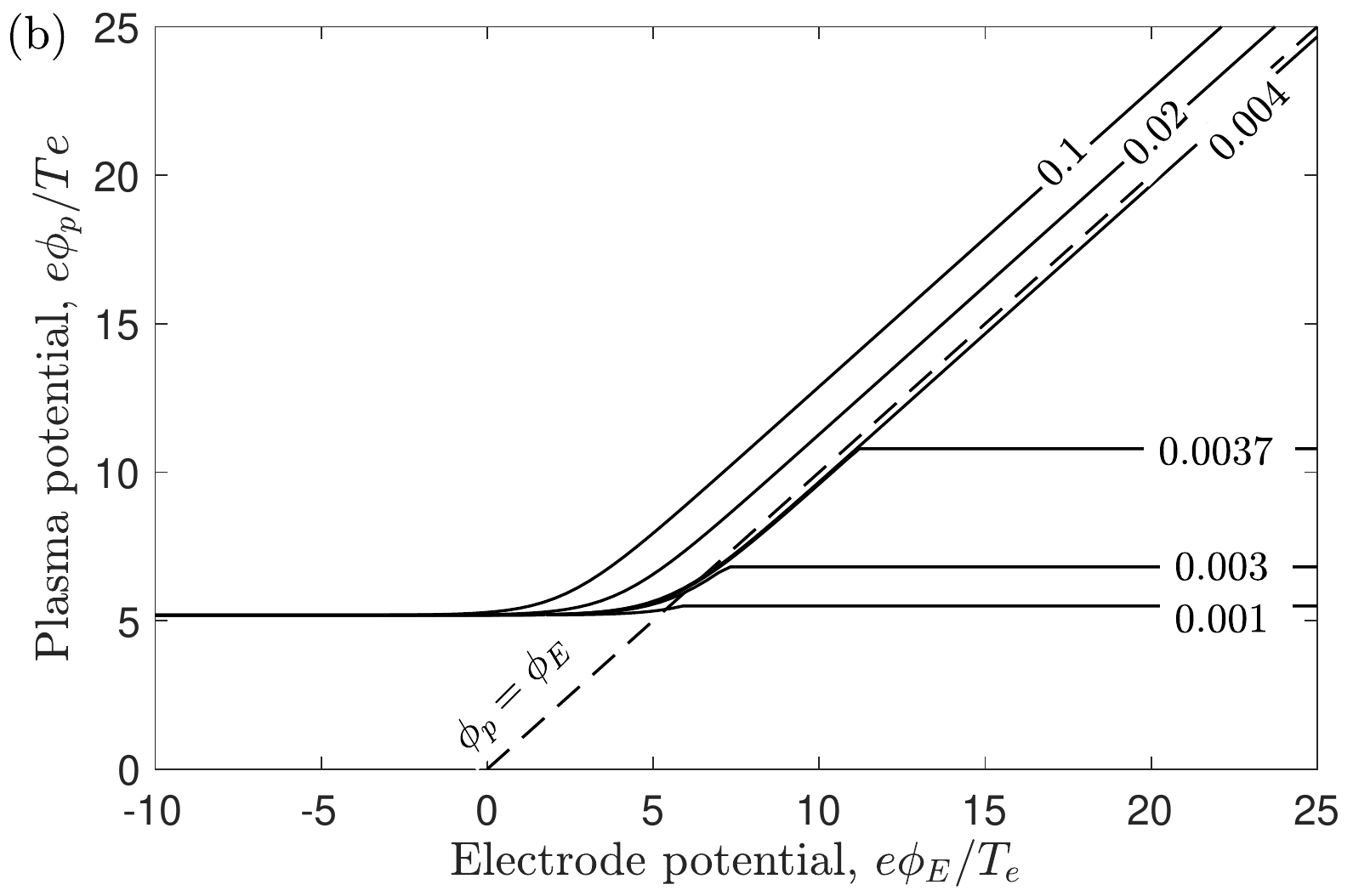}
\caption{Plasma potential predicted from equation~(\ref{eq:Vp_connect}) as a function of (a) area ratio for an electrode biased at $e\phi_\E/T_e=15$ and (b) electrode potential at a variety of area ratio $(A_\E/A_\w)$ values indicated by the numbers. The mass ratio is taken to be that of an argon plasma $m_i/m_e = 7.3\times 10^4$. Curves in the electron sheath region correspond to the solution of equation~(\ref{eq:Vp_es}) for $\alpha_e = 1.5$ (solid) and $\alpha_e = 1$ (dashed). The solution in the ion sheath region corresponds to equation~(\ref{eq:Vp_is}). }
\label{fg:area}
\end{center}
\end{figure}

As long as the electrode is biased at least a few $T_i/e$ above the plasma potential, $e(\phi_\E - \phi_p) \gtrsim T_e \gg T_i$, ions will be lost only to the chamber wall. 
The total ion current lost is then $I_i = \Gamma_{i,\bb} A_\w$, where $\Gamma_{i,\bb} = \exp(-1/2) e c_s n_o \approx 0.6 e c_s n_o$. 
Here the factor of $\exp(-1/2) \approx 0.6$ is due to the ion density drop in the ion presheath~\cite{hers:05}. 
Balancing the electron and ion losses determines the plasma potential 
\begin{equation}
\label{eq:Vp_es}
\phi_p = - \frac{T_e}{e} \ln \biggl(\mu - \alpha_e \frac{A_\E}{A_\w} \biggr) ,
\end{equation} 
where $\mu \equiv \sqrt{2.3m_e/m_i}$. 
The limit of small electrode area returns the floating potential limit $\phi_p = -(T_e/e) \ln (\mu)$. 
As the area of the electrode increases, the plasma potential gradually increases until the limit $A_\E \approx \mu A_\w/\alpha_e$ is approached, where the plasma potential diverges up to the electrode potential, no matter how high; see figure~\ref{fg:area}. 
Thus, the area ratio criterion
\begin{equation}
\label{eq:A_es}
\frac{A_\E}{A_\w} < \frac{\mu}{\alpha_e}
\end{equation} 
must be satisfied for an electron sheath to be present. 

In the opposite limit of a large electrode, current balance demands that the plasma potential be higher than the electrode potential, i.e., that an ion sheath forms at the electrode. 
If the electrode sheath is an ion sheath, ions are lost to both the electrode and wall with a total current of $I_i = \Gamma_{i,\bb} (A_\E + A_\w)$. 
Electrons are also lost to each boundary, with total current $I_e = e \Gamma_{e,\ths} \lbrace A_\w \exp(-e\phi_p/T_e) + A_\E \exp[-e(\phi_p-\phi_\E)/T_e]\rbrace$. 
Equating these, the plasma potential in the case of an ion sheath is  
\begin{equation} 
\label{eq:Vp_is}
\phi_p = \phi_\E -\frac{T_e}{e} \ln \biggl[ \frac{\mu (1 + A_\E/A_\w)}{A_\E/A_\w + \exp(-e\phi_\E/T_e)} \biggr] .
\end{equation} 
Conventionally, an ion sheath is expected to satisfy Bohm's criterion where ions are accelerated in a presheath with a potential drop of at least $T_e/(2e)$ (this is the argument that leads to the $\exp(-1/2) \approx 0.6$ factor for the density drop in the presheath). 
Thus, a minimum area ratio criterion for an ion sheath near the biased electrode is obtained by taking $e(\phi_p-\phi_\E) = T_e/2$ and $\phi_p \gg T_e$, leading to 
\begin{equation}
\label{eq:A_is}
\frac{A_{\E}}{A_\w} \geq \biggl( \frac{0.6}{\mu} -1 \biggr)^{-1} \approx 1.7 \mu .
\end{equation}

Figure~\ref{fg:area} shows the plasma potential obtained from equations~(\ref{eq:Vp_es}) and (\ref{eq:Vp_is}) within the range of values at which the expressions are expected to be valid, equations~(\ref{eq:A_es}) and (\ref{eq:A_is}) respectively. 
This illustrates that there is a gap between the area ratio at which an electron sheath or ion sheath is predicted. 
Multiple proposals have been made for how the sheath transitions from one solution to the other through this region. 
Some experiments have measured a double sheath of the form shown in figure~\ref{fg:type}c at conditions where the area ratio was predicted to be in, or near, this transition region; see figure~2 of~\cite{baal:07}. 
Earlier work has also documented similar double sheath structures in experiments at similar conditions~\cite{fore:86}. 
In this scenario, the virtual cathode (i.e., potential dip) regulates the electron current reaching the electrode to achieve global current balance, such that $\alpha_e \rightarrow \exp(-e\Delta \phi_\D/T_e)$ in equation~(\ref{eq:Vp_es}), where $\Delta \phi_\D = \phi_p-\phi_\D$ is the potential drop from the plasma to the dip minimum~\cite{hers:05}. 

A recognized challenge with a steady-state double sheath is that there must be some mechanism to pump ions that get trapped in the potential well (for instance, due to a collision with a neutral atom)~\cite{hers:87}. 
Otherwise, the positive space charge would build and flatten the potential well. 
Two possible explanations have emerged.  
One is that ions can be pumped to grounded or dielectric boundaries nearby the electrode (such as dielectric coatings on the back or sides of the electrodes)~\cite{fore:86}. 
This requires a two-dimensional description of the sheath potential where ions can ``slide'' out the sides of the one-dimensional potential well~\cite{hood:16}. 
Another is that the potential well oscillates at a timescale characteristic of the ion plasma frequency, which allows time-dependent pumping of otherwise trapped ions, and the double sheath potential profile emerges in the long-time average. 
Each of these possibilities has backing from experiments or simulations, and will be discussed in more detail in section~\ref{sec:ds}. 

Recent work has also shown that the transition from ion to electron sheath can be achieved without the formation of a local potential minimum (i.e., with monotonically increasing or decreasing potential profiles, that become nearly flat when $\phi_\E \approx \phi_p$)~\cite{sche:16,hopk:16}. 
These do not satisfy the conventional Bohm criterion or require the existence of a presheath. 
Such kinetic presheaths and Bohm criteria that emerge in this scenario have been discussed recently~\cite{baal:11c}, and will be reviewed in section~\ref{sec:transition}. 

In this situation, the electron and ion fluxes to the electrode transition to the random thermal flux, rather than the Bohm flux, in the transition region. 
A model for the plasma potential can be obtained by generalizing equations~(\ref{eq:Vp_es}) and (\ref{eq:Vp_is}) to account for this. 
Since $A_\E \ll A_\w$ in the transition region, we focus on this regime. 
In this case, the ion current lost to the electrode is negligible compared to the ion current lost to the chamber wall, regardless of the plasma potential. 
Accounting for this, the term $(1+A_\E/A_\w) \approx 1$ in equation~(\ref{eq:Vp_is}), and the ion sheath solution can be extended to its intersection with the electron sheath solution. 
This leads to the expression
\begin{eqnarray}
\label{eq:Vp_connect}
\phi_p = 
\left\lbrace \begin{array}{ll}
-\frac{T_e}{e} \ln \biggl(\mu - \alpha_e \frac{A_\E}{A_\w} \biggr) ,  \ \textrm{if} \ \frac{A_\E}{A_\w} \leq \frac{\mu}{\alpha_e} - e^{-e\phi_\E/T_e} \\
\phi_\E - \frac{T_e}{e} \ln \biggl[ \frac{\mu}{A_\E/A_\w + \exp(-e\phi_\E/T_e)} \biggr] ,  \ \textrm{otherwise}
\end{array} \right.
\end{eqnarray}
for the plasma potential that includes the ion and electron sheath limits and spans the transition region. 

This elementary analysis based on current balance demonstrates that a biased electrode significantly influences the plasma on a global scale when $A_\E/A_\w \gtrsim \sqrt{m_e/m_i} \ll 1$. 
The use of Langmuir probes in plasmas is predicated on the assumption that the diagnostic itself causes a negligible perturbation to the plasma~\cite{hers:89}. 
The current balance condition implies that the smallness of a Langmuir probe depends on the size of the probe itself, the size of the plasma chamber in which it is confined, as well as the mass ratio of ions and electrons in the plasma. 
A Langmuir probe must be very small to not perturb a plasma and one must think globally, not just locally, to understand the influence that the probe has on the plasma. 
Furthermore, we emphasize that the area $A_\E$ is the ``effective'' area for electron collection at the electrode. 
Sheaths cause the effective area of an electrode be larger than the geometric area~\cite{sher:00,lee:07b}. 
In fact, for a sufficiently small geometric probe size, such as a wire electrode, the effective size of the probe can be dominantly determined by the Debye length of the plasma. 
In this limit, the ratio of the Debye length to the plasma chamber size becomes the relevant scale comparison to assess the global influence of a probe. 

Finally, we note that this analysis has assumed that electrons or ions are absorbed by the boundary if they reach it. 
In fact, it is possible for a particle to reflect from the boundary back into the plasma. 
The probability of absorption of the charge, called the sticking coefficient, is a highly material dependent property~\cite{bron:15}. 
However, it can have a significant influence on the current balance and corresponding sheath structure.

\subsection{Tests of the geometric transitions} 

Experiments and simulations have been performed to explicitly test the predicted area ratio criteria from equations~(\ref{eq:A_es}) and (\ref{eq:A_is})~\cite{baal:07,hopk:16,barn:14}. 
One factor complicating such tests is that the effective areas for electron or ion collection are often expected to differ substantially from the geometric areas, and this can be difficult to quantify. 
For example, some experiments are conducted in multidipole confinement chambers where the effective loss area depends on the loss width of the magnetic cusps~\cite{limp:73,coop:16}. 
Another complicating factor is that the experiments often have more elaborate geometries than that described in figure~\ref{fg:electrode}, as well as electron sources used to generate the plasma that influence the current balance.
Each of these effects must be accounted for in the current balance. 
Despite these complications, progress has been made. 

\begin{figure}
\begin{center}
\includegraphics[width=6.5cm]{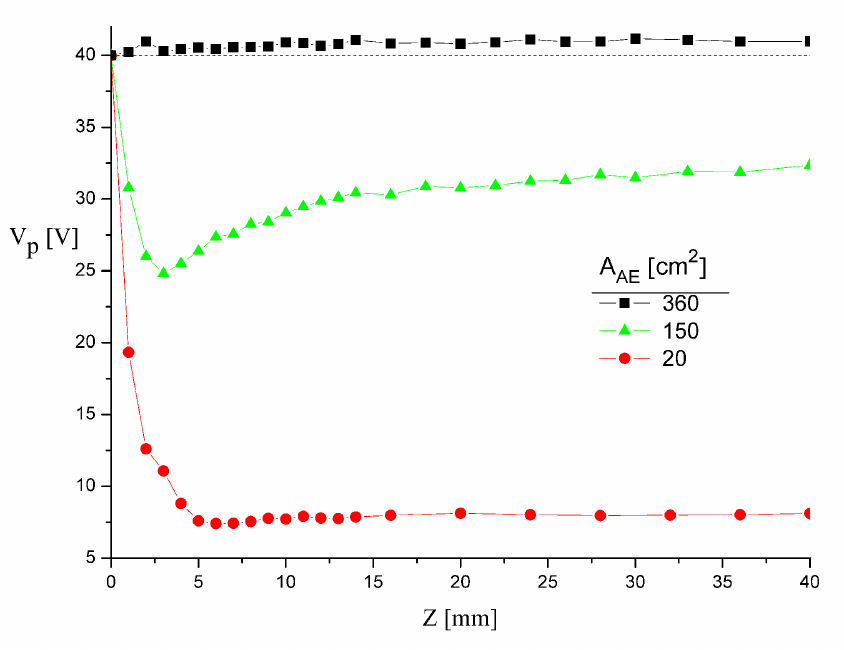}
\caption{Emissive probe measurements of the plasma potential in front of an electrode biased at +40 V with respect to a grounded chamber wall. The area of the conducting surface of each electrode was predicted by current balance to fall in the regime of an electron sheath (red circles), transition region, leading to the observed double sheath (green triangles), or ion sheath (black squares). Figure reprinted from reference~\cite{baal:07}. }
\label{fg:NAF_sheath}
\end{center}
\end{figure}

The first experimental tests were made in a multidipole confinement device with an electrode configuration similar to that depicted in figure~\ref{fg:electrode}~\cite{baal:07}. 
The electrodes were circular disks with the front side conducting and the back side covered with a dielectric coating. 
Electrodes with different exposed surface areas were tested. 
Figure~\ref{fg:NAF_sheath} shows plasma potential profile measurements, made with an emissive probe, in front of three electrodes of different surface areas. 
The surface areas were chosen to correspond to the predicted regimes of electron sheath, ion sheath, and near the transition region; though predicting a precise area ratio was difficult in this device because the effective wall area $A_\w$ was influenced by the local cusp magnetic field. 
Nevertheless, the figure shows an electron sheath in front of the small electrode, an ion sheath in front of the large electrode, and a double sheath in front of an electrode of intermediate size. 

\begin{figure}
\begin{center}
\includegraphics[width=6.5cm]{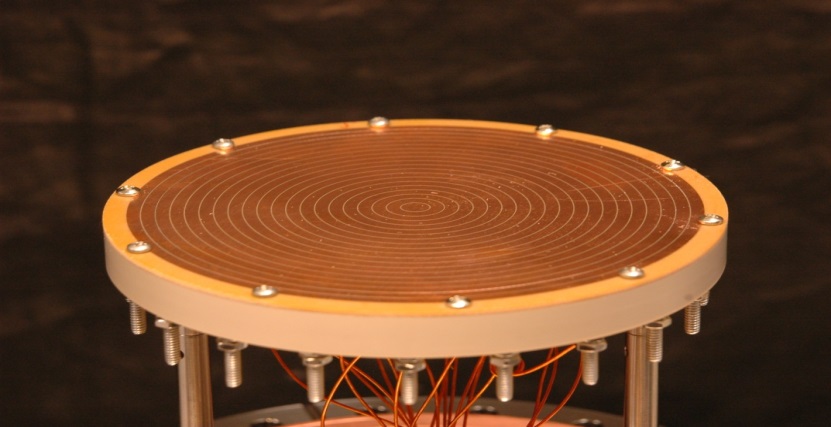}
\caption{Photograph of a segmented electrode used to test the area ratio criterion for electron versus ion sheath formation. Figure reprinted from reference~\cite{barn:14}.}
\label{fg:seg_photo}
\end{center}
\end{figure}

More detailed experimental tests were made using a segmented electrode to more sensitively vary the effective area of the electrode~\cite{barn:14}; see figure~\ref{fg:seg_photo}. 
The geometrical area of the electrode for collecting electrons was varied by positively biasing a subset of the segments, while electrically connecting the rest to the grounded chamber wall. 
An ion sheath was observed near the positive electrode when sufficiently many segments were biased positively, and an electron sheath was observed when sufficiently few were biased positively. 
The transition between the two regimes was found to be consistent with the predictions from current balance indicated by equations~(\ref{eq:A_es}) and (\ref{eq:A_is}) (with a minor modification to account for the electron current source in that experiment). 
Likewise, the relationship between the plasma potential and area ratio was consistent with that predicted from current balance. 
The measured sheath potential profile was observed to smoothly transition from an electron sheath to an ion sheath as the effective electrode size was increased in this experiment. 
The presence or absence of a double sheath is discussed further in section~\ref{sec:ds}. 

Further tests were performed with 2D PIC simulations in reference~\cite{hopk:16}. 
These used a rectangular 2D domain with a small portion of the boundary biased positive with respect to the rest of the boundary. 
The two boundaries were separated by a thin dielectric layer, which was included to remove strong electric fields associated with a sharp contact point; see figure 4 of \cite{hopk:16}. 
Since the electrode and wall areas (lengths in a 2D geometry) were set by the chosen computational domain, these simulations provided strict tests of the area ratio criteria. 
By changing the length of the biased segment of the boundary, the electron-to-ion sheath transition was explored. 
They were found to be in very close agreement with the predictions of equations~(\ref{eq:A_es}) and (\ref{eq:A_is}). 
However, unlike the experiment from reference~\cite{baal:07} that measured a double sheath in the transition region, a smooth transition from electron to ion sheath was observed as the electrode area increased. 
Similar smooth transitions from electron sheath to ion sheath were observed in combination with a loss-cone-like distribution for electrons in reference~\cite{sche:16}.


\subsection{Global non-ambipolar flow} 

The current balance arguments suggest that sufficiently large electrodes can be used to control the plasma potential and, in turn, the boundaries that electrons and ions are lost to. 
If the plasma potential is much larger than the electron temperature ($e\phi_p \gg T_e$), essentially all electrons will be blocked from reaching the chamber wall, and consequently will only be lost to the electrode. 
If the electrode is sized to be in (or near) the transition region, such that it can be biased more positive than the plasma potential by an amount that is much larger than the ion temperature [$e (\phi_\E - \phi_p) \gg T_i$], then essentially all ions will be blocked from reaching the electrode, and will consequently only be lost to the chamber wall. 
In reference~\cite{baal:07} this situation was dubbed ``global non-ambipolar flow''. 

\begin{figure}
\begin{center}
\includegraphics[width=6.5cm]{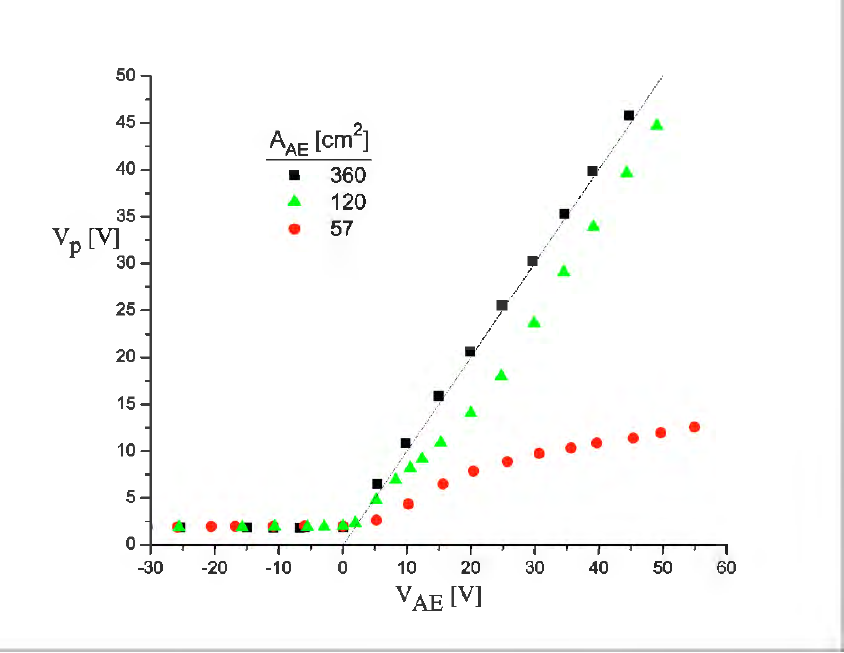}
\caption{Dependence of plasma potential ($\phi_p=V_p$) on the electrode potential ($\phi_\E = V_{\textrm{\scriptsize AE}}$) for electrode areas ($A_\E = A_{\textrm{\scriptsize AE}}$) predicted to be in each of the three sheath regimes: electron sheath (red circles), transition region leading to a double sheath (green triangles), and ion sheath (black squares). Measurements of the associated sheath potential profiles are shown in figure~\ref{fg:NAF_sheath}. 
The line indicates $\phi_\E = \phi_p$. This figure is reprinted with permission from reference~\cite{baal:07}.}
\label{fg:Vp_NAF}
\end{center}
\end{figure}

Figure~\ref{fg:Vp_NAF} shows measurements of the bulk plasma potential as electrode bias is varied. 
Electrodes with three surface areas were chosen to correspond to the three sheath regimes shown in figure~\ref{fg:NAF_sheath}, providing experimental measurements of the predicted plasma potential control shown in figure \ref{fg:area}b. 
The plasma potential is always slightly above that of a large electrode (ion sheath), varies little in response to a small electrode (electron sheath), and is proportional to (but slightly less than) that of an intermediate sized electrode (double sheath). 
By choosing an electrode area near the transition region, the plasma potential could then be raised far above the grounded chamber wall. 
Measurements of the current with electrodes in this regime confirmed global non-ambipolar flow~\cite{baal:07}. 

Global non-ambipolar flow provides an efficient way to extract the maximum electron current from a plasma. 
This was utilized in the design of the \emph{Non-Ambipolar Electron Source}~\cite{long:06,long:07,long:08}. 
To extract the electrons as a beam, the source design also made use of results from magnetic mirror experiments showing that the electrostatic potential inside a biased ring in a magnetized plasma is uniform (in other words, it spans the gap)~\cite{seve:92}. 
This enables one to construct a ``virtual electrode'' through which the electrons can pass and be extracted as a beam. 
By tailoring the size of this electrode to be near the transition region, the electron current extracted can be maximized. 
This enabled the continual extraction of several amps of electron current from a compact and reliable helicon plasma source~\cite{long:08}. 

\begin{figure}
\begin{center}
\includegraphics[width=6.5cm]{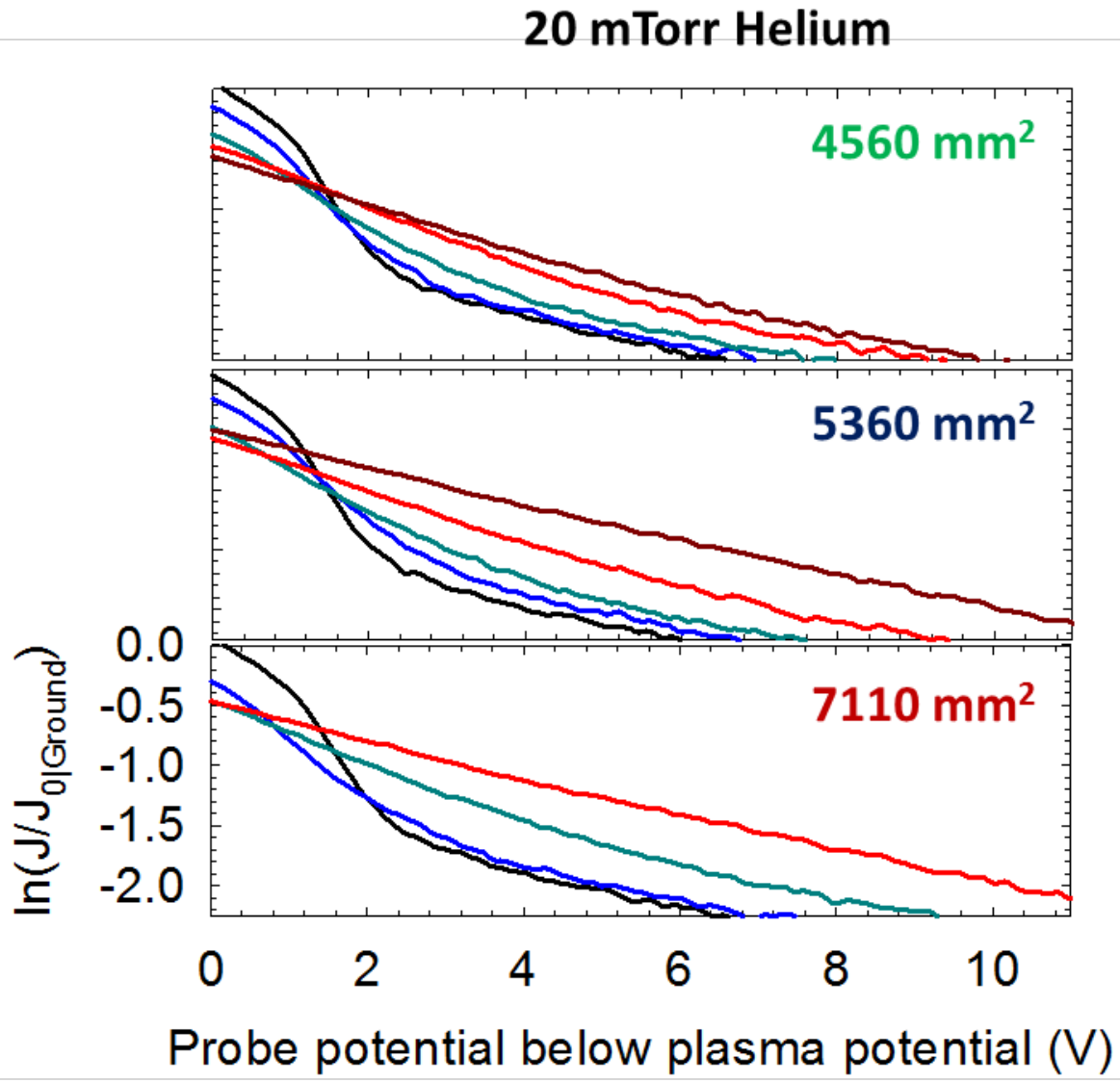}
\caption{Logarithm of the current collected by a Langmuir probe in a 20 mTorr helium discharge for five values of the electrode potential referenced to the plasma potential: $\phi_\E - \phi_p = -2.5$ V (black line), 0 V (blue line), 2.5 V (aqua line), 5 V (red line) and 10 V (maroon line). The numbers listed in the panel indicate the electrode surface area. This figure is reprinted from reference~\cite{barn:14}.}
\label{fg:EEDF}
\end{center}
\end{figure}

The size and bias of the electrode also influences the EEDF in the bulk plasma, and in turn the plasma density and electron temperature~\cite{schw:13}.
Figure~\ref{fg:EEDF} shows measurements from reference~\cite{barn:14} of the current collected by a Langmuir probe in the bulk plasma for three different electrode areas, and for biases spanning below to above the plasma potential. 
A Maxwellian EEDF would be expected to lead to a linear profile in this measurement in which the slope is proportional to the electron temperature. 
When the electrode is biased a few volts below the plasma potential, the EEDF is found to consist of a cool and dense population of thermal electrons and a hotter, but much less dense, population of electrons on the tail of the distribution. 
There is a sharp divide between these populations at an energy corresponding to the potential drop of the ion sheath at the chamber wall. 
This is the typical expectation for the EEDF in a plasma without an electrode: The dense and cool thermal population is confined by the ion sheaths at the boundaries, while the hotter but much less dense tail population is associated with degraded primary electrons injected from the electron source (filaments or hot cathode). 
In contrast, as the electrode bias approaches or exceeds the plasma potential, the dense and cool population disappears and the entire EEDF has a temperature that is closer to that of the original higher energy tail population. 
This is interpreted to be a result of the electrode collecting electrons indiscriminate of energy; i.e., the ion sheaths at the chamber wall can no longer confine a low energy electron population because these electrons are rapidly lost to the electrode~\cite{baal:07}. 

\begin{figure}
\begin{center}
\includegraphics[width=5.0cm]{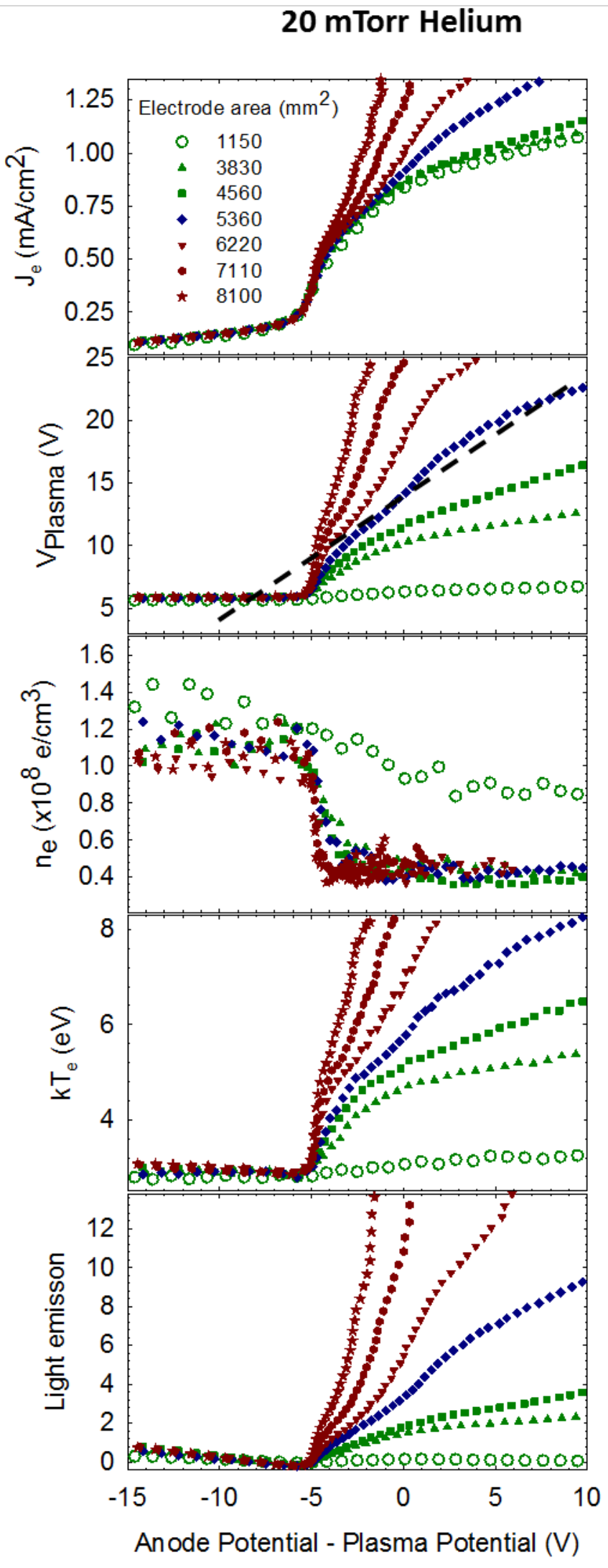}
\caption{Measurements of bulk plasma parameters as a function of the electrode bias referenced to the plasma potential ($\phi_\E - \phi_p$) in a 20 mTorr helium plasma. Numbers in the legend indicate the electrode surface area. Color of the data points indicates the predicted sheath regime: electron sheath (green), transition region (blue) and ion sheath (maroon). This figure is reprinted from reference~\cite{barn:14}.}
\label{fg:nT_profiles}
\end{center}
\end{figure}

This change of the EEDF leads to changes in the bulk plasma parameters. 
Figure~\ref{fg:nT_profiles} shows the corresponding current collection, plasma potential, electron density, electron temperature and light emission profiles as a function of electrode potential and for several electrode surface areas in the same discharge used to obtain the data for figure~\ref{fg:EEDF}. 
Data are shown from electrodes of seven different surface areas. 
For the smallest electrode size, the plasma potential, density and temperature were essentially constant regardless of the electrode bias. 
However, when the electrode area approached the transition region, the plasma potential rose along with the electrode potential, and a corresponding decrease of the electron density and increase of the electron temperature was measured. 
Each of these is a direct consequence of the observed EEDF behavior from figure~\ref{fg:EEDF}; the electron temperature increases because the cool confined electron population is lost, and the density decreases for the same reason. 

Biased electrodes can thus be used to control (within limits) the plasma density and temperature~\cite{demi:14}. 
Based on this principle, it has been shown that plasma parameters can be varied mechanically by moving a biased electrode into or out of a localized cusp magnetic field and thus changing the effective surface area of the electrode~\cite{hers:92}. 
The mechanism is also similar to MacKenzie's Maxwell demon, which has been used to control electron temperature in a plasma~\cite{mack:71,yip:13,yip:15a}. 
In MacKenzie's work, it was argued that a thin positively biased wire preferentially collects low energy electrons because of orbital motion effects, leading to an effective heating effect~\cite{mack:71}. 
However, it was also shown that a biased planar electrode leads to essentially the same result due to the mechanism described above~\cite{yip:13}. 

\subsection{Influence of increased ionization due to strong sheath fields} 

\begin{figure}
\begin{center}
\includegraphics[width=7cm]{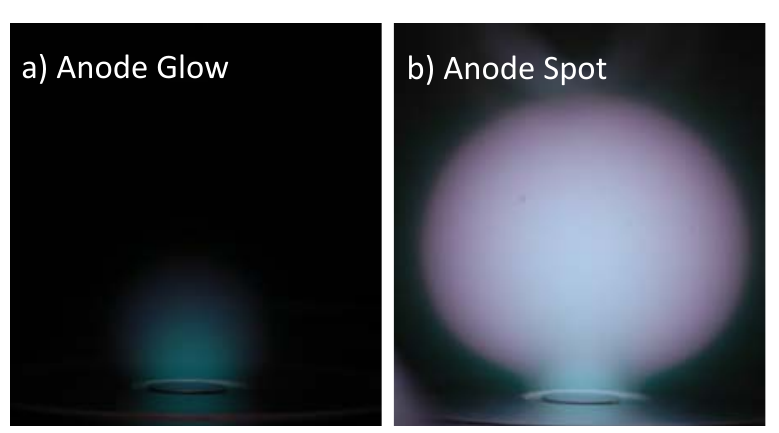}
\caption{Photographs of an anode glow (a) and an anode spot (b). Figure reprinted from reference~\cite{sche:17}.}
\label{fg:spot_photo}
\end{center}
\end{figure}

An electrode of sufficiently small surface area can be biased far above the plasma potential. 
As the energy of the sheath-accelerated electrons approaches the ionization potential of the neutral gas in a partially ionized plasma, a thin region of increased ionization will form. 
This thin region glows due to increased atomic excitation from the energetic electrons, as shown in figure~\ref{fg:spot_photo}a, and is called ``anode glow''. 
Since the ions born from ionization in this region are much more massive than electrons, they have a much longer residence time than electrons in this region, before being swept into the plasma by the electron sheath electric field. 
This causes a positive space charge near the electrode surface, and the potential profile in this region to flatten, as depicted in figure~\ref{fg:type}d. 
When the electron sheath potential drop is sufficiently large, and the neutral pressure is sufficiently high, enough ion space charge can build up that it causes a pressure imbalance between this region and the bulk plasma. 
This causes the plasma to rapidly self-organize into a new ``anode spot,'' or ``fireball'' state, as pictured in figure~\ref{fg:spot_photo}b. 
In this configuration, the potential profile takes the form of a double layer with a large (typically several centimeter) quasineutral fireball discharge separated from the bulk plasma by a potential drop that is approximately the ionization potential of the neutral gas; see figure~\ref{fg:type}e~\cite{torv:79,cart:87,song:91,song:92d}. 

This section has introduced four types of sheaths that can form near a biased electrode in a low-pressure plasma: ion sheath, electron sheath, double sheath and fireball. 
The following four sections provide a more detailed summary of the recent progress in understanding each of these configurations.

\section{Ion Sheaths\label{sec:is}} 

\subsection{Conventional ion sheath properties\label{sec:is_standard}} 

The basic properties of ion sheaths have been well characterized theoretically and experimentally. 
The topic has been summarized in several reviews and monographs~\cite{riem:91,fran:03a,robe:13,hers:05}. 
This section does not attempt to review this literature. 
Instead, it recalls a few of the main results regarding steady-state ion sheath properties in order to compare and contrast with them when discussing other sheath types in later sections.

Ion sheaths can often be modeled from a one-dimensional steady-state two-fluid description, consisting of the continuity equation 
\begin{equation}
\label{eq:continuity}
\frac{d}{dx} (n_s V_s) = S_{s},
\end{equation}
where $S_s$ is a source term, the force balance equation 
\begin{equation}
\label{eq:fb}
m_s n_s V_s \frac{d V_s}{dx} = - n_s q_s \frac{d\phi}{dx} - \frac{d p_s}{dx} - \frac{d \Pi_{xx,s}}{dx} + R_{x,s}
\end{equation} 
and Poisson's equation 
\begin{equation}
\label{eq:poisson}
\frac{d^2 \phi}{dx^2} = - \frac{\rho_q}{\epsilon_o} = - \frac{1}{\epsilon_o} (q_i n_i - e n_e).
\end{equation}
Here, $p_s = n_s T_s$ is the scalar pressure, $\Pi_{xx}$ the stress tensor component, $R_x$ the friction force density due to collisions, $\rho_q$ is the charge density and the subscript $s$ denotes the species type (either electrons or ions). 

In the typical circumstance that the ion sheath potential drop exceeds the electron temperature, the dominant terms of the electron force balance are the electric field and the scalar pressure gradient. 
The electron flux is approximately the ion flux $n_e c_s$, or less, so the inertia term is approximately $m_e/m_i$ smaller than the electric field or pressure terms. 
Also, the stress tensor is negligible in this case because only tail electrons lead to a stress gradient. 
In this case, the electron density obeys the Boltzmann density relation $n_e = n_o \exp(-e\phi/T_e)$. 

Ions are usually well modeled as a drifting Maxwellian distribution, in which case the ion stress tensor can also be neglected. 
The ion scalar pressure is also typically negligible in low pressure discharges because $T_i \ll T_e$. 
Each of the other terms will contribute in some portion of the plasma boundary transition region, but the problem can be simplified by noting that Debye shielding limits most of the potential drop to a sheath region of a few Debye lengths from the boundary. 
At low-pressure conditions, this justifies a scale separation between a collisionless sheath and a weakly-collisional presheath. 
In the thin sheath region, collisions are negligible as long as $\lambda_\D \ll \lambda_{\tiny \textrm{in}}$, where $\lambda_\D$ is the Debye length and $\lambda_{\tiny \textrm{in}}$ is the ion-neutral collision mean free path. 
Here, the ion force balance then predicts ballistic motion, in which case $n_i(x) = J_o/[eV_i(x)]$, where $J_o$ is the ion current density at the sheath boundary. 
In the limit that there are no electrons in the ion sheath, using this in Poisson's equation, multiplying by $d\phi/dx$ and integrating twice leads to the Child-Langmuir law~\cite{lieb:05} 
\begin{equation}
\label{eq:cl}
J_o = \frac{4}{9} \sqrt{\frac{2e \epsilon_o^2}{m_i}} \frac{[\phi(x) - \phi_\w]^{3/2}}{x^2} 
\end{equation} 
describing the electrostatic potential profile in the sheath. 

The boundary between the non-neutral sheath and the quasineutral presheath can be described via an expansion in the charge density: $d^2\phi/dx^2 = -[\rho_q(\phi_o) + d\rho_q/d\phi|_{\phi_o} (\phi - \phi_o) + \ldots]/\epsilon_o$, where $\phi_o$ is the plasma potential at the ``sheath edge''.  
The sheath edge is associated with the breakdown of charge neutrality, and can be identified as the location where the first order term in this expansion is the largest: $d^2\phi/dx^2 = -d \rho_q/d\phi |_{\phi_o} (\phi - \phi_o)$.  
Multiplying by $d\phi/dx$ and integrating leads to the condition $\epsilon_o E + d\rho_q/d\phi|_{\phi_o}(\phi - \phi_o)^2 = C$, where $C$ is a constant. 
Here $x=x_o$ is the sheath edge location. 
Since $\phi \rightarrow \phi_o$ on the sheath scale in the limit of small Debye length, $(x-x_o)/\lambda_\D \rightarrow \infty$, $C$ must be zero~\cite{riem:95}.  
We are then left with $d\rho_q/d\phi|_{\phi_o} = -\epsilon_o E^2/(\phi - \phi_o)^2$, which implies $d\rho_q/d\phi |_{\phi_o} \leq 0$, as the sheath edge criterion. 
Using $dn_s/d\phi = -(dn_s/dx)/E$ shows that this is the location where the charge density gradient first becomes positive  
\begin{equation}
\label{eq:edge_criterion}
\sum_s q_s \frac{dn_s}{dx} \biggl|_{x_o} \geq 0 .
\end{equation}
Combining the continuity~(\ref{eq:continuity}) and force balance~(\ref{eq:fb}) equations, this implies~\cite{baal:11c} 
\begin{equation}
\label{eq:gen_bohm}
\sum_s q_s \biggl[ \frac{q_s n_s - (n_s dT_s/dx + d \Pi_{xx,s}/dx - R_{x,s})/E}{m_s V_{s}^2 - T_s} \biggr]_{x_o} \leq 0 
\end{equation}
at the sheath edge. 

For the typical case considered above the sheath potential drop is at least as large as the electron temperature ($e\Delta \phi_s \gtrsim T_e$), the term in parenthesis in equation~(\ref{eq:gen_bohm}) is small for both electrons and ions. 
Here, $\Delta \phi_s$ is the potential drop in the sheath. 
Also making use of the conditions $m_i V_i^2 \gg T_i$ and $m_e V_e^2 \ll T_e$ in this situation, leads to the Bohm criterion for the minimum ion speed at the sheath edge~\cite{bohm:49} 
\begin{equation}
V_i \geq c_s .
\end{equation}
Here, $c_s = \sqrt{k_BT_e/m_i}$ is the ion sound speed. 
Thus, the conventional Bohm criterion that ions must flow supersonically into the sheath is obtained in this limit. 
Usually, this criterion is met via the minimum speed $V_i = c_s$, in which case the ion current density at the sheath edge is $J_o = en_{i,o} c_s$. 
The Child-Langmuir law~(\ref{eq:cl}) then implies that the total sheath thickness is~\cite{lieb:05} 
\begin{equation}
\label{eq:st}
\frac{l_s}{\lambda_\D} = \frac{\sqrt{2}}{3} \biggl( \frac{2e \Delta \phi_s}{T_e}\biggr)^{3/4}  .
\end{equation} 
Thus, the sheath scale is characterized by the Debye length, but a sheath can be several Debye lengths thick if $\Delta \phi_s \gg T_e$. 
The sheath potential drop is determined from the global current balance, as described in section~\ref{sec:structures}. 

\begin{figure}
\begin{center}
\includegraphics[width=8cm]{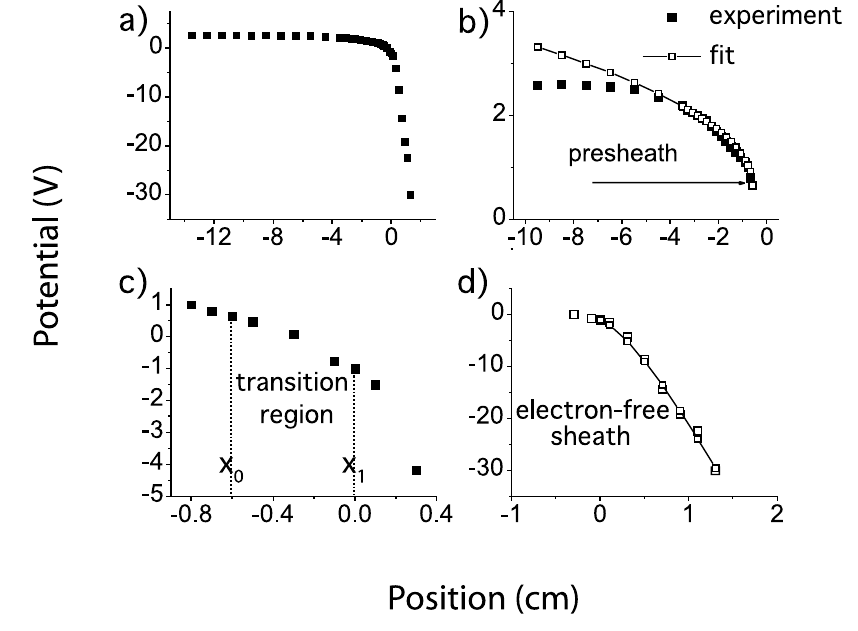}
\caption{Emissive probe measurements of the plasma potential in front of an electrode biased at -30 V referenced to ground. Each panel shows the same data set of a different spatial scale with 0 referenced to the sheath edge. The neutral pressure in the discharge was 0.44 mTorr and the electron density $2.1 \times 10^7$ cm$^{-3}$. 
Panel (b) shows a fit to the $\sqrt{x-x_o}$ scaling from equation~(\ref{eq:phi_sr}). 
Figure reprinted from reference~\cite{oksu:02}.}
\label{fg:oksuz}
\end{center}
\end{figure}

The acceleration of ions required to meet the Bohm criterion occurs in a long ($\lambda_{\tiny \textrm{in}}$-scale) quasineutral region called the presheath. 
A key aspect of the transition between the sheath and the bulk plasma is the generation of particle flux, which is zero in the bulk but equal to the Bohm flux at the sheath edge. 
A full solution of the plasma potential profiles throughout this transition typically requires a numerical solution of equations~(\ref{eq:continuity})--(\ref{eq:poisson}). 
Approximate solutions can be made by dividing the domain into regions and applying multiscale analysis. 
Much has been written about how to properly match the approximate solutions obtained this way~\cite{riem:97,fran:02,kaga:02,gody:90,fran:00b,fran:70,bene:00,riem:05,gody:03}. 

Here, we briefly summarize the modified mobility-limited flow presheath model~\cite{riem:97,sher:89,sher:91}. 
Consider a region near the sheath edge where collisions (presumed to be dominantly ion-neutral collisions) cause a friction force $R_{x,i} = - \nu_{\tiny \textrm{in}} V_i$, but do not generate significant flux $S_{i} = 0$, the ion continuity, force balance (with negligible ion pressure) and quasineutrality relation $dn/dx = -(e/T_e)En$ imply
\begin{equation}
\label{eq:Vi_E}
V_i = \mu E \biggl(1 - \frac{V_i^2}{c_s^2} \biggr)
\end{equation} 
where $\mu = e/(m_i \nu_{\tiny \textrm{in}})$ is the ion mobility. 
Quasineutrality also implies $n(x) = n_o c_s/V_i = n_o \exp(-e\phi/T_e)$, so 
\begin{equation}
\label{eq:phi_cs}
\frac{e(\phi-\phi_o)}{T_e} = \ln \biggr(\frac{c_s}{V_i} \biggl). 
\end{equation}
Using $E = -d\phi/dx = (T_e/e) (dV_i/dx)/V_i$ in equation~(\ref{eq:Vi_E}) provides a first order differential equation for the ion flow speed in the presheath 
\begin{equation}
\label{eq:dVi}
\biggl( \frac{c_s^2 - V_i^2}{V_i^2 \nu_{\tiny \textrm{in}}} \biggr) d V_i = dx .
\end{equation}
Equations~(\ref{eq:phi_cs}) and (\ref{eq:dVi}) describe the potential and ion flow speed profiles in the presheath if the speed dependence of the collision frequency $\nu_{\tiny \textrm{in}}(V_i)$ can be specified. 

Analytic solutions can be obtained in two common limits. 
In the constant mean free path limit ($\nu_{\tiny \textrm{in}} =V_i/\lambda_{\tiny \textrm{in}}$), 
\begin{equation}
\label{eq:vi_mfp}
\frac{V_i}{c_s} = \exp \biggl\lbrace \frac{1}{2} - \frac{x-x_o}{\lambda_{\tiny \textrm{in}}} + \frac{1}{2} W_{-1} \biggl[ - \exp \biggl(2 \frac{x-x_o}{\lambda_{\tiny \textrm{in}}} - 1 \biggr) \biggr] \biggr\rbrace 
\end{equation} 
and 
\begin{equation}
\label{eq:phi_mfp}
\frac{e(\phi-\phi_o)}{T_e} = - \frac{1}{2} + \frac{x-x_o}{\lambda_{\tiny \textrm{in}}}  - \frac{1}{2} W_{-1} \biggl[-\exp \biggl( 2 \frac{x-x_o}{\lambda_{\tiny \textrm{in}}} - 1 \biggr) \biggr] . 
\end{equation} 
Here, $W_{-1}$ is the Lambert-W function. 
In the constant collision frequency limit $\nu_{\tiny \textrm{in}} \approx c_s/\lambda_{\tiny \textrm{in}}$, 
\begin{equation}
\label{eq:vi_nu}
\frac{V_i}{c_s} = \biggl[1 - \frac{x-x_o}{2\lambda_{\tiny \textrm{in}}} \biggl(1 - \sqrt{1 - \frac{4\lambda_{\tiny \textrm{in}}}{x-x_o} } \biggr) \biggr]
\end{equation}
and 
\begin{equation}
\label{eq:phi_nu}
\frac{e(\phi - \phi_o)}{T_e} = \textrm{arccosh} \biggl( 1 - \frac{x-x_o}{2\lambda_{\tiny \textrm{in}}} \biggr) .
\end{equation} 

In either the constant mean free path model [equation~(\ref{eq:phi_mfp})] or the constant collision frequency model [equation~(\ref{eq:phi_nu})], the potential profile scales as the square root of distance for $(x-x_o)/ \lambda_{\tiny \textrm{in}} \ll 1$~\cite{riem:97}
\begin{equation}
\label{eq:phi_sr}
\frac{e(\phi-\phi_o)}{T_e} = \sqrt{\frac{x-x_o}{\lambda_{\tiny \textrm{in}}}} 
\end{equation} 
from either model. 
We note that although this provides a prediction for the electrostatic potential that continually matches sheath and presheath, it predicts that the electric field diverges as $x\rightarrow x_o$. 
Godyak has analyzed this region, showing that the electron density must be accounted for in this transition region~\cite{gody:82}. 
Since the electron density obeys the Boltzmann relation, $dn_e/dx = -en_e E/T_e$, and the scale for variation of this density is the Debye length, the electric field at the sheath edge is actually expected to have a value of $E = T_e/(e\lambda_\D)$~\cite{gody:82}. 
Riemann has also predicted that the length of this ``transition region'' between sheath and presheath scales as $\lambda_{\tiny \textrm{in}}^{1/5} \lambda_\D^{4/5}$. 

The above predictions have been experimentally verified using emissive probe and laser-induced fluorescence (LIF) measurements~\cite{sher:89,sher:91,oksu:02,oksu:05,clai:06,jaco:07,hers:05b,goec:92,moor:13}. 
An example is shown in figure~\ref{fg:oksuz}. 
These experiments validated a number of the features predicted above, including the Child-Langmuir law [equation~(\ref{eq:cl})] relating the sheath potential to distance from the electrode, the square root scaling of the presheath potential [equation~(\ref{eq:phi_sr})], Godyak's prediction of an electric field of $E=T_e/(e\lambda_\D)$ at the sheath edge, and the predicted $\lambda_{\tiny \textrm{in}}^{1/5} \lambda_\D^{4/5}$ scaling of the transition region. 
Experiments using LIF to measure the ion velocity distribution function also validated the ion flow speed profiles in the presheath and that the Bohm criterion was met at its minimum value ($V_i=c_s$) at the sheath edge~\cite{goec:92}. 
Similar tests of these analytic predictions have also been made using PIC and other kinetic simulations~\cite{schw:90,proc:90,proc:91,tsus:98,jeli:07,milo:11,caga:17,hara:18}. 
In addition to confirming aspects of the above model, its limitations have also been explored. 
Measurements have been made showing ion heating in the direction perpendicular to the sheath electric field~\cite{lee:08}, and the effect has also been observed in PIC simulations~\cite{meig:07}. 
Much related work has also been done exploring sheaths in rf discharges~\cite{kawa:99}.

\subsection{Drift-induced instabilities\label{sec:is_i}} 

Traditional sheath models are based on steady-state kinetic or fluid descriptions in which the plasma smoothly transitions to the boundary. 
However, research over the past decade has revealed that instabilities can arise in the plasma-boundary transition region, particularly the presheath of low pressure and low temperature discharges~\cite{hers:05,baal:16,hers:05b,hala:00,baal:09a,baal:09b,baal:11,yip:10,hers:11,baal:15a,sher:17,adri:17}. 
These instabilities are driven by the presheath electric field, which generates a relative drift between species (either electron-ion or ion-ion). 
Since this is a weak driving force, the instabilities are typically of a kinetic nature in which depressions in velocity phase-space lead to Landau growth of thermal fluctuations.  
However, in the case of multiple ion species the instabilities can transition to a two-stream fluid instability~\cite{baal:11}. 
The basic instability properties can be understood using a linear stability analysis based on the steady-state plasma parameter profiles discussed in section~\ref{sec:is_standard}. 
The linear dispersion relation can be derived from the roots of the plasma dielectric function
\begin{equation}
\hat{\varepsilon} (\vc{k}, \omega) = 1 + \sum_s \frac{q_s^2}{\epsilon_o k^2 m_s} \int d^3v \frac{\vc{k} \cdot \partial f_{s,o}(\vc{v})/\partial \vc{v}}{\omega - \vc{k} \cdot \vc{v}} , \label{eq:ephat1}
\end{equation}
where $\omega$ is the complex frequency, $\vc{k}$ the wavenumber and $f_{s,o}$ is the steady-state velocity distribution function of species $s$. 
The instabilities can, but do not always, feed back to cause observable changes in the steady-state properties. 
Examples from plasma with single or multiple ion species are considered below. 

\subsubsection{Plasmas with one ion species\label{sec:1is}} 

In a plasma with one ion species, the presheath electric field causes the ions to drift toward the boundary until they reach a flow speed near the sound speed at the sheath edge ($V_i \lesssim c_s$ in the presheath). 
Meanwhile, the sheath causes some depletion in the tail of the electron distribution function. 
This generates a net flux of electrons to the boundary, but the peak of the electron distribution function is not shifted significantly in comparison to the background plasma. 
This situation can be unstable to ion-acoustic instabilities if the ratio of electron-to-ion temperature is sufficiently high ($T_e/T_i \gg 1$) and the neutral pressure is sufficiently low that collisions do not damp the excited waves. 
These conditions are often met in low-temperature plasmas. 

\begin{figure}
\begin{center}
\includegraphics[width=7.5cm]{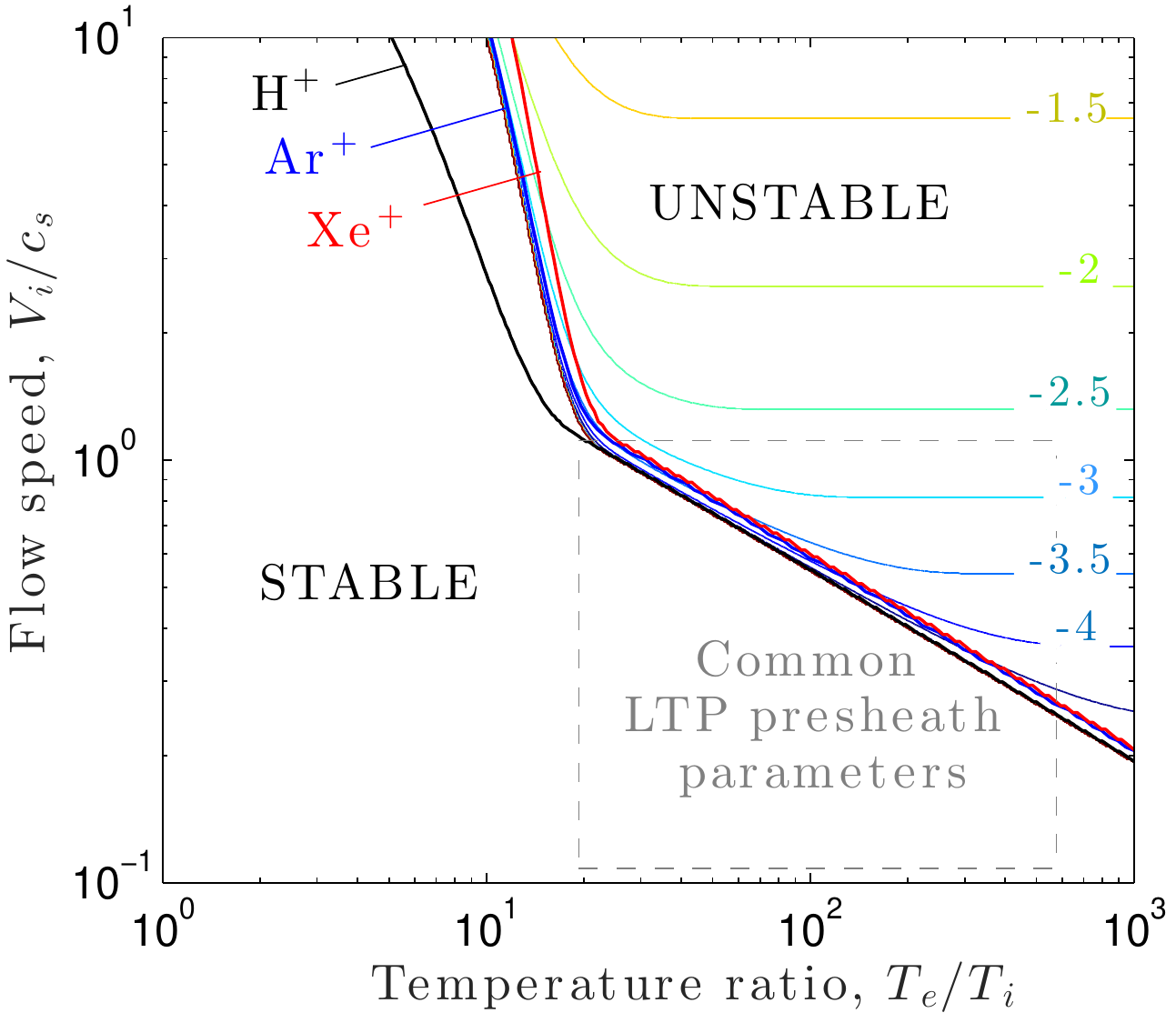}\\
\includegraphics[width=7.5cm]{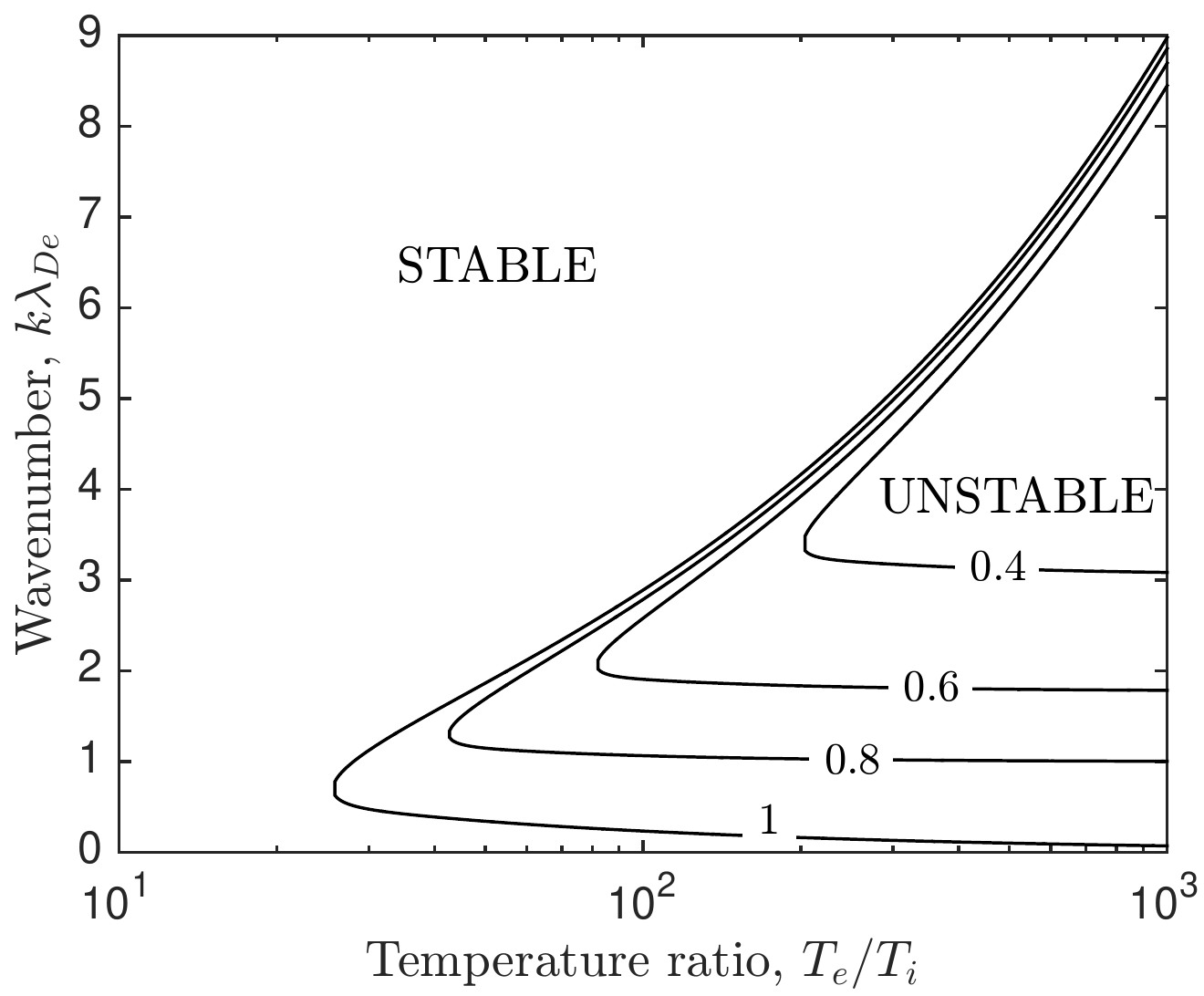} 
\caption{(top) Ion-acoustic stability boundaries for H$^+$ (black), Ar$^+$ (blue) and Xe$^+$ (red) ions. 
Contours with number labels show $\textrm{log}_{10}(\gamma_{\max}/\omega_{pi})$  for the H$^+$ plasma computed from equation~(\ref{eq:gama}). 
(bottom) Stability boundaries for the ion-acoustic instability in a H$^+$ plasma for four values of the ion drift speed: $V_i/c_s$ = 0.4, 0.6, 0.8 and 1.
 Figure reprinted from reference~\cite{baal:16}.}
\label{fg:ia_stability}
\end{center}
\end{figure} 
 
Figure~\ref{fg:ia_stability} shows the ion-acoustic instability boundaries for common noble gas plasmas as a function of the ion flow speed and the temperature ratio, as well as the stability bounds in terms of wavenumber. 
These stability boundaries were obtained from equation~(\ref{eq:ephat1}) assuming that ions and electrons both have Maxwellian distribution functions, but with a flow shift indicated by the ion flow speed. 
The solutions were computed using the Penrose criterion~\cite{penr:60}, as described in~\cite{baal:16}. 
At temperature ratios common to the presheath of low-temperature plasmas, ion-acoustic instabilities are expected even though the ion flow speed is subsonic. 
The wavelengths associated with these instabilities are predominately on the Debye length scale and shorter. 
Ion-acoustic waves convect in the direction of the ion flow. 
They are thus expected to be excited in the presheath, and grow in both amplitude and growth rate as they convect toward the sheath. 

The standard approximation of the ion-acoustic dispersion relation has a real component 
\begin{equation}
\label{eq:ia_real}
\omega_r = \vc{k} \cdot \vc{V}_i - \frac{kc_s}{\sqrt{1 + k^2\lambda_{De}^2}}
\end{equation}
and a growth rate 
\begin{eqnarray}
\nonumber
\gamma &=& - \frac{k c_s \sqrt{\pi/8}}{(1+k^2\lambda_{De}^2)^{2}} \biggl\lbrace \biggl( \frac{T_e}{T_i} \biggr)^{3/2} \exp \biggl[ - \frac{T_e/T_i}{2 (1 + k^2 \lambda_{De}^2)} \biggr] \label{eq:gama} \\ 
& + & \sqrt{ \frac{m_e}{m_i}} \biggl( 1 - \frac{V_i}{c_s} \sqrt{1 + k^2 \lambda_{De}^2} \biggr) \biggr\rbrace  .
\end{eqnarray}
The growth rate predicted by equation~(\ref{eq:gama}) is shown as contours in figure~\ref{fg:ia_stability}. 
This shows that the growth rate is a small fraction of the ion plasma frequency (typically $~\sim 10^{-3}\omega_{pi}$) over the range of conditions relevant to the presheath. 
Although the growth rate is small compared to the plasma frequency, the wavelength is much smaller than the presheath length ($\lambda_\D \ll \lambda_{\tiny \textrm{in}}$), so the excited waves can grow over several e-folding distances before reaching the sheath edge. 
It has been proposed that these excitations can cause wave-particle scattering that effectively enhances the Coulomb collision rate in the plasma-boundary transition region~\cite{baal:08,baal:10}. 
The increased effective collision rate can feed back to influence aspects of the ion and electron velocity distribution functions near the sheath edge. 

Often, the rate of ionization and charge-exchange collisions is sufficiently rapid in the presheath that the ion velocity distribution function (IVDF) is not simply a flowing Maxwellian, but also consists of a ``slow tail'' associated with ions born within the presheath~\cite{clai:06,bach:95,manf:04,bilo:10,pige:19}. 
This tail is an important part of the IVDF in many models of the plasma-boundary transition, including in the classical Tonks-Langmuir model~\cite{sher:00,tonk:29,robe:09}. 
Its presence can influence applications, such as the aspect ratio achieved in reactive ion etching of semiconductors~\cite{gott:93} because slow ions do not penetrate as deeply inside the trench. 
It has been proposed that the increased collisions associated with ion-acoustic instabilities may act to ``thermalize'' the ion distribution function as it progresses towards the sheath~\cite{baal:11c}. 
This causes the distribution to approach a flowing Maxwellian and the ``slow tail'' feature to be reduced. 
An important aspect of this theory is that the ion-acoustic instabilities remain in a linear growth regime from the location of their excitation in the presheath until they are lost to the wall along with the ions. 
In this regime, a quasilinear kinetic equation has been developed that describes the resultant wave-particle scattering~\cite{baal:08,baal:10}. 
The collision operator in this regime can be thought of as occurring via ``dielectrically dressed'' (or quasiparticle) Coulomb collisions, for which the dielectric dressing includes the possibility of linear wave growth of the potential associated with the discrete particles. 
A consequence of this is that the associated collision operator obeys the Boltzmann H-theorem, and that the plasma evolves to a Maxwellian distribution due to the scattering of particles with the linear waves~\cite{baal:10}. 

The proposal that instabilities can thermalize the IVDF has recently been tested experimentally using LIF by Yip \etal~\cite{yip:15c}. 
These experiments were conducted in a low-pressure ($p=0.1-0.3$ mTorr) low temperature ($T_e = 1-2.5$ eV) plasma. 
They observed that at sufficiently low neutral pressure, the IVDF was well approximated by a Maxwellian at the entrance to the presheath, gained a non-Maxwellian tail in the mid presheath, then became more Maxwellian near the sheath edge. 
The re-thermalization near the sheath edge was only observed at sufficiently low neutral pressure that ion-neutral collisions could not damp the ion-acoustic instabilities. 
Measurements of the critical pressure necessary to damp the instabilities were found to correspond well with the observed pressure threshold for the re-thermalization effect. 
Each of these predictions was found to be consistent with the model of reference~\cite{baal:11c}. 
Recent Vlasov simulations have also pointed out that ion acceleration by the sheath electric field alone leads to the appearance of a ``collisionless thermalization'' effect that is akin to velocity bunching in charged particle beams~\cite{coul:15}. 
They conclude that although wave-particle collisions are likely responsible for much of the experimentally observed thermalization, the collisionless mechanism also plays a role in understanding the observations. 

The Tonks-Langmuir model is the seminal kinetic theory for the IVDF in the plasma-boundary transition~\cite{tonk:29}. 
It explicitly models a low-speed tail associated with ions born in the presheath. 
It has motivated many subsequent generalizations and extensions~\cite{harr:59,caru:62,self:63}, including ion source models~\cite{proc:90,emme:80,biss:87,sche:88}, finite source temperature~\cite{robe:09,sher:01,tskh:14}, asymmetric plasmas~\cite{vand:91}, collisional plasmas~\cite{riem:81}, extended electron models~\cite{sato:92,gody:95} and electronegative discharges~\cite{sher:99}. 
Since the Tonks-Langmuir model is a steady-state model, it does not consider the stability of the plasma, but one may question if a time-dependent generalization of the model (such as proposed in~\cite{robe:09,sher:01}) is stable. 

A recent numerical solution of Sheridan's time-dependent generalization~\cite{sher:01} has shown that, in fact, the Tonks-Langmuir model is unstable~\cite{sher:17}. 
However, the nature of the instability is different than the classical ion-acoustic instability. 
Electrons are considered adiabatic, being modeled solely via the Boltzmann density relation in this model $n_e = n_{eo} \exp(e\phi/k_BT_e)$, so the inverse Landau damping mechanism responsible for the ion-acoustic instability is not accessible. 
The instabilities are also observed to have a much lower frequency than the ion-acoustic instability ($\approx 0.1 \omega_{pi}$) and a longer wavelength than what would be the most unstable ion-acoustic mode (several $\lambda_\D$)~\cite{sher:17}. 
This work showed that the instability was consistent with a type of instability first predicted by Fried \etal~\cite{frie:60} in which the instability derives energy directly from the equilibrium electric field. 
Of course, in a real experiment electrons have a non-adiabatic response, and the usual ion-acoustic instability is accessible. 
It is still an open question if this low-frequency instability can exist in nature, or if it is particular to this mathematical model~\cite{sher:17}. 
Numerical models including electron dynamics do model ion-acoustic instabilities~\cite{kosh:15}. 

\begin{figure}
\begin{center}
\includegraphics[width=7.5cm]{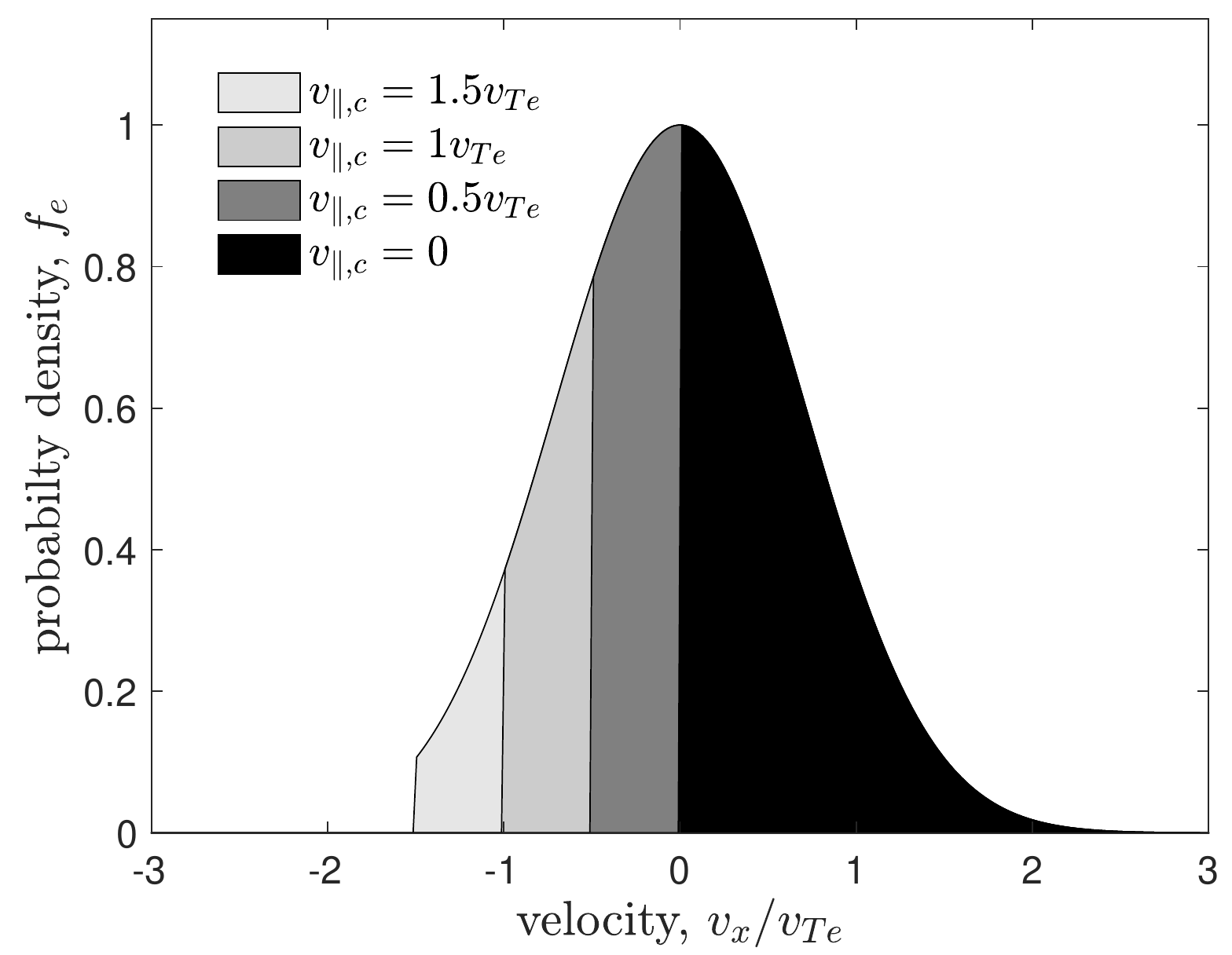}
\caption{Plot of the truncated electron velocity distribution function from equation~(\ref{eq:fe_trun}) for four values of the critical velocity. Darker colors overlap the distribution function associated with lighter colors.}
\label{fg:evdf_draw}
\end{center}
\end{figure}

It has been predicted that excitation of ion-acoustic instabilities in the presheath may scatter electrons in addition to ions, and that the enhanced electron scattering (in comparison to the Coulomb collision level) can lead to a rapid thermalization of the electron distribution~\cite{baal:09a}. 
Measurements of anomalously fast thermalization of electrons near plasma boundaries date to the earliest days of plasma physics research~\cite{lang:25,ditt:26}. 
Specifically, at a distance much smaller than the electron collision mean free path from the sheath, one would expect the EVDF in the direction perpendicular to the boundary to be devoid of electrons in the region of phase space corresponding to the population that is lost to the wall; i.e., the EVDF would have a truncated Maxwellian of the form illustrated in figure~\ref{fg:evdf_draw}. 
Instead, the EVDF is often measured to have some tail population (even if it is not a full Maxwellian) at a distance from the boundary that is sufficiently short that it cannot be explained by standard Coulomb collisions alone. 
This observation has come to be known as Langmuir's paradox~\cite{gabo:55,gody:15,gorg:98,tsen:03,tsen:09}. 
In a wave-particle scattering model, such as quasilinear theory, the largest effect on scattering occurs for particles that have velocities resonant with the phase-velocity of the excited wave~\cite{vede:62}. 
Since $v_{Te}/c_s \simeq \sqrt{m_i/m_e} \gg 1$, the ion-acoustic phase-speed is much slower than the speed associated with tail electrons, so one expects much less scattering of high energy electrons than low energy electrons. 
Nevertheless, applying the kinetic theory described above~\cite{baal:08,baal:10} shows that the effective Coulomb collision rate decreases as $v^{-3}$ from the phase-speed of the wave, and that even accounting for this decay, ion-acoustic instabilities are expected to significantly enhance electron scattering on the tail of the distribution~\cite{baal:09a}. 
This proposal remains untested experimentally, however. 
No direct measurement of ion-acoustic instabilities in the presheath has yet been reported, and it has also been suggested that the sensitivity of the early EVDF measurements (which are made using Langmuir probes) were not sufficient to prove that there is a paradox~\cite{gody:15,gody:11}.

\subsubsection{Plasma with multiple ion species~\label{sec:is_mis}} 

If the plasma contains multiple species of singly charged ions with different masses, each species will be accelerated to a different speed as it traverses the presheath electric field. 
The speed that each species obtains by the time it reaches the sheath edge is constrained by the sheath criterion from equation~(\ref{eq:gen_bohm}).
As in the single species case, the kinetic terms in parenthesis are expected to be small if the ion sheath potential drop is larger than the electron temperature. 
This leads to a generalization of the Bohm criterion for multiple ion species~\cite{riem:95,cook:80,bene:96,vale:96}  
\begin{equation}
\frac{n_1}{n_e} \frac{c_{s1}^2}{V_1^2} + \frac{n_2}{n_e} \frac{c_{s2}^2}{V_2^2} \leq 1 . \label{eq:bohm2}
\end{equation}
As in the single species Bohm criterion, equality is expected to hold in this condition~\cite{hers:11}. 
Even so, equation~(\ref{eq:bohm2}) alone does not uniquely specify the speed of each ion species at the sheath edge. 
Here, $c_{s,i} \equiv \sqrt{T_e/m_i}$ is the sound speed associated with species $i$. 

It is often expected that the mean free path for Coulomb collisions between the ion species is much longer than the presheath length scale. 
In this case, the force balance for each species can be analyzed from equation~(\ref{eq:fb}) neglecting the Coulomb contribution to the friction force, analogously to what was done for a single species plasma in section~\ref{sec:is_standard}. 
If the collision rate between each ion species and neutrals are not dramatically different, it is expected that the presheath potential drop $\Delta \phi_{\tiny \textrm{ps}}$ imparts the same kinetic energy to each species $\frac{1}{2}m_1 V_1^2 = \frac{1}{2}m_2 V_2^2$~\cite{fran:00}. 
Using this in equation~(\ref{eq:bohm2}) leads to the prediction that each species leaves the ion sheath at its individual sound speed: $V_1 = c_{s1}$ and $V_2 = c_{s2}$. 
This is the commonly quoted expectation for a multiple-ion-species plasma~\cite{lieb:05}

Recent experiments using LIF measurements of the IVDFs throughout the presheath revealed the surprising result that ions often reach the sheath edge with a speed much closer to a common speed than is predicted by the individual sound speed solution~\cite{hers:05,hers:05b,seve:03,wang:06,lee:07,oksu:08}. 
If one considers the limit that the ions are strongly collisionally coupled with one another ($V_1=V_2$), equation~(\ref{eq:bohm2}) predicts that each reaches the sheath edge at the system sound speed~\cite{lee:07c}
\begin{equation}
c_s = \sqrt{\frac{n_1}{n_e} c_{s1}^2 + \frac{n_2}{n_e} c_{s2}^2 } ,
\end{equation} 
which is close to the measured values in an Ar-Xe plasma from~\cite{lee:07}. 
However, this does not explain why the ions are collisionally coupled when Coulomb collisions are expected to be infrequent. 

\begin{figure}
\begin{center}
\includegraphics[width=7.5cm]{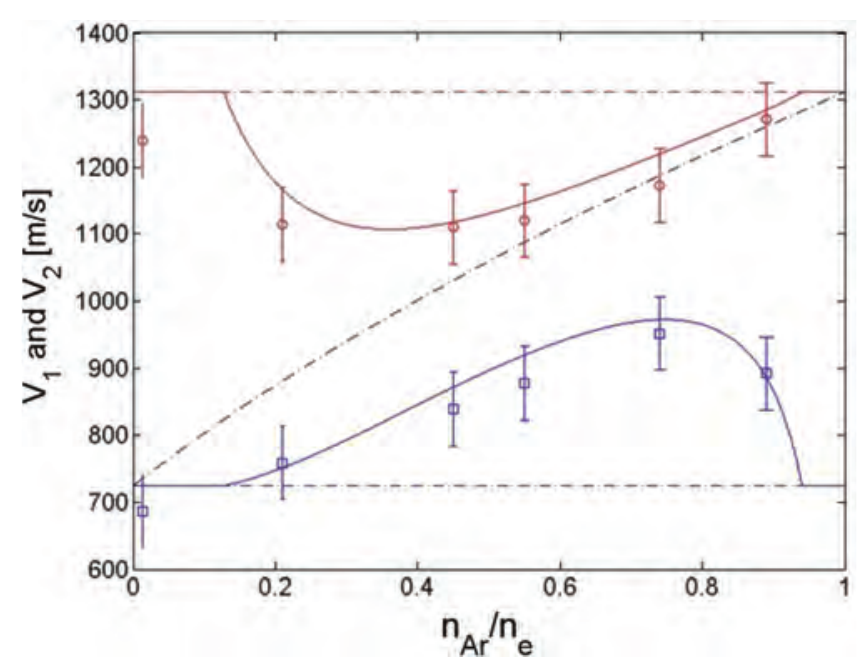}
\caption{Experimental measurement (using LIF) of the speed of argon ions (circles) and xenon ions (squares) at the sheath edge, as a function of the argon ion concentration. The electron temperature in the plasma was $T_e = 0.68$ eV. 
Solid lines show the theoretical prediction from equation~(\ref{eq:dvc}), the top and bottom lines indicated the individual sound speeds of argon and xenon and the black dash-dotted line indicates the system sound speed. 
 Figure reprinted from reference~\cite{hers:11}.}
\label{fg:ar_xe_speeds}
\end{center}
\end{figure} 

An explanation for these measurements has since been established~\cite{baal:09b,baal:11,yip:10,hers:11,baal:15a,zhan:12}. 
If the difference between the flow speed of each ion species exceeds a threshold value $|V_1 - V_2| \geq \Delta V_c$, ion-ion two-stream instabilities will be excited. 
Quickly after onset, increased scattering due to these instabilities rapidly increases the friction force between ion species so that the relative drift cannot significantly exceed the threshold condition for instability onset~\cite{baal:09b,baal:11,baal:15a}. 
This leads to the prediction that 
\begin{equation}
|V_1 - V_2| = \min \lbrace \Delta V_c, |c_{s1} - c_{s2}| \rbrace \label{eq:dvc}
\end{equation}
at the sheath edge. 
Equation (\ref{eq:dvc}) along with the Bohm criterion from equation (\ref{eq:bohm2}) combine to determine the speed of each ion species. 
The critical relative drift for instability onset can be predicted by solving for the dispersion relation from equation~(\ref{eq:ephat1}). 
Assuming that the ion species have flow-shifted Maxwellian distribution functions and that the wave phase speed of interest is much smaller than the electron thermal speed, equation~(\ref{eq:ephat1}) can be written 
\begin{equation}
\hat{\varepsilon} = 1 + \frac{1}{k^2 \lambda_{De}^2} \biggl[ 1 - \frac{z_1^2}{2} \frac{T_e}{T_1} \frac{n_1}{n_e} Z^\prime (\xi_1) - \frac{z_2^2}{2} \frac{T_e}{T_2} \frac{n_2}{n_e} Z^\prime (\xi_2) \biggr]  \label{eq:ephat}
\end{equation}
where $\xi_1 = \hat{\vc{k}} \cdot \Delta \vc{V} (\Omega -1/2)/v_{T1}$, $\xi_2 = \hat{\vc{k}} \cdot \Delta \vc{V} (\Omega + 1/2)/v_{T2}$, and $z_i$ is the ionic charge. The parameter $\Omega$ has been defined by the substitution 
\begin{equation}
\omega = \frac{1}{2} \vc{k} \cdot (\vc{V}_1 + \vc{V}_2) + \vc{k} \cdot \Delta \vc{V} \Omega . \label{eq:osub}
\end{equation}
The critical speed $\Delta V_c$ can be obtained by numerically solving equation~(\ref{eq:ephat}) for the dispersion relation as a function of relative ion drift (at fixed plasma parameters) and determining the lowest value for which the growth rate of the most unstable mode becomes positive~\cite{baal:15a}. 

This theory has been tested experimentally~\cite{yip:10,hers:11}. 
Figure~\ref{fg:ar_xe_speeds} shows a comparison between the theoretical predictions of this model and LIF measurements in an Ar-Xe discharge as a function of the argon ion concentration. 
When the concentration is in either the dilute or saturated limit, the speed of each species approaches the traditional expectation of individual sound speeds. 
No instability is predicted in these limits because $\Delta V_c > |c_{s1} - c_{s2}|$. 
In contrast, the speed of each species is found to approach the system sound speed at intermediate concentration, and the observed speeds to agree well with the model predictions. 
In addition to this evidence provided by the ion speed measurements, the presence of two-stream instabilities have been directly measured using Langmuir probes~\cite{hers:05b,hers:11}.  
The observed modification of ion speeds at the sheath edge has important implications for plasma boundary interactions, as well as global plasma models~\cite{kim:15b}. 

The theory and experiments have been extended to plasmas containing three ion species~\cite{yip:16,seve:17}. 
These consist of argon, xenon, krypton mixtures~\cite{yip:16} as well as argon, xenon, neon mixtures~\cite{seve:17}. 
Initial tests indicate that the theory can be extended to three (or more) ion species, but the analysis becomes more complicated because there are more possible unstable modes to track. 
Here, two-stream instabilities between each possible combination of species must be considered and must include the presence of the third species. 
If any combination causes instability, it leads to an enhanced friction force between each species. 

\begin{figure}
\begin{center}
\includegraphics[width=7.5cm]{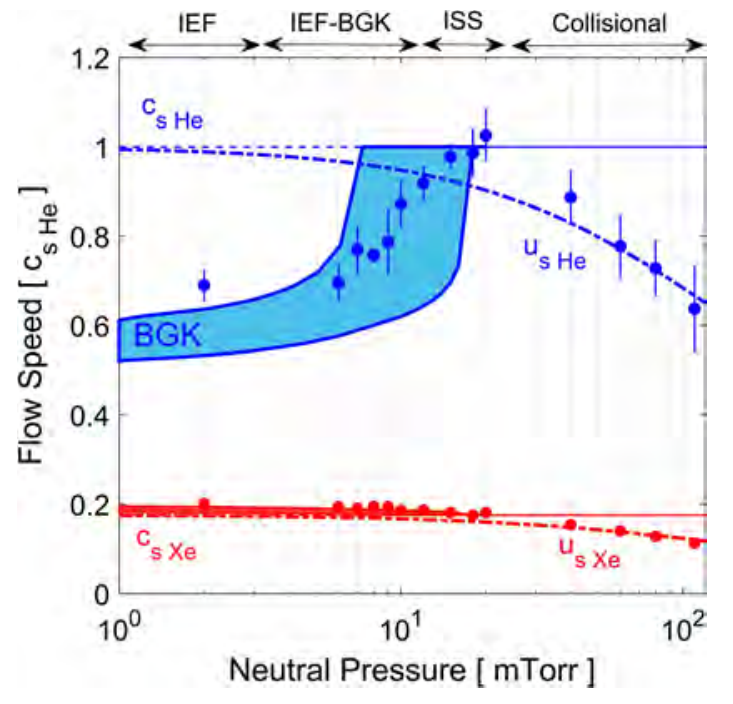} 
\caption{Ion speed at the sheath edge of a He-Xe discharge from PIC simulations (circles) and theory predictions (shaded regions). 
The extrema of the shaded regions are from the theory predictions at the extrema of the plasma parameters encountered in the simulations: $T_e=4.0-3.8$ eV, $T_i = 0.40-0.45$ eV, $n_{\tiny \textrm{He}}/n_e = 0.035-0.07$, and $n_e = 1\times 10^{10}$ cm$^{-3}$. 
Dashed-dotted lines show the prediction of the collisional Bohm criterion from equation~(\ref{eq:gody_cb}). 
 Figure reprinted from reference~\cite{adri:17}.}
\label{fg:speeds_adrian}
\end{center}
\end{figure} 

This model has also been tested using PIC simulations~\cite{baal:15a,adri:17,gudm:11}. 
The first simulations appeared to contradict the model because the ion speeds were observed to enter the ion sheath with their individual sound speeds~\cite{gudm:11}, even though the prediction for $\Delta V_c$ of an early analytic model predicted instability~\cite{baal:11}. 
It was subsequently questioned whether PIC simulations are capable of simulating the predicted instability-enhanced ion-ion friction force~\cite{hers:12}. 
However, later analysis showed that the discrepancy was actually due to inaccuracies of the early analytic approximation for $\Delta V_c$, which didn't apply at the simulated plasma conditions~\cite{baal:15a}. 
Direct numerical solutions of equation~(\ref{eq:ephat}) showed that the full solutions of linear theory actually predicts stability at the conditions of the earlier simulations. 
Further PIC simulations were conducted for conditions where instability was predicted by linear theory, and the simulations observed both the presence of the instabilities and that the simulated ion speeds at the sheath edge agreed with the theoretical predictions (see figure 8 of \cite{baal:15a}). 
Furthermore, the simulations enabled one to directly simulate the instability-enhanced friction force, and associate the merging of ion speeds with this force. 
An explanation for why PIC simulations are able to capture the instability-enhanced friction force has recently been provided~\cite{sche:19b}. 

The experiments and simulations described above pertain to low-pressure discharges (around 1 mTorr or less), but many plasmas of interest operate at higher neutral pressures. 
At sufficiently high neutral pressure, one would expect that ion-neutral collisions cause the ion-ion two-stream instability to damp, which will alter the predicted threshold condition ($\Delta V_c$). 
Recent work has extended the above analysis to account for ion-neutral collisions by including a BGK collision model in the linear dielectric function~\cite{adri:17,kim:17}.  
Experiments revealed good agreement with the predictions of the extended theory~\cite{kim:17}. 
The extended model has also been shown to agree well with PIC simulations~\cite{adri:17}. 
An example is shown in figure~\ref{fg:speeds_adrian}, which shows the speed of He and Xe ions in a mixture as a function of the neutral pressure. 
At low pressure, ion-neutral collisions are sufficiently rare that they can be ignored, but at pressures of a few mTorr (for these plasma conditions), collisions lead to a predicted increase in the relative ion drift at the sheath edge. 
For sufficiently high neutral pressure, the instability is completely absent and the ions enter the sheath with their individual sounds speeds. 
This figure also demonstrates that at even higher pressure the Bohm criterion itself [equation (\ref{eq:bohm2})] breaks down.
The figure shows a comparison with the collisionally modified Bohm criterion proposed by Godyak~\cite{gody:90}
\begin{equation}
\label{eq:gody_cb} 
V_{i} = \frac{c_{si}}{\sqrt{1 + \frac{\pi}{2} \frac{\lambda_{De,s}}{\lambda_{\tiny \textrm{in}}}}} .
\end{equation} 
Although this was developed in the context of a single ion species plasma, it accurately models the speed of each species in this mixture at high pressure. 
This is likely because the two ion species are collisionally decoupled from one another at this pressure, so the assumptions of the model apply to each species individually. 
Describing how collisions modify the Bohm criterion remains a topic of continuing research~\cite{fran:03b,brin:11}.

\subsection{Weak ion sheaths and the transition to electron sheath\label{sec:transition}} 

The standard ion sheath properties summarized in section~\ref{sec:is_standard} pertain to sheaths with a potential drop that is at least as large as the electron temperature. 
This is a common situation because it applies to floating boundaries and electrodes biased negatively with respect to the wall.  
However, it is often not satisfied for positively biased electrodes~\cite{lunt:08}. 
Even if the electrode is large enough that current balance demands that the plasma potential is more positive than it (ion sheath regime), the sheath potential drop may be small. 
In fact, figure~\ref{fg:area} shows that the electrode area must be much larger than the transition area before the associated ion sheath potential drop is significantly larger than the electron temperature. 
Kinetic effects influence the plasma-boundary transition for these ``shallow sheaths''. 

Conservation equations and Poisson's equation [(\ref{eq:continuity})--(\ref{eq:poisson})] can still be useful to analyze this situation, but additional information must be provided to close the equations based on non-local kinetic theory arguments. 
For instance, if the sheath potential drop is shallow, a large fraction of the electron distribution will be lost through the ion sheath, creating a large absence of particles in the EVDF corresponding to those electrons that reach the wall. 
Since the electron collision mean free path is often expected to be much larger than the scale of the plasma-boundary transition layer, electrons can be expected to have a truncated Maxwellian of the form~\cite{baal:11c,loiz:11}
\begin{equation}
\label{eq:fe_trun}
f_e = \frac{\bar{n}_e}{\pi^{3/2} \bar{v}_{Te}^3} e^{-v^2/\bar{v}_{Te}^2} H(v_z + v_{\parallel,c})
\end{equation}
where $v_{\parallel,c} = - \sqrt{2e(|\phi_\E | + \phi)/m_e}$ is the speed associated with the cutoff velocity at a potential $|\phi_\E| + \phi$ from the electrode. 
This distribution is illustrated schematically in figure~\ref{fg:evdf_draw}.
Here, $\bar{n}_e$, and $\bar{T}_e$ are parameters that characterize the distribution, whereas the density $n_e$ and temperature $T_e$ are defined via moments of the distribution function. 

\begin{figure}
\begin{center}
\includegraphics[width=7cm]{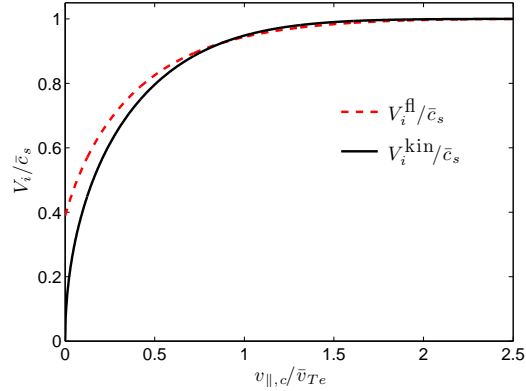}
\caption{Ion velocity in the sheath edge compared to $\bar{c}_s$ as a function of the cutoff velocity predicted by equation~(\ref{eq:g_bohm_trunc}) (black line). Also shown is the corresponding ``fluid'' result, which would be obtained by the same procedure, but neglecting the temperature and stress gradient terms in equation~(\ref{eq:gen_bohm}). Figure reprinted from reference~\cite{baal:11}.}
\label{fg:eBohm}
\end{center}
\end{figure}

The assumption that electrons are collisionless in the plasma boundary transition provides a closure via the expression for $v_{\parallel,c}(\phi)$. 
The plasma-boundary transition based on this model was analyzed in Refs.~\cite{baal:11c,loiz:11}. 
Expressions for the density, temperature, flow velocity, heat flux and stress tensor associated with this distribution (provided in~\cite{baal:11c,loiz:11}) can be used directly in equation~(\ref{eq:gen_bohm}) to derive a kinetic generalization of the Bohm criterion  
\begin{equation} 
\label{eq:g_bohm_trunc}
V_i \geq \bar{c}_s \biggl\lbrace 1 + \frac{\bar{v}_{Te}}{v_{\parallel,c}} \frac{\exp (-v_{\parallel,c}^2/\bar{v}_{Te}^2)}{\sqrt{\pi} [ 1+ \textrm{erf} (v_{\parallel,c}/\bar{v}_{Te})]} \biggr\rbrace^{-1/2} .
\end{equation} 
Here, $\bar{c}_s \equiv \sqrt{\bar{T}_e/m_i}$ and $\bar{v}_{Te} \equiv \sqrt{2 \bar{T}_e/m_e}$. 
The solution of equation~(\ref{eq:g_bohm_trunc}) is shown in figure~\ref{fg:eBohm}. 
This shows that the ion flow at the sheath edge is predicted to be subsonic when the ion sheath potential drop is small compared to the electron temperature. 
Also shown is the ``fluid'' solution obtained by following the same procedure, but excluding the terms in parenthesis in equation~(\ref{eq:gen_bohm}). 
This corresponds to what the standard Bohm criterion would predict if the electron flow velocity were included, and the electron density, flow velocity and temperature were interpreted via the appropriate moments of equation~(\ref{eq:fe_trun})~\cite{baal:11c}. 
The difference between the curves illustrates the importance of temperature and stress gradients for a sheath with a small potential drop.  

\begin{figure}
\begin{center}
\includegraphics[width=8.5cm]{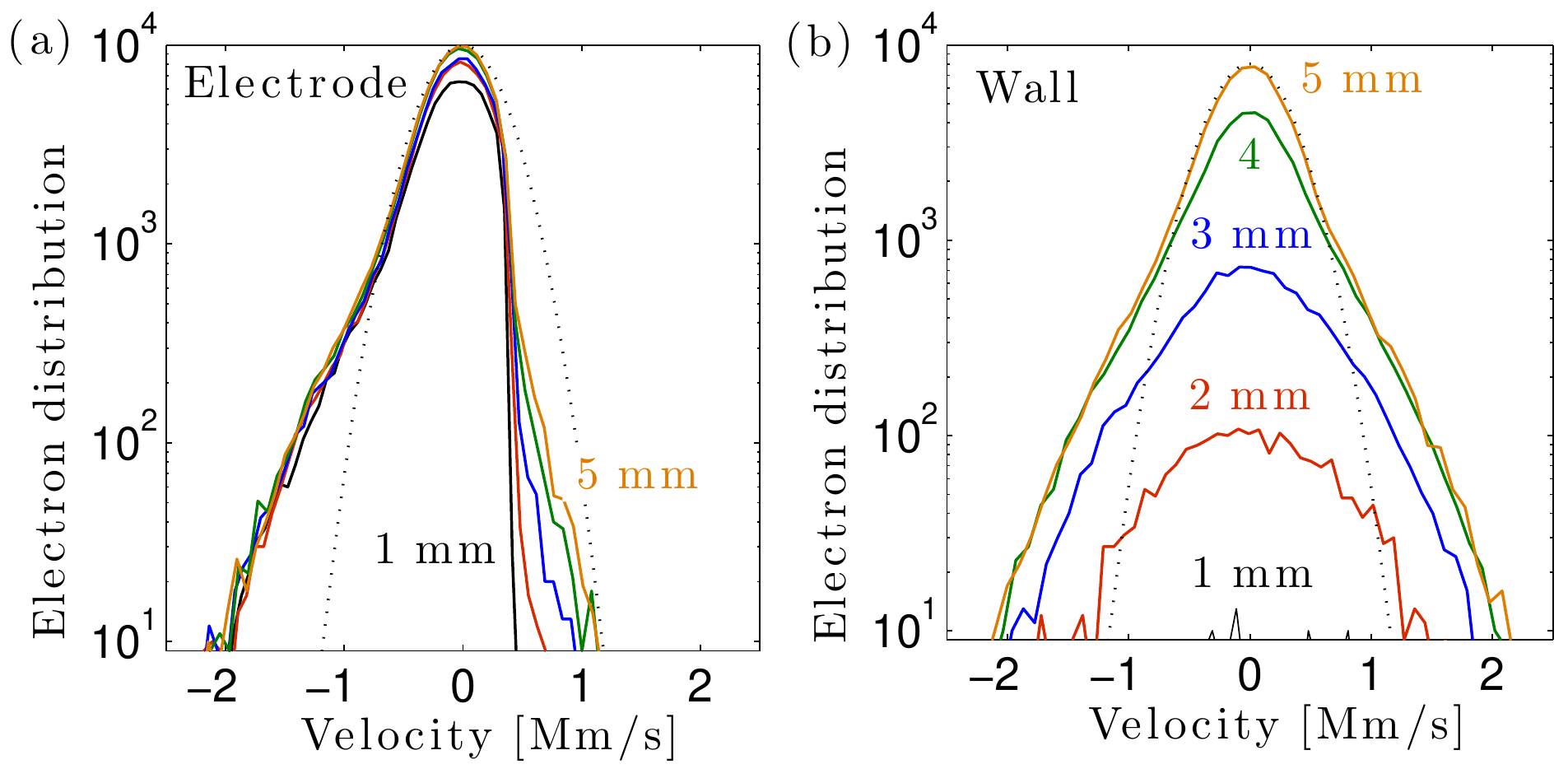}\\
\includegraphics[width=7cm]{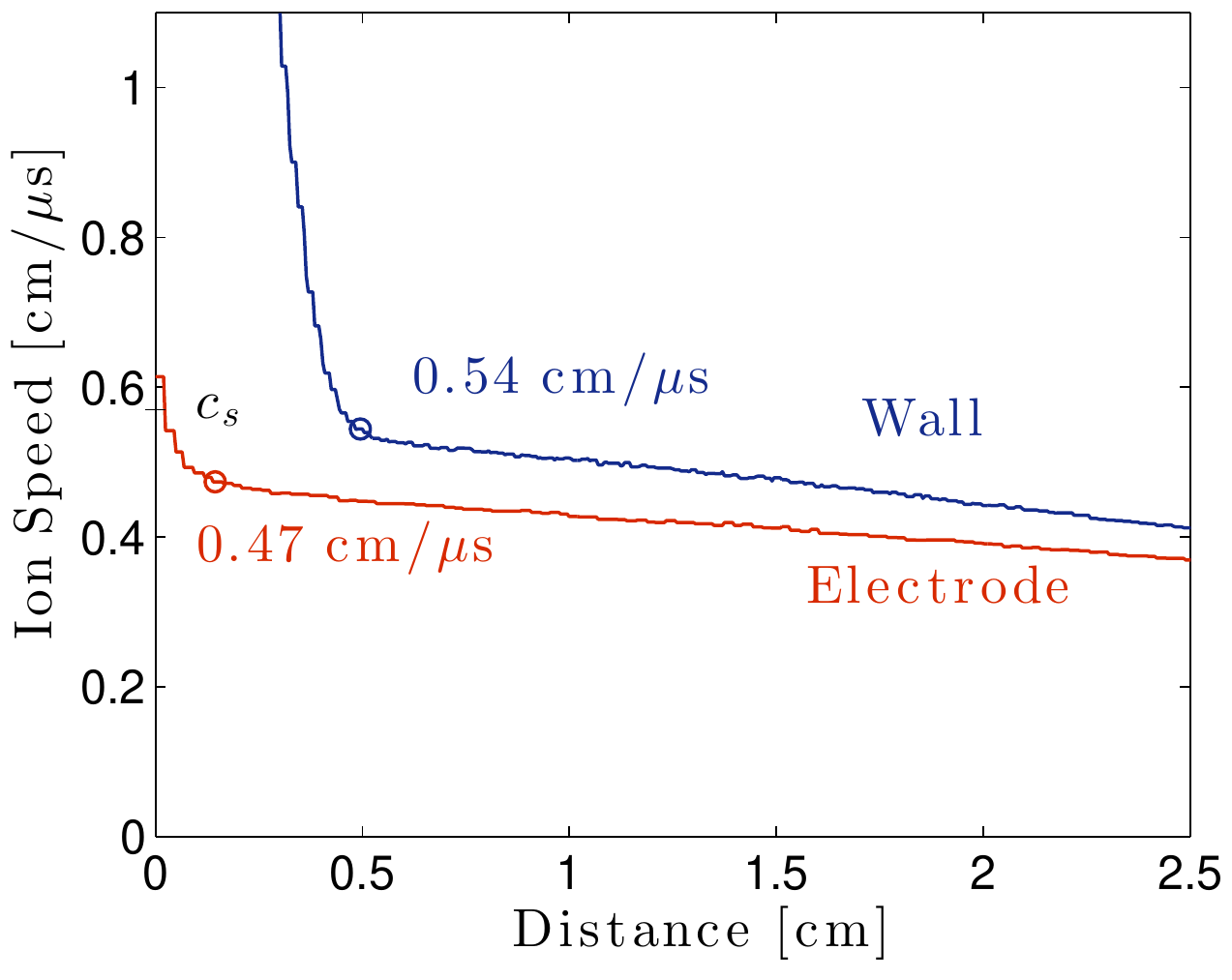}
\caption{(top) (a) Electron velocity distribution function in front of a biased electrode (0.5 V below the plasma potential), and (b) in front of a grounded wall nearby (20.5 V below the plasma potential). 
(bottom) The associated ion flow velocity profiles, with circles indicating the sheath edge location. Data from 2D PIC simulations. Figure reprinted with permission from Ref.~\cite{baal:15b}.}
\label{fg:k_bohm}
\end{center}
\end{figure}

Alternative approaches to a generalization of the Bohm criterion for arbitrary ion and electron velocity distribution functions have also been explored~\cite{riem:91,harr:59,riem:81}. 
The result is typically called the ``kinetic Bohm criterion''. 
For the model electron distribution function of equation~(\ref{eq:fe_trun}) it leads to the same prediction as equation~(\ref{eq:g_bohm_trunc}); see~\cite{baal:11c}. 
However, for other model distribution functions the two approaches lead to vastly different predictions~\cite{baal:11c,baal:15b,baal:12}. 
For example, the conventional kinetic Bohm criterion predicts that the low-speed part of the IVDF contributes disproportionately to the restriction on the total ion fluid speed at the sheath edge~\cite{riem:12}, and even diverges if any part of the IVDF corresponds to ions leaving the sheath~\cite{hall:62,fern:05}.  
This distinction with the above model has been thoroughly discussed in~\cite{baal:15b,baal:12,riem:12}, where the two models have been compared with experimental and simulation data. 
The main point of this discussion is that in real plasmas ionization and charge exchange causes a small population of ions to traverse from the sheath, across the sheath edge, into the plasma. 
This small population is physically insignificant, but it causes the conventional kinetic Bohm criterion to break down. 
Comparison with a generalization of the Tonks-Langmuir model~\cite{robe:09,sher:01} to include a warm (finite temperature) ion source clearly illustrates the point; the small population of ions exiting the sheath does not significantly influence the Bohm criterion~\cite{baal:15b,baal:12}. 
Recently, Tsankov and Czarnetzki extended the kinetic Bohm criterion to account for charge exchange collisions and ionization, revealing a new term that connects the fluid-moment and conventional kinetic pictures~\cite{tsan:17}. 
This work also shows consistency with measured IVDFs.

\begin{figure*}
\begin{center}
\includegraphics[width=16cm]{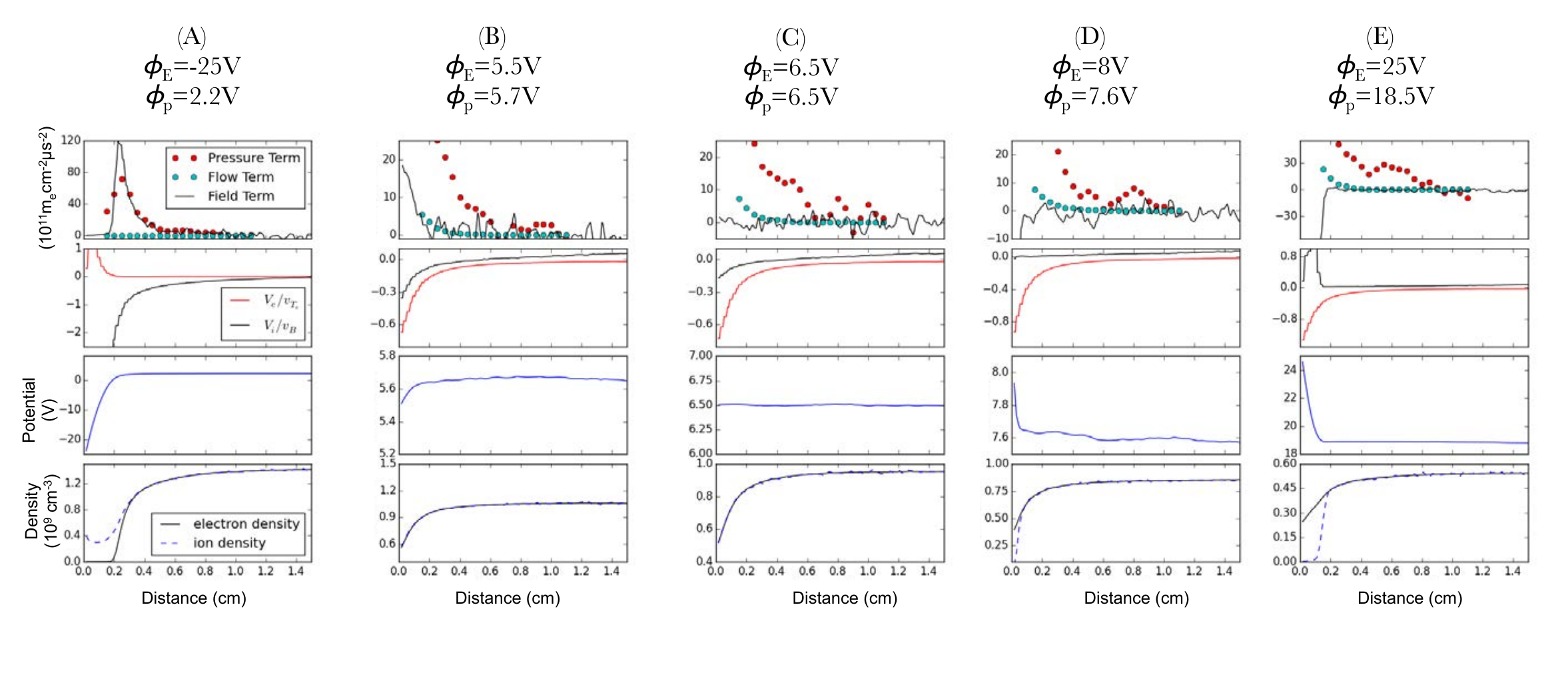}
\caption{Plasma parameter profiles in front of an electrode biased at a potential ($\phi_\E$) (a) far below, (b) just below, (c) equal to, (d) just above and (e) far above the plasma potential ($\phi_p$). The profiles shown include terms of the momentum balance (first row) [electric field $-neE_x$, flow $m_en_e V_{e,x}dV_{e,x}/dx$ and pressure $\nabla \cdot \mathcal{P}_e$], electron and ion flow (second row), plasma potential (third row) and electron and ion density (fourth row). Data is from 2D PIC simulations.  Figure reprinted from reference~\cite{sche:16}.}
\label{fg:transition}
\end{center}
\end{figure*}

The predictions of equation~(\ref{eq:g_bohm_trunc}) have also been tested using PIC simulations of biased electrodes~\cite{baal:15b,loiz:12}. 
In this case, the modification from the fluid Bohm criterion arises due to the non-local kinetic character of the electron distribution. 
Figure~\ref{fg:k_bohm} shows an example from 2D PIC simulations in which part of the boundary was a positively biased electrode~\cite{baal:15b}. 
The electrode size was sufficiently large that the plasma potential was higher than the electrode potential; as discussed in section~\ref{sec:structures}. 
The figure shows the EVDF at five locations in front of a biased electrode, in comparison with the grounded wall. 
The EVDF is significantly depleted beyond energies corresponding to the sheath potential drop (of approximately 0.5 V) in the case of the biased electrode; a feature that is not observed in front of the grounded wall. 
Equation~(\ref{eq:fe_trun}) provides an accurate representation of this truncated distribution function. 
The figure also shows that flow speed of ions at the sheath edge is sonic near the wall, but subsonic near the electrode. 
A comparison of the simulated speed was found to be consistent with the prediction of equation~(\ref{eq:g_bohm_trunc})~\cite{baal:15b}.

Using a similar 2D PIC simulation, a detailed study of the ion-to-electron sheath transition was carried out in Ref.~\cite{sche:16}. 
In this case, the biased electrode was placed interior to the simulation domain (rather than as part of the boundary) and it was sufficiently small that it could be biased above or below the plasma potential (due to the constraints of global current balance discussed in section~\ref{sec:structures}). 
Figure~\ref{fg:transition} shows the results, which indicate that the electrode could be biased essentially equal to the plasma potential, in which case the potential profile was observed to be flat. 
The associated ion flow is far below the sound speed in this case, and transitions to zero net flow as the electrode becomes slightly positive; in agreement with equation~(\ref{eq:g_bohm_trunc}). 
It is also observed that the net electron flow becomes comparable to the electron thermal speed when the sheath potential is slightly below the plasma potential. 
This is due to the expected truncation of the EVDF by absorption from the electrode. 
For a small electrode, the EVDF was observed to have a loss-cone type distribution due to ``shadowing'' by the electrode, rather than the truncated distribution shown in figure~\ref{fg:evdf_draw}, which is based on a 1D picture. 

The instabilities described in section~\ref{sec:is_i} would also be expected for weak ion sheaths near the transition to electron sheaths, but the details of the dispersion relation and stability boundaries are modified by the non-Maxwellian electron distribution function. 
If the projection of the electron distribution into the direction normal to the boundary is of the truncated form described by equation~(\ref{eq:fe_trun}), the linear dielectric response function can be written similarly to the standard form, but where the plasma dispersion function is replaced by the incomplete plasma dispersion function~\cite{baal:13,xie:13}. 
Reference~\cite{baal:13} showed that an electron distribution function with a depleted tail modifies both the Langmuir wave and ion-acoustic wave dispersion relations in non-trivial ways. 
It shifts the real frequency of the waves to lower frequencies, and reduces the magnitude of Landau damping. 
For the ion-acoustic instability, the linear growth rate is observed to increase when the electron distribution function is depleted, and the most unstable wavenumber is observed to shift to longer wavelengths.

\section{Electron Sheaths\label{sec:es}} 

\subsection{Steady-state properties}

\begin{figure}
\begin{center}
\includegraphics[width=7cm]{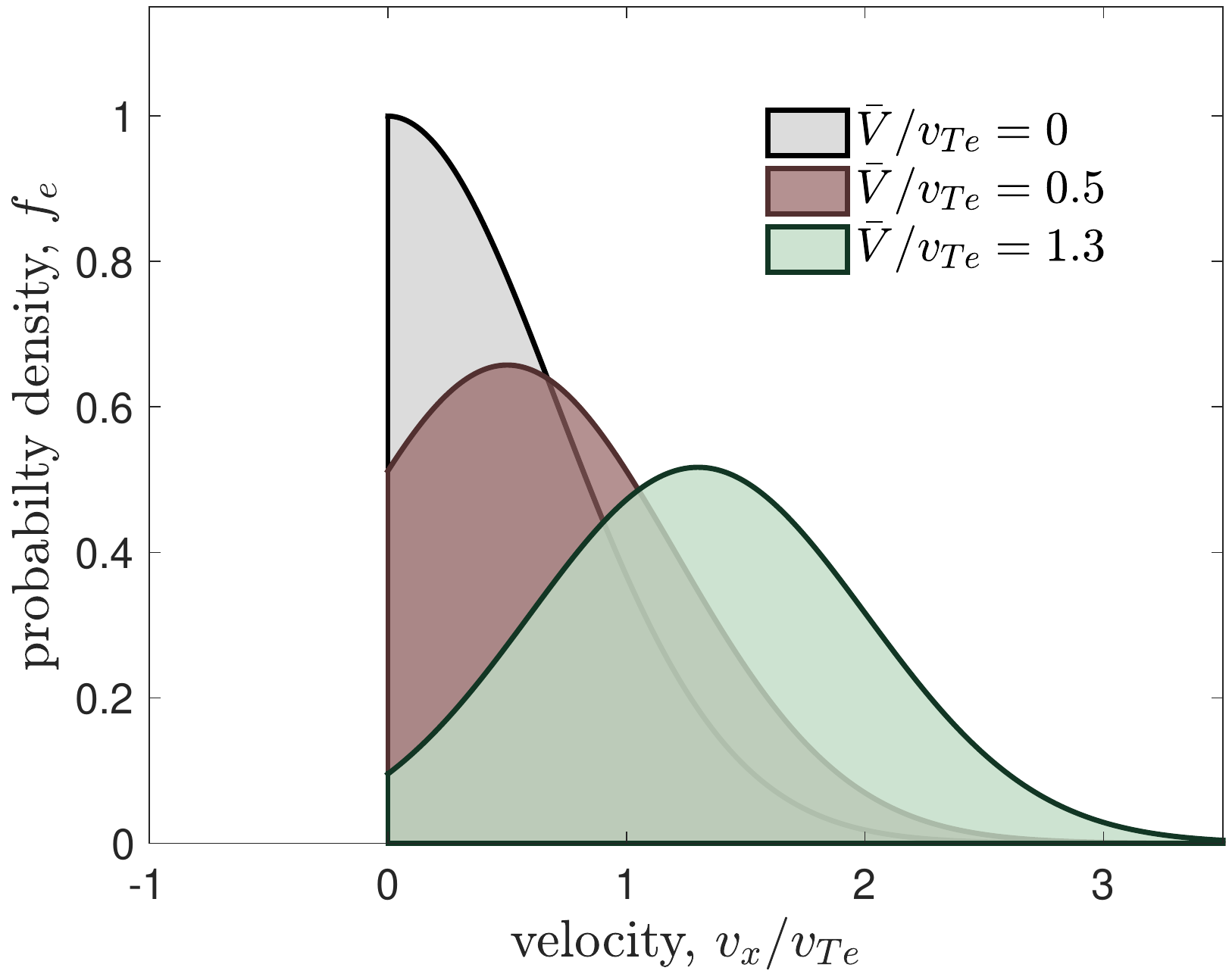}
\caption{Illustration of possible EVDF configurations near the edge of an electron sheath. Here, $\bar{V}/v_{Te}$ locates the peak of the distribution function. }
\label{fg:evdf_drift_fig}
\end{center}
\end{figure}

Electron sheaths are thin regions of negative space charge in which the electric field is directed from the electrode toward the plasma, as shown in figure~\ref{fg:type}~\cite{hers:05,baal:07,fred:07}. 
As discussed in section~\ref{sec:structures}, an electron sheath will form near a positively biased electrode if its effective surface area is small enough to satisfy equation~(\ref{eq:A_es}). 
A common example is the electron saturation regime of a Langmuir probe trace.  
Until recently, the description of electron sheaths (arising from the theory of Langmuir probes) was thought to be quite different from the description of ion sheaths. 
The difference stemmed from the assumption that the electrode collects the random thermal flux of electrons incident on the sheath edge. 
Correspondingly, the EVDF near the electron sheath edge was expected to be truncated, such as the half-Maxwellian depicted by the grey curve in figure~\ref{fg:evdf_drift_fig} and equation~(\ref{eq:fe_trun}) with $v_{\parallel,c}=0$. 
This is a natural expectation. 
The electron collision mean free path is typically orders of magnitude larger than the electron sheath thickness, so the problem may be expected to be one characterized by the collisionless process of ``effusion,'' in which the region of velocity phase-space corresponding to electrons that have escaped to the electrode is missing. 

The traditional expectation that the electrode collects the random thermal flux of electrons leads to different predictions for the electron sheath than an analogous ion sheath.
For instance, applying the general form of a Bohm criterion associated with equation~(\ref{eq:gen_bohm}) to the electron sheath, with the terms in parenthesis assumed negligible, would predict that the electron flow speed satisfies 
\begin{equation}
\label{eq:e_bohm}
V_e \ge v_{e\bb},
\end{equation}
where $v_{e\bb} = \sqrt{k_\bb (T_e + T_i) / m_e} \approx \sqrt{k_\bb T_e/m_e}$ is the electron-equivalent Bohm speed.
However, this is essentially satisfied by the electron flux associated with a truncated distribution. 
Taking equation~(\ref{eq:fe_trun}) with $v_{\parallel,c}=0$, the moment definitions give $V_e = \sqrt{2 k_\bb \bar{T}_e/(\pi m_e)}$ and $T_e = \bar{T}_e[1-2/(3\pi)]$, so $V_e \approx 0.9 \sqrt{k_\bb T_e/m_e}$. 
Perhaps for this reason, an electron presheath was not thought to be necessary to satisfy the equivalent Bohm condition~\cite{riem:91,chen:06}. 

Recently, the electron sheath was analyzed in more detail~\cite{sche:15}, indicating that it has more in common with ion sheaths than was previously thought.  
One surprising result is that the electron flux associated with the EVDF was observed to be primarily associated with a flow-shift, rather than a truncation~\cite{yee:17}. 
That is, the peak of the distribution was observed to be shifted, as depicted by the maroon or green curves in figure~\ref{fg:evdf_drift_fig}. 
This indicates that the electron behavior is more akin to a collisional diffusive process, rather than a collisionless effusive process. 
Diffusive processes generally rely on collisions to maintain a near-equilibrium configuration. 
A detailed description of how sufficient collisions can be established remains an open question, but two observations have been made that suggest a source of this collisionality.  
First, the presheath is expected to be a long region where the electron pressure gradient establishes a flow shift (as described below). 
The electron collision mean free path should be compared with this comparatively long presheath length scale, rather than the thin electron sheath.  
Second, the flow shift excites instabilities that can significantly increase the effective electron collision rate (as described in sections~\ref{sec:es_lf} and \ref{sec:es_hf}). 

\begin{figure}
    \centering
    \includegraphics[width=7cm]{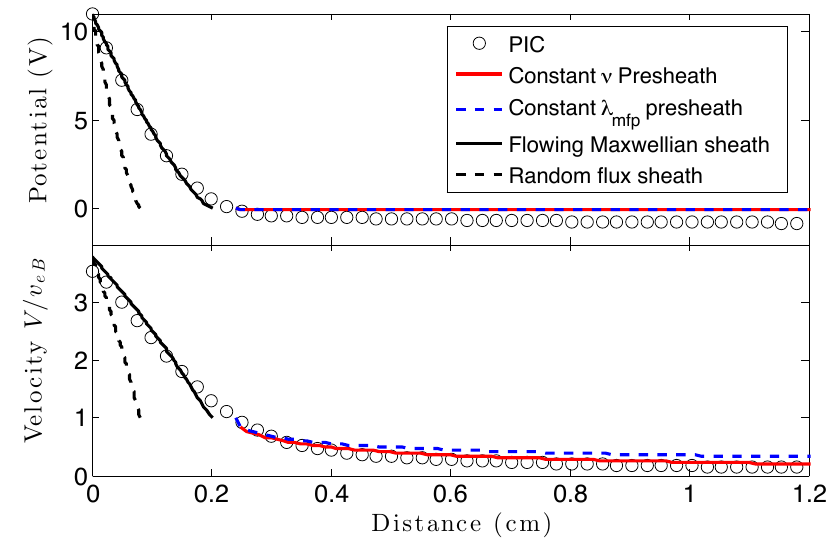}
    \caption{Electrostatic potential and velocity profile in front of a biased electrode. Circles show results of PIC simulations, and lines show predictions of the theoretical models.  Figure reprinted from reference~\cite{sche:15}.}
    \label{fg:es_profile}
\end{figure}

These observations suggest that the electron sheath can be described using a two-fluid analysis akin to the ion sheath, as described in section~\ref{sec:is}. 
Analogous to the ion sheath, the natural Debye scale of the electron sheath justifies a scale separation between collsionless sheath and weakly-collisional presheath. 
Assuming there are no ions in the electron sheath, and that the collisionless nature of electrons in this region implies that they traverse ballistically, $n_e(x) = J_o/[eV_e(x)]$ where $J_o$ is the electron current density at the sheath edge. 
Using this in Poisson's equation, multiplying by $d\phi/dx$ and integrating leads to a Child-Langmuir law 
\begin{equation}
\label{eq:cl_e}
    J_o = \frac{4}{9} \sqrt{\frac{2e\epsilon_0}{m_e}}
          \frac{\left[\phi_\E - \phi(x)\right]^{3/2}}{x^2}.
\end{equation}
This is analogous to equation~(\ref{eq:cl}), which was obtained for ion sheaths, but where the electron mass replaces the ion mass. 
Applying the electron Bohm criterion from equation~(\ref{eq:e_bohm}), we expect that the electron flux at the sheath edge is $J_o = en_{e,o} v_{e,\bb}$. 
Using this leads to an expression for the electron sheath thickness 
\begin{equation}
\label{eq:st_e}
\frac{l_s}{\lambda_{De}} = \frac{\sqrt{2}}{3} \biggl( \frac{2e\Delta \phi_s}{T_e} \biggr)^{3/4} .
\end{equation}
This is the same expression as was obtained for the ion sheath thickness in equation~(\ref{eq:st}). 
It is over twice as large as what is obtained based on the random flux model~\cite{sche:15}. 
Simulation results are consistent with equation~(\ref{eq:st_e}), as shown in figure~\ref{fg:es_profile}. 

The most important distinction between ion sheaths and electron sheaths arises in the presheath. 
Consider the force balance from equation~(\ref{eq:fb}). 
In the mobility-limited ion presheath, the electric field balances the ion inertia and ion friction terms, as described in section~\ref{sec:is_standard}. 
In this case, the ion pressure is negligible because $T_e \gg T_i$, so the Boltzmann electron density relation (assuming constant temperature) and quasineutrality relation imply $dp_i/dx = enE(T_i/T_e) \ll enE$. 
The electron presheath is different. 
Here, if ions are assumed to obey the Boltzmann density relation in the presheath $n_i = n_o \exp(-e\phi/T_i)$, then $dp_e/dx = enE(T_e/T_i) \gg enE$. 
Thus, one expects that the electron pressure gradient, rather than the electric field, is the dominant force driving the electron drift. 
This is a qualitative difference with the ion presheath: in an electron presheath the electric field is weak, but the pressure gradient is strong enough to drive a large electron drift. 
Figures~\ref{fg:transition} and \ref{fg:es_profile} confirm that the general behavior predicted by this simple analysis is observed in PIC simulations; see Refs.~\cite{sche:16,yee:17,sche:15} for details.

A modified mobility limited flow model for the electron presheath follows in an analogous fashion to that outlined in section~\ref{sec:is_standard} for the ion presheath. 
In this case, equation~(\ref{eq:Vi_E}) is replaced by 
\begin{equation}
\label{eq:mlf_es}
    V_e = -\mu_e E \left(1 - \frac{V_e^2}{v_{e\bb}^2}\right),
\end{equation}
where $\mu_e = e(1 + T_e/T_i) / ( m_e \nu_e)$ is the electron mobility. 
Here, one may distinguish the electron collision processes in terms of a momentum transfer rate, $\nu_c$, and an ionization rate, $\nu_S$: $\nu_e = \nu_c+2\nu_S$. 
The electron mobility, $\mu_e$, generally greatly exceeds the ion mobility in low temperature plasmas. 

Profiles for the electrostatic potential and electron flow velocity in the electron presheath, similar to equations~(\ref{eq:phi_cs})--(\ref{eq:phi_nu}) for the ion presheath, can be derived if one assumes that ions obey the Boltzmann density relation in the electron presheath $n_i(x) = n_o \exp(-e\phi/T_i)$. 
Based on this, quasineutrality $n(x)=n_ov_{e\bb}/V_e=n_o\exp(-e\phi/T_i)$ implies 
\begin{equation}
\label{eq:phi_es}
- \frac{e(\phi - \phi_o)}{T_i} = \ln \biggl( \frac{v_{e\bb}}{V_e} \biggr) .
\end{equation}
This corresponds to equation~(\ref{eq:phi_cs}) for the ion sheath, but where $T_e, c_s$ and $V_i$ are replaced by $T_i, v_{e\bb}$ and $V_e$, respectively. 
Equation~(\ref{eq:phi_es}) highlights the expectation that the potential drop in an electron presheath is expected to be of order $T_i/e$, which is much smaller than the order $T_e/e$ potential drop of an ion presheath.  
Using $E=-d\phi/dx=(T_i/e)(dV_e/dx)/V_e$ in equation~(\ref{eq:mlf_es}) provides 
\begin{equation}
\label{eq:es_ode}
\biggl(1+\frac{T_i}{T_e} \biggr) \biggl( \frac{v_{e\bb}^2 - V_e^2}{V_e^2 \nu_e} \biggr) dV_e = dx.
\end{equation}
Typically $T_e\gg T_i$, so the first term in parenthesis is approximately unity. 
Equation~(\ref{eq:es_ode}) provides the electron presheath analog of equation~(\ref{eq:dVi}).

Like the ion presheath, analytic solutions can be derived for the electron fluid velocity in the presheath in the constant mean free path or constant collision frequency limits. 
Assuming $T_e \gg T_i$ and the constant mean free path limit [$\nu_e(x) = V_e(x)/\lambda_{e}$] 
\begin{equation}
\frac{V_e}{v_{e\bb}} =  \exp \left\{ \frac{1}{2} - \frac{x-x_o}{\lambda_{e}}
          + \frac{1}{2} W_{-1} \left[-\exp\left(2 \frac{x-x_o}{\lambda_{e}}
          - 1 \right) \right] \right\}.
\end{equation}
and
\begin{equation}
   - \frac{e(\phi - \phi_o)}{T_i} = - \frac{x-x_o}{\lambda_{e}} + \frac{1}{2}
        + \frac{1}{2} W_{-1} \left[ -\exp \left(2\frac{x-x_o}{\lambda_{e}} - 1
        \right) \right].
\end{equation}
These correspond to equations~(\ref{eq:vi_mfp}) and (\ref{eq:phi_mfp}) from the ion presheath. 
Similarly, in the constant collision frequency limit ($\nu_e \approx v_{e\bb}/\lambda_e$) 
\begin{equation}
\frac{V_e}{v_{e\bb}} =  1 - \frac{x-x_o}{2\lambda_e} \biggl(1 - \sqrt{1 - \frac{4\lambda_e}{x-x_o}} \biggr)
\end{equation}
and 
\begin{equation}
-\frac{e(\phi - \phi_o)}{T_i} = \textrm{arccosh} \biggl(1 - \frac{x-x_o}{2\lambda_e} \biggr) .
\end{equation}
These correspond to equations~(\ref{eq:vi_nu}) and (\ref{eq:phi_nu}) from the ion presheath. 
A comparison of these models to PIC simulations is shown in figure~\ref{fg:es_profile}. 
As in the ion presheath from equation~(\ref{eq:phi_sr}), both the constant mean free path and constant collision frequency models reduce to a square root potential profile 
\begin{equation}
- \frac{e(\phi - \phi_o)}{T_i} = \sqrt{\frac{x-x_o}{\lambda_e}}
\end{equation}
in the neighborhood of the sheath edge $(x-x_o)/\lambda_e \ll 1$. 

\begin{figure*}
    \centering
    \includegraphics[width=14cm]{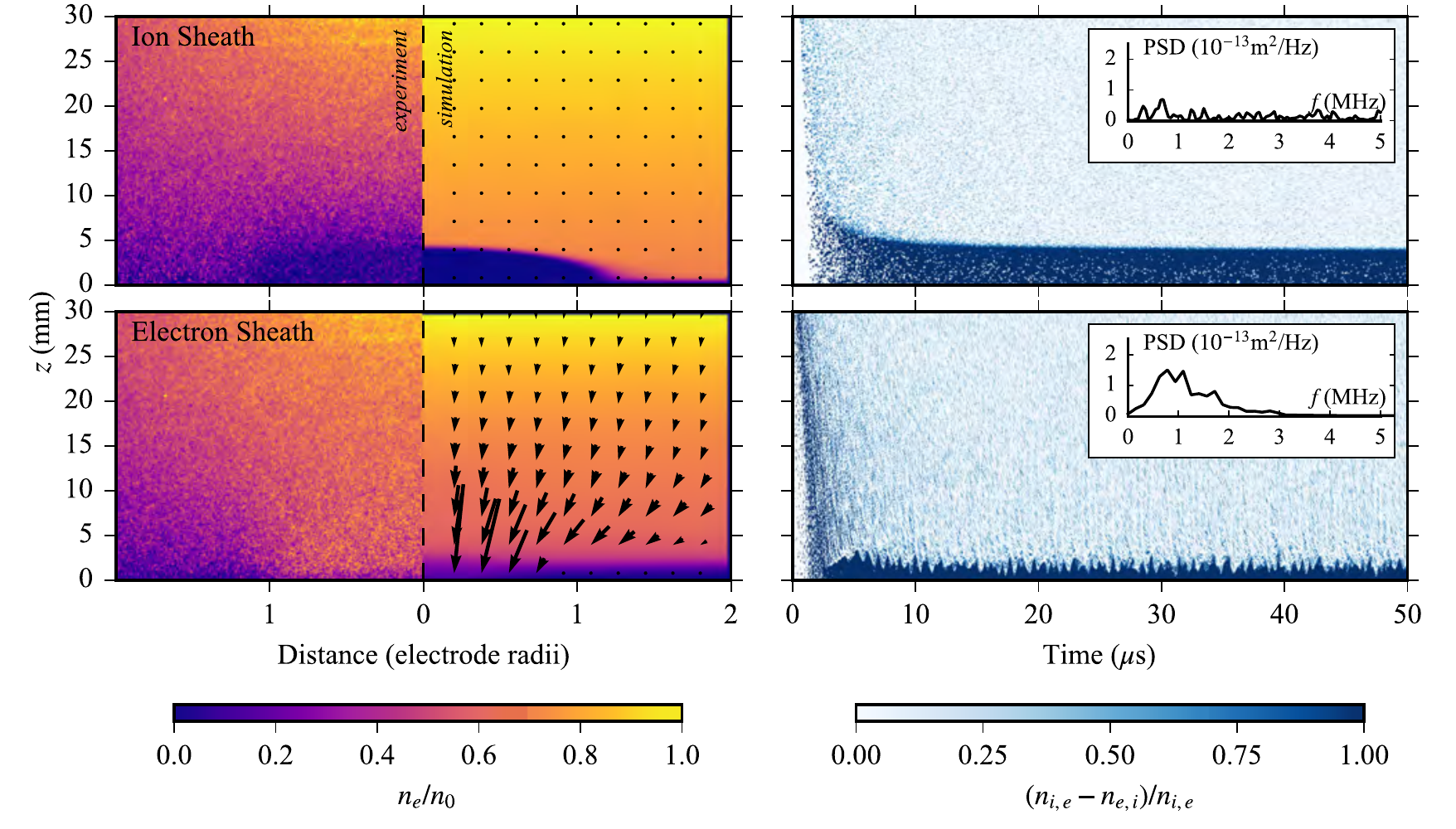}
    \caption{(left) Comparison of electron densities in experiment to a 2D PIC simulation. (right) Plot of space charge fluctuations at the sheath edge from the simulation. Top row is for an ion sheath, with the bottom row is for an electron sheath. Arrows indicate vectors of the electron particle flux. The magnitude of the arrows ranges from 0.1 to 1.2$\times 10^{10}$ m$^{-2}$s$^{-1}$ for the electron sheath and all vectors are less than $5\times 10^8$ m$^{-2}$s$^{-1}$for the ion sheath. Figure reprinted from reference~\cite{yee:17}. }
    \label{fg:e_sheath_fig}
\end{figure*}

In comparison to an ion presheath, three important distinguishing features of the electron presheath are: (1) The electron flow speed is much larger in an electron presheath than the ion flow is in an ion presheath ($v_{e\bb}/c_s \approx \sqrt{m_i/m_e} \gg 1$).  
(2) The change of the electrostatic potential is much smaller in the electron presheath than it is in an ion presheath (since $T_e \gg T_i$). 
(3) The length scale associated with the electron presheath is much larger than the scale associated with an ion presheath at similar discharge conditions (since $\lambda_e \gg \lambda_{\textrm{\scriptsize{in}}}$). 
In low-temperature plasmas in which the dominant electron collision process is due to interactions with neutrals, the constant collision frequency model is often expected to apply. 
In this limit, reference~\cite{sche:15} has shown that the expected electron presheath length would be approximately 5-10 times longer than an ion presheath in a helium discharge at common low-temperature plasma conditions. 
The presence of collisions due to instabilities may modify the electron presheath length scale. 
Evidence for this is provided by the PIC simulations of \cite{sche:15}, which observed a finite-scale electron presheath even though they did not include electron neutral collisions; see figure~\ref{fg:es_profile}. 

This connection between the expected flow and potential profiles of 1D ion and electron presheaths relies on the assumption of a Boltzmann density relation for the species that is reflected back toward the plasma (electrons in an ion presheath or ions in an electron presheath). 
Such a 1D model is predicated on the assumption that the boundary can be treated as an infinite plane, which is justified only if the characteristic scale of the boundary is much larger than the presheath length. 
This is often justified when ion sheaths are encountered because the boundary surface is usually at least several cm, whereas the ion presheath scale is usually less than a few cm. 
However, expectations change for an electron sheath. 
As described in section~\ref{sec:structures}, electron sheaths can form only near sufficiently small electrodes. 
In the common laboratory experiments described above, this often limits the maximum electrode size to a few cm in diameter. 
It was also just shown that the electron presheath is usually at least as long as an ion presheath. 
Because there is not a large scale separation between the size of the electrode and the size of the electron presheath, the infinite planar model, and hence the 1D ion Boltzmann density relation, do not apply.

This geometrical effect associated with the finite electrode size has been studied using 2D PIC simulations~\cite{sche:16,yee:17,sche:15} and experiment~\cite{yee:17,hood:16}. 
Figure \ref{fg:e_sheath_fig} shows a comparison of measured and modeled properties of ion and electron sheaths. 
The left panel shows the electron density and electron flux vectors.  
The electron sheath is observed to cause electrons to ``funnel'' into the electrode, influencing the electron flow far beyond the thin region of negative space charge. 
The measurements and simulations both support the prediction that the electron sheath and plasma are adjoined by a long presheath region 
Although the electric field is weak in this region, the pressure gradient is large enough to drive a fast electron drift~\cite{sche:16,yee:17}. 
The funneling effect causes the presheath to have a two-dimensional nature. 

As a consequence, it has been shown that the ion density relation depends on the ion inertia~\cite{sche:15}. 
This work provided a model for the ion density that generalizes the Boltzmann density relation to include ion inertia, which was found to provide fair agreement with the simulation results when the simulated plasma parameters were used as the boundary conditions. 

These simulations also showed a counterintuitive observation where ions were found to flow toward the electron sheath. 
This was counterintuitive because the electron sheath electric field points from the electrode into the plasma, and should therefore reflect ions back to the plasma. 
The observation was later found to be associated with the boundary conditions nearby the electrode: different results were observed if the electrode was embedded in a surrounding dielectric or if it was ``free''~\cite{hood:16}. 
Here, ion drift was actually due to the ion sheath associated with the surrounding dielectric, rather than the electron sheath itself. 
This aspect is discussed in more detail in section~\ref{sec:ds_ss}. 
The conclusion is that, while a one-dimensional analysis can serve to guide a physical understanding of the electron sheath, one should be cautious that multi-dimensional effects are important when the electron presheath scale is larger than the electrode scale, which is a common situation. 

\begin{figure}
    \centering
    \includegraphics[width=7cm]{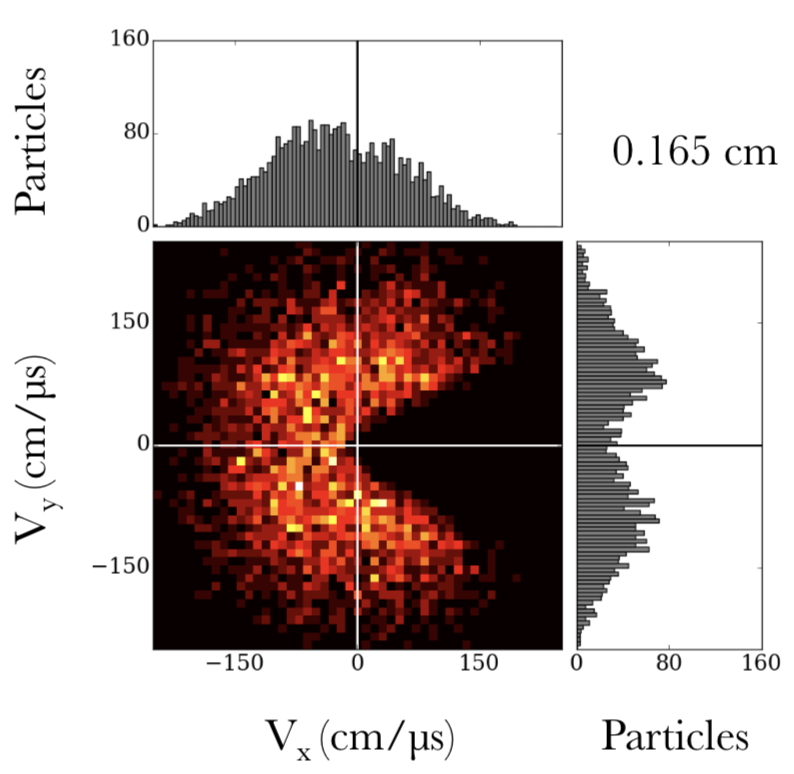}
    \caption{PIC simulation of the two-dimensional electron velocity distribution function at the edge of an electron sheath. Figure adapted from reference~\cite{sche:16}.}
    \label{fg:loss_cone}
\end{figure}

Finite electrode size also influences the structure of the EVDF. 
The model EVDF shown in figure~\ref{fg:evdf_drift_fig} is based on a 1D picture where no electrons have a negative velocity in the direction normal to the sheath, corresponding to the expectation that all electrons that reach the infinite planar boundary are collected. 
Figure~\ref{fg:loss_cone} shows the 2D EVDF from a simulation where the electrode was located interior to the plasma (a free electrode)~\cite{sche:16}. 
This shows that the electron distribution has both a flow-shift and a loss-cone type truncation. 
The conical feature is explained as shadowing due to the electrode: all electrons that reach the electrode are collected, but electrons can also enter the presheath from oblique angles and fill a region with a velocity component directed normal to and facing away from the electrode~\cite{sche:16}. 
The angle of the loss cone is influenced by the size of the electrode, as well as distance from the electrode surface. 
Projecting the EVDF in the direction normal to the sheath shows a distribution resembling a drifting Maxwellian, but the projection parallel to the sheath shows a distribution with a depleted interior associate with the loss cone. 
These results show that the experimental reality often does not conform to either of the simple assumed models (truncated or drifting), but is a combination of them and is often fundamentally two-dimensional. 
Similar simulations of the EVDF have also been made in the case of electrodynamic tether simulations using a Vlasov-Poisson approach~\cite{sanc:13,sanc:14}.

\subsection{Low frequency (ion) instabilities\label{sec:es_lf}} 

The fast electron drift toward the electrode in an electron presheath generates current-driven instabilities. 
The right side of figure~\ref{fg:e_sheath_fig} illustrates the normalized space charge above both an electron and an ion sheath, obtained from 2D PIC simulations. 
A substantial fluctuation at a frequency of approximately 1 MHz is observed in the case of an electron sheath, but is either absent or of a much lower power for the ion sheath. 
Reference \cite{sche:15} provides a detailed examination showing that these are current-driven ion-acoustic instabilities. 
Since the differential flow between electrons and ions is near the electron thermal speed, the usual ion-acoustic instability dispersion relation from equations~(\ref{eq:ia_real}) and (\ref{eq:gama}) does not apply (the conventional relation assumes a sufficiently small relative drift). 
However, the wave is on the same ion-acoustic branch, and a simple extension for the dispersion relation was obtained
\begin{equation}
\label{eq:ia_dr}
    \frac{\omega}{\omega_{pi}} \approx \frac{k\lambda_{De}}
    {\sqrt{k^2\lambda_{De}^2 - \frac{1}{2}Z'(-V_e/v_{T_e})}},
\end{equation}
where $Z'$ is the derivative of the plasma dispersion function. 

\begin{figure}
    \centering
    \includegraphics[width=7cm]{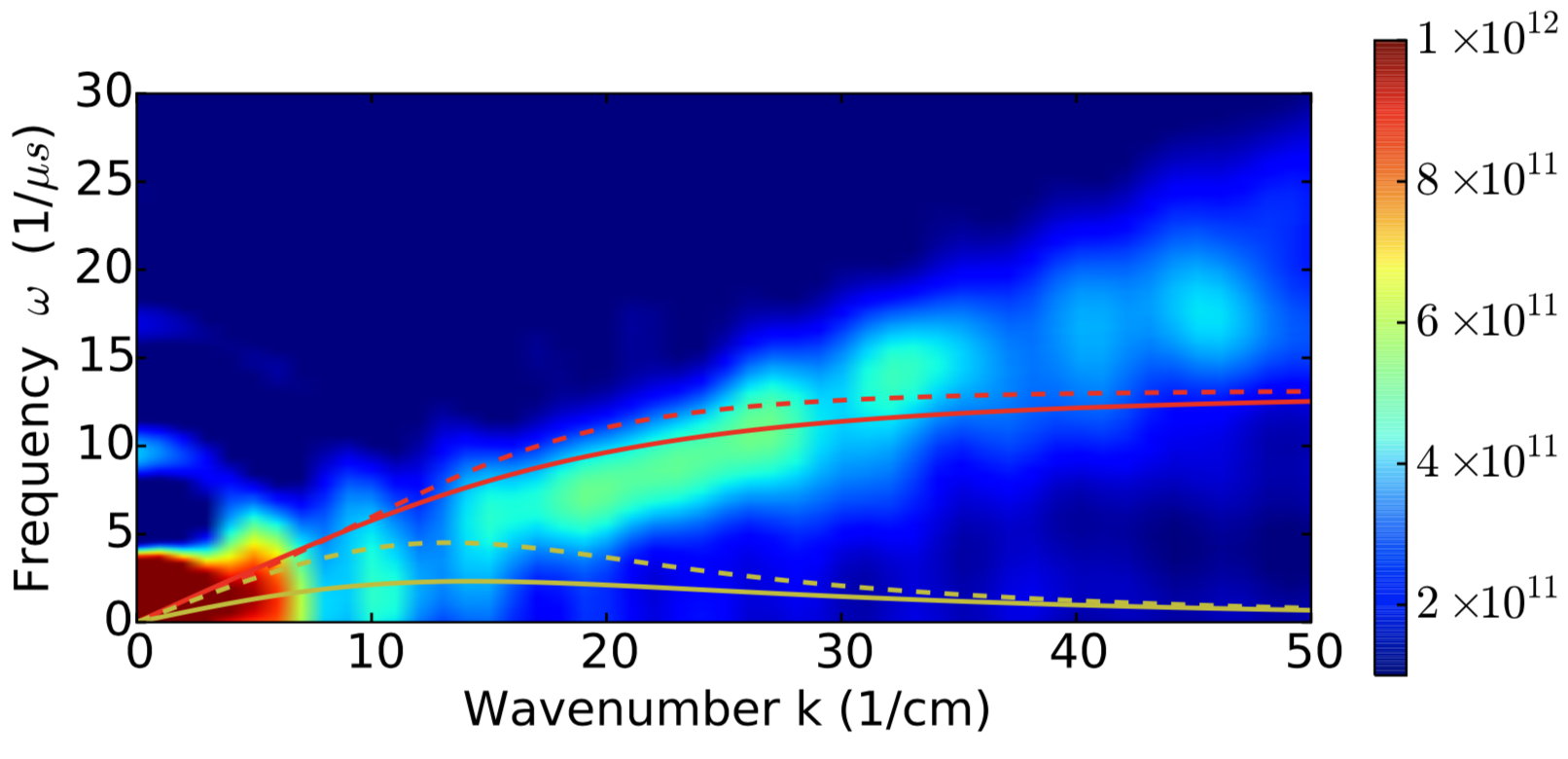}
    \caption{A 2D Fourier transform of ion densities in the vicinity of an electron sheath with real (red) and imaginary (yellow) components of the dispersion relation from equation~(\ref{eq:ia_dr}) overlaid. The two lines show the predictions at electron flows of $0.5v_{eB}$ and $0.9_{eB}$, which is a characteristic range encountered near the electron sheath. Figure reprinted from reference \cite{sche:15}. }
    \label{fg:e_inst_fig}
\end{figure}

Figure~\ref{fg:e_inst_fig} shows a 2D Fourier transform of the modeled ion density in the vicinity of an electron sheath, with the real and imaginary components of the dispersion relation predicted by equation~(\ref{eq:ia_dr}) overlaid.
There is good qualitative agreement with the predicted wave frequency, indicating that the flow of electrons is indeed inducing a streaming instability. 
The cascading of power to higher wavenumbers than the most linearly unstable modes may provide evidence for a nonlinear character of the instability. 
Only wavenumbers up to 30 cm$^{-1}$ were resolved in this simulation. 
The ion waves generated by this instability convect toward the electron sheath, but since ions are turned around by the electric field in the sheath, some of the wave power is reflected. 
The reflection of electrostatic waves by an electron sheath was first studied theoretically by Baldwin~\cite{bald:67}. 
Recently, experimental measurements have been presented for the reflection coefficient using LIF diagnostics~\cite{beru:18}. 

Measurements of ion-acoustic noise associated with an electron sheath were first made by Glanz \emph{et al}~\cite{glan:81} using a spectrum analyzer to measure oscillations in the current collected by the electrode.  
The measured spectra were peaked slightly below the ion plasma frequency, which is consistent with the expected wave frequency of ion-acoustic instabilities excited by the relative electron-ion drift. 
A range of plasma densities was explored. 
The spectra were also observed to sharpen when a negatively biased probe was brought within several centimeters of the positive electrode. 

\begin{figure}
    \centering
    \includegraphics[width=7cm]{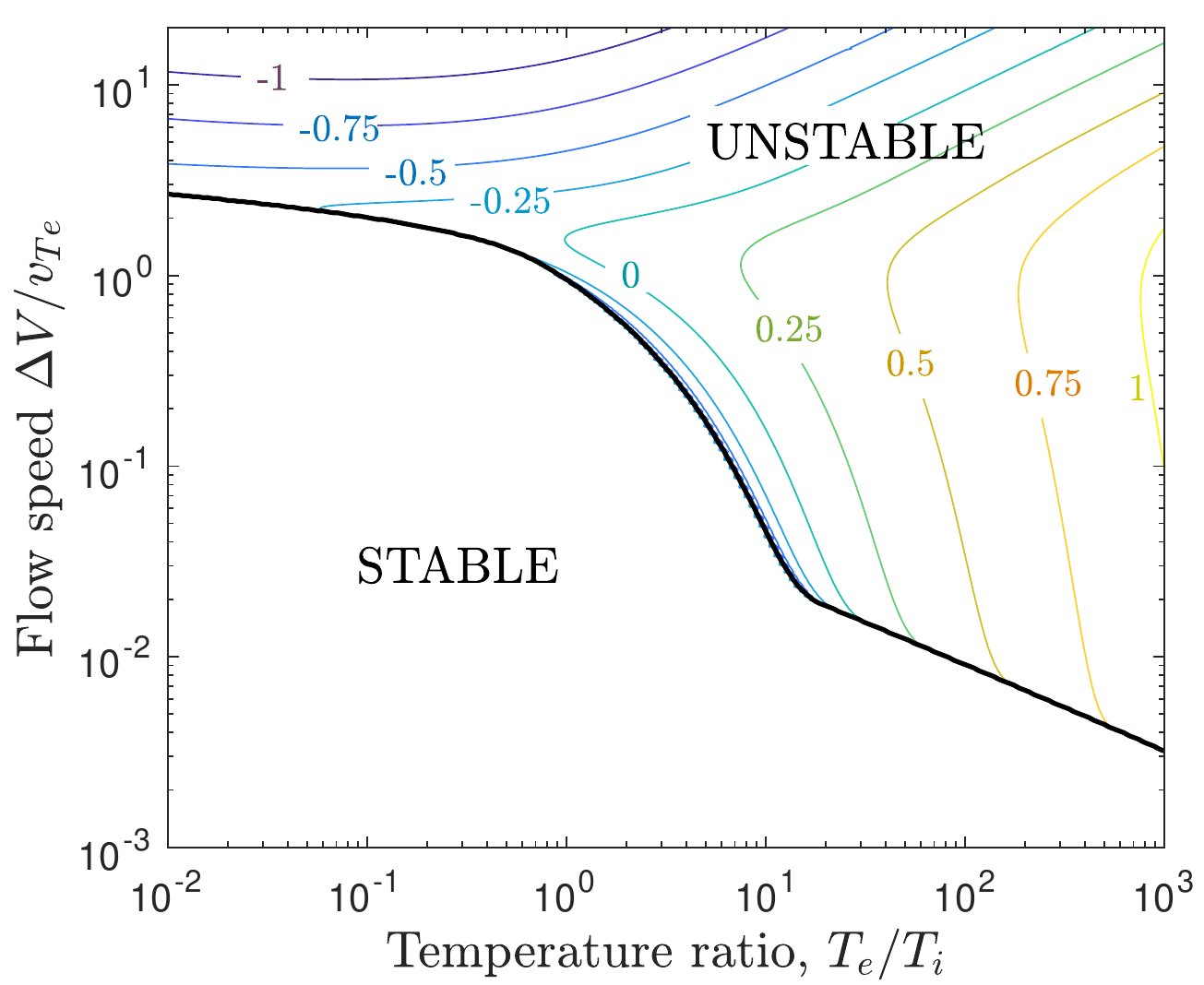}
    \caption{Ion-acoustic instability boundaries for a H$^+$ plasma computed from equation~(\ref{eq:ephat1}) for Maxwellian distributions. Contours show lines of constant $\textrm{log}_{10} (k_{\tiny \textrm{max}} \lambda_{De})$, where $k_{\tiny \textrm{max}}$ is the maximum unstable wavenumber. This figure corrects that shown in figure 1 of reference~\cite{baal:16}, where it was incorrectly stated that there is a second region of stability at sufficiently high temperature ratio and flow speed.}
    \label{fg:ia_vt}
\end{figure}

The existence of ion-acoustic instabilities in the electron sheath and presheath is likely to contribute an increase in the effective collision rate, akin to that seen in the ion presheath \cite{baal:11}; see section~\ref{sec:1is}. 
In fact, since the differential flow between electrons and ions is much larger in the case of the electron sheath, the unstable waves have a much larger growth rate and the associated fluctuations have a much large amplitude (as shown in figure~\ref{fg:e_sheath_fig}). 
Correspondingly, the predicted wave-particle scattering will be larger. 
Figure~\ref{fg:ia_vt} shows the stability boundaries for the ion-acoustic wave computed from equation~(\ref{eq:ephat}) assuming that both ions and electrons have Maxwellian distribution functions, but with a differential flow speed $\Delta V = |V_e-V_i|$. 
This shows that for a common temperature ratio $T_e/T_i \gtrsim 10$, the instability is excited at a drift speed that is much smaller than the electron thermal speed. 
This implies that much of the electron presheath is expected to be unstable. 
Consequently, the waves will have room to undergo several exponentiations of amplitude before reaching the boundary. 
It is thus likely that the waves become saturated due to nonlinear effects. 
No detailed study of the saturation mechanism has yet been presented, but it is known that ion trapping is a common mechanism influencing ion-acoustic waves of this type as they grow to sufficiently large amplitude~\cite{onei:65}.

Wave-particle scattering by these instabilities may be a contributing factor to the observed flow-shift in the presheath. 
As stated at the beginning of this section, such diffusive (rather than the commonly assumed effusive) behavior relies on sufficient collisions to establish a pressure gradient and transition to the bulk plasma. 
This remains an open question, but the instabilities are clearly observed and the fluctuation amplitudes are large, which provides strong indirect evidence for the importance of wave-particle scattering. 
Such scattering may also contribute to filling in the loss cone observed in figure~\ref{fg:loss_cone} as the electron flow transits through the presheath. 
This would have repercussions for the presheath length scale and the evolution of the electron velocity distribution as the flow transits the presheath. 

\subsection{High frequency (electron) instabilities\label{sec:es_hf}} 

In addition to the low frequency ion-acoustic branch instabilities, a series of measurements of high-frequency fluctuations on the collected current have been observed for positively biased electrodes~\cite{sten:88,sten:89}. 
The frequency of these fluctuations is near the electron plasma frequency. 
Stenzel proposed an explanation as a sheath-plasma resonance instability arising from the negative resistance associated with the finite transit time of electrons through the sheath~\cite{sten:88}. 
A circuit model was developed, and basic features of the model were shown to agree with measurements~\cite{sten:89}. 
In this explanation, the fluctuations are caused by wave evanescence, rather than a plasma instability. 
Harmonics are radiated as electromagnetic waves, such that the electrode acts as an antenna. 

\begin{figure}
\begin{center}
\includegraphics[width=8.5cm]{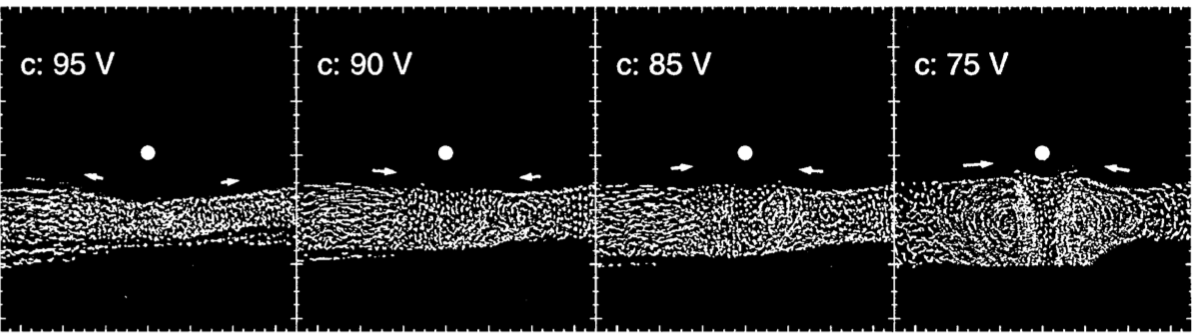}
\caption{\label{law98fig3} The disruption of circulation in dust grains induced by the positive bias of an electrode. Figure reprinted from reference~\cite{law:98}.}
\end{center}
\end{figure}

More recent work has also explored high-frequency instabilities near the electron sheath of spherical electrodes in magnetized plasmas~\cite{sten:10}. 
These instabilities were observed to propagate with the average $\vc{E} \times \vc{B}$ drift and to form toroidal eigenmodes. 
They were also attributed to the electron inertia in the electron sheath.

\subsection{Applications}

There are several potential applications of a better understanding of electron sheaths. 
The most commonly encountered situation is the collection of the electron saturation current of a Langmuir probe. 
Because this occurs in a regime in which no electrons are repelled, the amount of current collected is related to the plasma density and electron temperature. 
A detailed understanding of electron sheaths may open new possibilities for the extraction of these values in the electron saturation of a Langmuir probe trace. 
Hass~\emph{et al}~\cite{haas:05} have improved the fidelity of microwave ``hairpin'' probe measurements by developing a method to account for the influence of the electron sheath~\cite{sten:76}. 
Another possible application in the diagnostics realm is in the production of X-band microwaves from a biased electrode~\cite{blio:12}. 
Experiments have indicated the possibility of the production of microwaves at $\sim10$ GHz at 10s of mW in association with the sheath-plasma resonance instability~\cite{blio:12}.

Electron sheaths have also demonstrated their utility in flow control. $\vc{E}\times\vc{B}$ flows have been demonstrated using different combinations of positively and negatively biased electrodes in scrape-off-layer plasmas of tokamaks~\cite{zweb:09,thei:12}. 
These flows have been utilized to manipulate the density and temperature profiles in such plasmas~\cite{zweb:09}. 
Electron sheaths have demonstrated the ability to manipulate the circulation in a dusty plasma~\cite{law:98} such as the one shown in figure~\ref{law98fig3}. 
They also arise near spacecraft in the ionosphere~\cite{sing:96}. 

Finally, we note that Schiesko~\emph{et al}~\cite{schi:08} have recently modeled, and studied experimentally, the influence of secondary electron emission on electron sheath properties. 
The dynamics of emitted electrons within the electron sheath may also be utilized in new applications, such as emission of electromagnetic radiation.

\section{Double Sheaths\label{sec:ds}} 

\subsection{Steady-state properties\label{sec:ds_ss}}

Double sheaths, also referred to as virtual cathodes, are potential structures in which a significant potential drop forms in front of the electrode surface, before rising again in the bulk plasma.
They can arise near biased electrodes due to current balance requirements associated with size, geometry or material properties~\cite{baal:07,hood:16,cho:90,bail:06,sun:15,yip:15b}, or due to electron emission from the electrode~\cite{hobb:67,guer:70,chen:11,li:12,schw:15a}. 
The plasma potential may be above or below the electrode potential, depending on the experimental circumstances. 
A double sheath is a type of double layer in that it consists of adjacent regions of positive and negative space charge, as indicated in figure~\ref{fg:type}. 
A double sheath is distinguished from a fireball double layer by the feature that the region of strong electric field is adjacent to the surface of the electrode, rather than separated from it by a quasineutral region of plasma~\cite{hers:85,char:07}. 
However, historically the term has sometimes also been used to describe double layers in the latter context~\cite{andr:71}. 
Double sheaths were first observed, and named, by Langmuir~\cite{lang:29}. 
They can exist in transient~\cite{intr:88} or steady~\cite{hers:87} states. 

Double sheaths may be present in unexpected circumstances, such as near Langmuir probe diagnostics. 
Yip and Hershkowitz have shown that the presence of a virtual cathode (double sheath) can flatten the measured I-V trace, leading to an overestimation of the measured electron temperature~\cite{yip:15b}. 
The effect was observed to have an especially pronounced influence at low pressure, where the virtual cathode was deeper. 
This example emphasizes the importance of understanding when double sheaths form. 
In this experiment, the formation was thought to be associated with the electrode size, in which the criterion based on the area ratio of electrode to chamber wall was near the transition region discussed in section~\ref{sec:gc}. 
This effect arising from the size of the Langmuir probe depends not only on the probe itself, but also on the global geometry of the plasma confinement device. 
To avoid the potential for misinterpretation of probe characteristics, the probe must be sufficiently small. 

\begin{figure}
\begin{center}
\includegraphics[width=8.5cm]{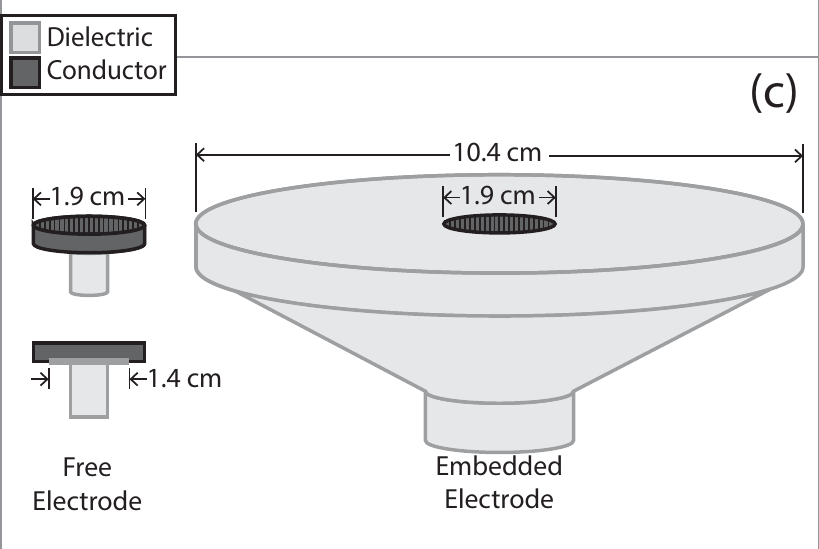}
\caption{\label{fg:hood_electrode} Schematic drawing of the free electrode (left) and embedded electrode (right) configurations used in experiments studying electron sheaths.  Figure reprinted from reference~\cite{hood:16}.}
\end{center}
\end{figure}

Predicting when double sheaths form can be complicated, and they can lead to counterintuitive effects on the plasma behavior. 
For example, a recent experiment studied the ion flow pattern in front of a small positively biased electrode in either a free or embedded configuration~\cite{hood:16}; see figure~\ref{fg:hood_electrode}. 
The ``free'' electrode was nearly a completely exposed conducting electrode, with just a small area of dielectric on the back side used to hold the electrode in place. 
The ``embedded'' electrode consisted of a similar conducting disk, but this time embedded in a larger dielectric disk. 
Both configurations were implemented in an experiment and simulated using 2D PIC simulations. 
In each case, the electrode was biased at 10V above ground, which was approximately 5V above the plasma potential in these experiments. 
An electron sheath formed, and it was expected that the ions would not be significantly influenced by the electrode, except for the expected density drop in the electron sheath close to the electrode. 

\begin{figure}
\begin{center}
\includegraphics[width=8.5cm]{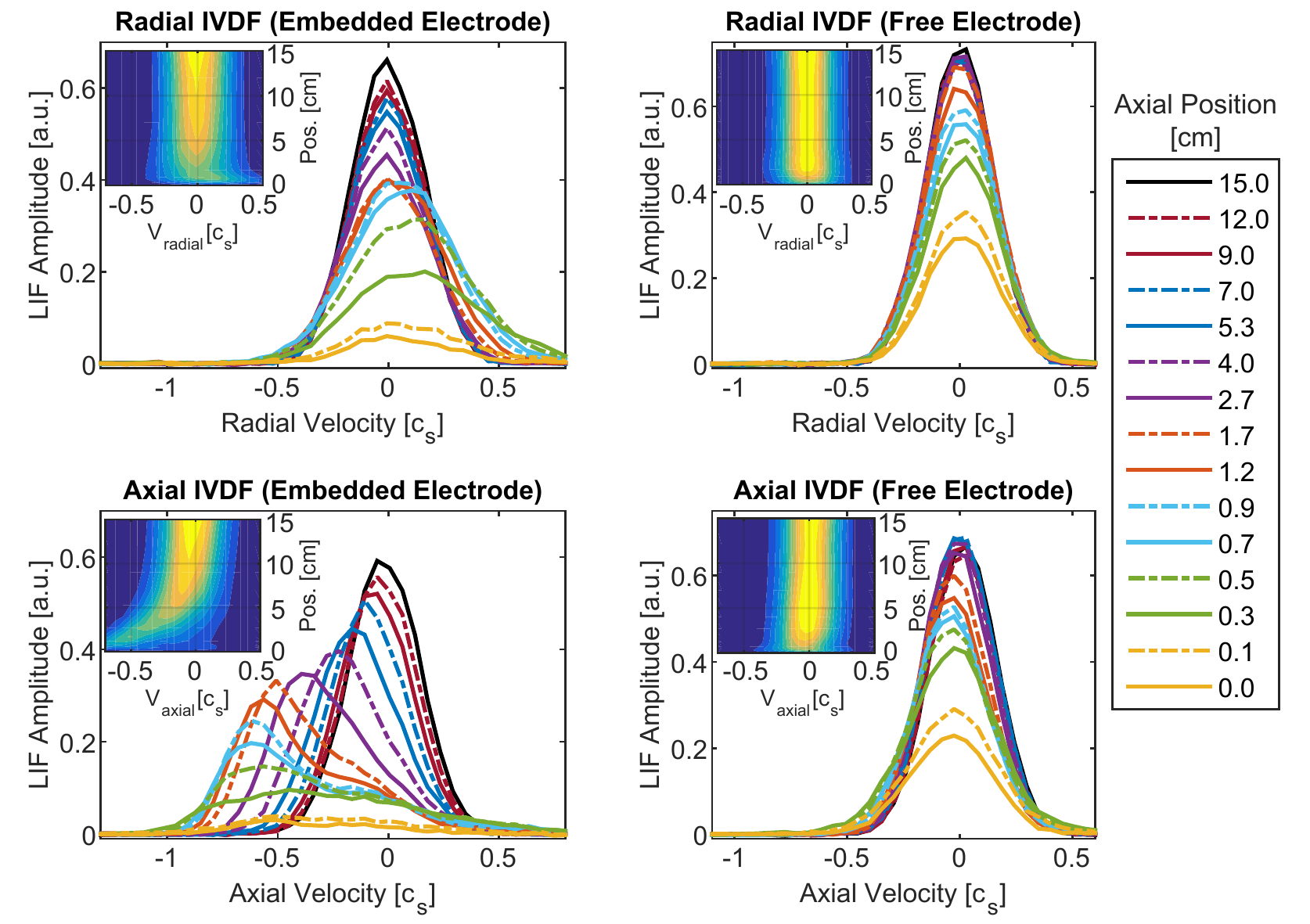}
\caption{\label{fg:hood_ivdf} Ion velocity distribution functions (IVDFs) measured in front of the free and dielectric-embedded electrodes shown in figure~\ref{fg:hood_electrode}. 
Speed is indicated in units of the sound speed, $c_s$. Figure reprinted from reference~\cite{hood:16}. }
\end{center}
\end{figure}

Figure~\ref{fg:hood_ivdf} shows that a counterintuitive effect was observed in the ion flow. 
The figure shows measurements of the radial and axial components of the IVDF made along a line extending from the center of the electrode at several axial positions. 
As expected, both the radial and axial components of the IVDF do not change significantly with position for the free electrode, showing only the expected density drop in the sheath region.  
In contrast, a strong ion drift was measured moving \emph{toward} the embedded electrode. 
A slight radial component to the drift was also measured in the 0.5 cm region above the electrode. 
This is counterintuitive because an electron sheath is expected to repel ions back toward the plasma, not accelerate them toward the electrode. 

\begin{figure}
\begin{center}
\includegraphics[width=6cm]{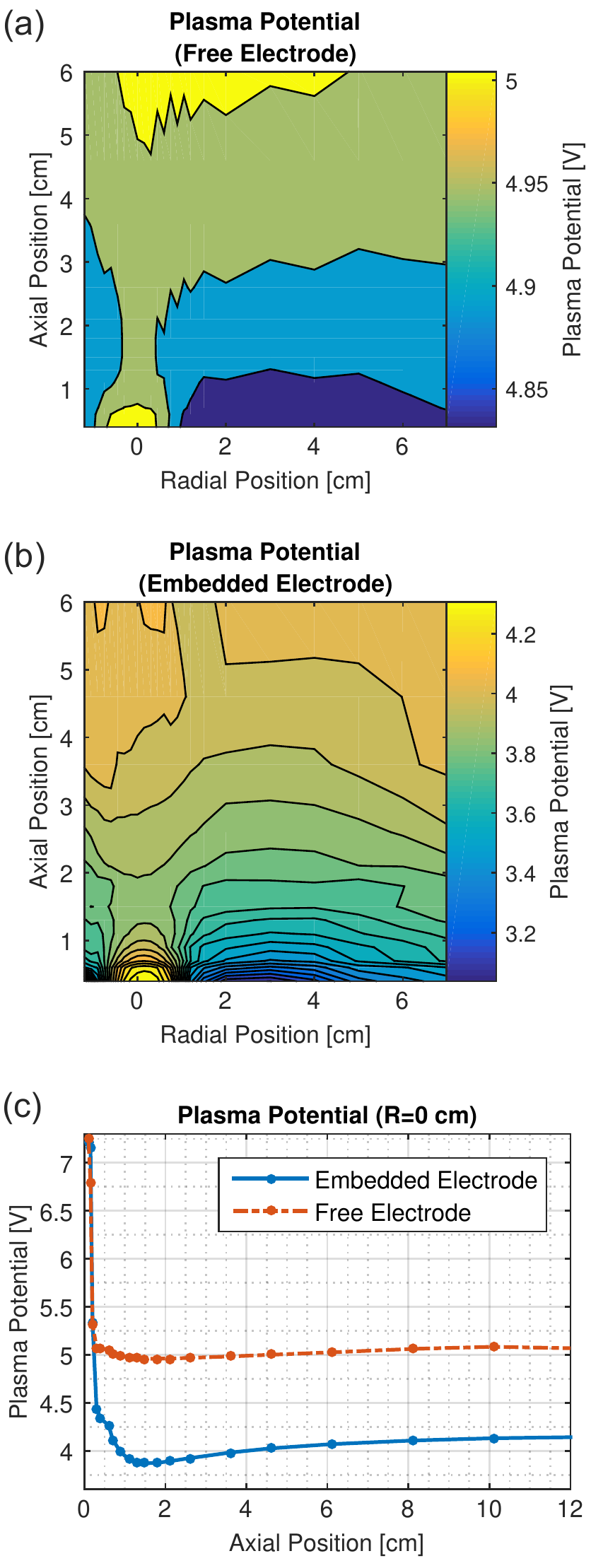}
\caption{\label{fg:hood_probe} Potential profiles from free and embedded electrode configurations obtained using an emissive probe. The electrode radius was 1 cm. Note that the scale used to describe the potential differs in each plot.  Figure reprinted from reference~\cite{hood:16}.}
\end{center}
\end{figure}

The explanation for this effect is revealed by the electrostatic potential maps shown in figure~\ref{fg:hood_probe}. 
In the case of the free electrode, the electrostatic potential drops essentially monotonically from the electrode into the plasma, as expected. 
In contrast, a saddle point is observed in front of the embedded electrode, which translates to a double sheath along the one-dimensional axial cut in front of the electrode. 
The mechanism for formation in this case is not global current balance, but the presence of the dielectric. 
As discussed in sections~\ref{sec:is} and \ref{sec:es}, the plasma potential drops $T_e/2e$ or more in an ion presheath, whereas the plasma potential rises by a much smaller amount characteristic of $T_i/e$ in an electron presheath. 
For this reason, the presheath associated with the ion sheath in front of the dielectric spreads in front of the electrode, causing the ion presheath to ``shadow'' the electrode. 
The superposition of electron sheath and ion presheath can take the form of a double sheath. 
This can be seen in the 2D maps of the potential profile, as well as the axial cutaway shown in figure~\ref{fg:hood_probe}. 
Furthermore, a similar series of IVDF measurements along the axis were made at different radial positions of the electrode. 
The 2D map of the ion flow vectors shows precisely this effect; the ion flow is ``diverted'' around the electrode and into the surrounding dielectric, see figure 5 of \cite{hood:16}. 

The 2D PIC simulations were found to agree quantitatively well with the measured IVDFs. 
The detail afforded by the simulations also revealed features of the 2D IVDF profile that were not possible to measure experimentally; see figure 7 of \cite{hood:16}. 
These show that the IVDF in the case of an embedded electrode obtains a drift nearly directly toward the electrode along the axis extending from the center of the electrode, but that the drift has nearly equal components in the axial and radial directions for radial positions toward the edge of the electrode. 
In contrast, the IVDF is observed to be nearly Maxwellian without a drift at all radial locations in front of the free electrode. 
The axial electrostatic potential profiles from the 2D simulations did not appear to display a virtual cathode, but the 2D potential structure had a saddle-point shape similar to what was measured in the experiment. 
 
\subsection{Ion trapping\label{sec:it}}

Perhaps the most persistent open question regarding double sheaths is understanding when they can be steady-state solutions in partially ionized plasmas. 
From the viewpoint of the one-dimensional potential profile, it has been noted for a long time that any type of ion-neutral collision, such as charge exchange or ionization, would cause a population of ions to become trapped in the potential well~\cite{fore:86}. 
Trapped ions would quickly fill the potential well, altering the potential profile leading to its disappearance~\cite{hers:87}. 
This has led to the thought that there must either be a mechanism present to ``pump'' these ions from the potential well~\cite{hers:87,fore:86}, or that double sheaths are a transient phenomenon~\cite{intr:88}. 
Another possibility may be that the potential profile fluctuates at a frequency that is low enough to pump ions, perhaps due to an instability, and that the double sheath potential results only in the time-averaged sense. 
There seems to be evidence for each of these possible mechanisms. 

The previous example showed a double sheath resulting from the presheath of a surrounding dielectric material shadowing the electrode. 
However, double sheaths have also been measured near free electrodes~\cite{baal:07,fore:86}, such as that shown in figure~\ref{fg:NAF_sheath}. 
In both of these cases, one side of a disk shaped electrode was conducting, while the back side was covered with dielectric material. 
Forest and Hershkowitz showed that ion pumping in this configuration can be provided by a saddle shaped potential profile that extends around the side of the electrode providing a path for ions to reach the dielectric~\cite{fore:86}. 
The double sheath in front of the electrode itself does not seem to be associated with the dielectric in this case, yet the dielectric is necessary to provide the ion pumping. 
This is an example of the plasma self-organizing to create a complex potential structure that maintains a steady-state. 
The work showed that even something as innocuous as a fingerprint on the electrode surface can provide a means of pumping ions~\cite{fore:86}. 
Increased neutral pressure was observed to decrease the depth of the potential well because the source of trapped ions increased at a faster rate than the ability for ions to be pumped from the potential well. 
Double sheaths have similarly been observed in other experiments with a dielectric backing on the biased electrode~\cite{baal:07,bail:06}.

A condition for the neutral density required to maintain sufficient ion pumping to sustain the double sheath was developed by applying current balance arguments within the potential well: trapped ion production by ionization and charge exchange must be less than the loss out the edges of the potential well. 
Combining this with a global current balance condition led to the prediction that the neutral density must satisfy~\cite{fore:86} 
\begin{equation}
n_n \leq \frac{n_t/n_i}{\sqrt{A_\E/\pi} [\sigma_{cx} + A_\w/(V2n_o)]}
\end{equation}
to maintain the double sheath. 
Here, $n_t$ is the trapped ion density, $n_i$ is the background ion density, $\sigma_{cx}$ is the charge-exchange cross section and $V$ is the plasma volume.  

Forest and Hershkowitz also derived an expression for the location of the minimum of the double sheath based on the Child-Langmuir law, as described in section~\ref{sec:is_standard}, and an assumption that the electron flux is the thermal flux reduced by the Boltzmann factor: $\Gamma_{e,\ths} \exp(-e\phi/T_e)$.  
The result is 
\begin{equation}
d_\textrm{\tiny min}^2 = 1.0 \times 10^6 \frac{(\phi_p - \phi_{\textrm{\tiny dip}})^{3/2}}{n_e \sqrt{T_e}} \biggl( 1 + \frac{2.66}{\sqrt{eV_p/T_e}}\biggr) .
\end{equation} 
This was found to agree well with the experiments in \cite{fore:86}, and was later validated independently by Bailung \etal~\cite{bail:06}. 
The latter work also showed that an applied ion beam increases the rate of ion trapping, rather than ion leaking due to pumping. 
It was observed that the threshold ion beam energy required to suppress double sheath formation was nearly equal to the electrode potential. 

Wang \etal~\cite{wang:86} have shown that a double sheath can be maintained in the time-averaged sense if the applied electrode bias includes an rf component at the appropriate frequency (1-500 kHz in this case). 
In this experiment the electrode was entirely conducting, preventing the possibility of ion pumping to nearby dielectric. 
Ion pumping was naturally provided in this case during the portion of the rf cycle in which the plasma potential was above the electrode potential. 
Electric field reversals that appear similar to double sheaths in the time-averaged potential structure in rf capacitively coupled discharges have also been observed~\cite{schu:08}. 

Fluctuations of the electrode potential suggest that plasma instabilities may also provide a natural mechanism by which ions could be pumped from the potential well~\cite{barn:14}. 
In this case, ions may either be scattered out of the potential well by wave particle interactions or the fluctuation may be of sufficiently high amplitude that the structure of the sheath itself is altered in a time-dependent manner. 
In fact, ion-acoustic frequency fluctuations have been observed in both experiments~\cite{yip:17} and simulations~\cite{sche:15}. 
Although no study has yet been done that attempts to identify if the instabilities can provide sufficient ion pumping, measurements have revealed an influence on the IVDF~\cite{yip:17}, which will be discussed in the next section. 


\subsection{Stability} 

The IVDF in a typical double sheath may be expected to consist of two components. 
One is the population entering from the plasma that is accelerated toward the electrode by the potential drop from the plasma to the minimum of the virtual cathode. 
After reaching the potential minimum, ions are then turned around by the larger electron sheath potential rise between this minimum and the electrode. 
This reflected population would be expected to create a second contribution to the IVDF associated with ions at essentially the same energy at any location, but moving in the opposite direction. 
Adding these two populations, the total IVDF near the dip minimum may thus be expected to consist of oppositely directed ion beams. 

Yip, Hershkowitz and Severn~\cite{yip:17} tested this hypothesis by measuring the IVDF throughout the double sheath using LIF. 
The results revealed a very different behavior than the above expectation. 
Although both incoming and reflected ion populations were observed, the energy of the incoming population was measured to be far less than what would be expected for ions freely falling through the measured potential drop from the plasma to the potential minimum. 
The reflected population was also low energy as well as a much lower density. 
Counter-streaming ion beams are expected to be unstable to the ion-ion two-stream instability, as described in section~\ref{sec:is_mis}. 
The authors proposed that ion-ion two-stream instabilities may be responsible, and invoked the instability-enhanced friction mechanism described in section~\ref{sec:is_mis} to explain the low energy difference between the counterstreaming populations. 

This hypothesis was tested by carrying out a similar series of LIF measurements at a variety of electrode biases, ranging from values where the electrode potential was above to below the plasma potential. 
This allowed control over the density of the reflected population because some of the ions were absorbed by the electrode when the electrode potential was near the plasma potential. 
Since the ion-ion two-stream instability threshold depends on the relative concentration of each population, this was expected to control the energy at which the instability onsets, and hence the expected energy of the ions as they traverse the double sheath; this is analogous to the velocity locking that results from the two-stream instability in the multiple ion species case described in section~\ref{sec:is_mis}. 
Indeed, the ion energy was observed to be fast and near the ballistic expectation when the reflected population was not present, but to be much lower than the ballistic expectation when the reflect population was present. 
A mapping for the expected ion energy in terms of concentration and electrode bias were provided, showing qualitative agreement with the experimental measurements. 

This work provides strong evidence that ion-ion two-stream instabilities influence the IVDF in double sheaths, but several open questions remain. 
No direct measurement of the instability dispersion relation has yet been provided. 
It is also interesting to question how ion scattering by the ion-ion two-stream instabilities might influence ion trapping in the double sheath potential well. 
Does the presence of instabilities feedback to influence the depth of the potential well? 
Can this also be a source for pumping ions out of the well?

\subsection{Electron emitting surfaces} 

One of the most common situations in which double sheaths arise is near electron-emitting surfaces~\cite{hobb:67,guer:70,chen:11,li:12,schw:15a,fang:69}. 
A variety of mechanisms can be responsible for electron emission, including secondary emission, photoemission or thermionic emission. 
Sheath structure near electron emitting surfaces is of interest in several applications such as thermionically emitting cathodes~\cite{geke:16}, discharges sustained by electron emission~\cite{schw:15b}, tokamak divertors~\cite{komm:17,camp:19}, hypersonic vehicles~\cite{hanq:17}, spacecraft applications~\cite{whip:81}, the Moon~\cite{popp:11}, meteoroids~\cite{sora:01}, as well as surrounding dust particles in space and the laboratory~\cite{delz:14,autr:18}. 
It is particularly important for emissive probe diagnostics because the double sheath influences the interpretation of the current-voltage trace~\cite{kemp:65,ye:00}. 
Methods, such as the inflection point method~\cite{smit:79}, have been developed to improve the accuracy of emissive probe measurements associated with this effect~\cite{shee:11}. 
It has also been demonstrated experimentally~\cite{oksu:11} and studied using PIC simulations~\cite{camp:13,camp:15,camp:16} that plasmas bounded by strongly emitting boundaries can be made to have a negative plasma potential with respect to the conducting boundaries. 
Here, double sheaths similar to that depicted in figure~\ref{fg:NAF_sheath} were observed~\cite{oksu:11}, but where the bulk plasma potential was below the electrode potential. 
This is a very different confinement state than the typical situation of a positive plasma potential that confines electrons via an ion sheath, emphasizing that secondary emission can fundamentally reconfigure a plasma~\cite{camp:15}. 

The question of a floating emitting surface was first considered by Hobbs and Wesson~\cite{hobb:67}, and has subsequently been studied by many others~\cite{schw:93,tier:16}. 
Other early work on this topic was motivated by diodes~\cite{fang:69} and Q machines~\cite{rynn:66}, where dc double sheaths were observed and theoretically modeled near strongly emitting thermionic cathodes.  
In addition to dc double sheaths, Braithwaite and Allen \cite{brai:81} studied the rapid formation of a double sheath following a voltage step in thermally produced diode plasma. 
There is a rich and broad literature on double sheaths near electron emitting boundaries. 
The following discussion focuses on recent studies that aim to determine if a sheath near an emitting surface will be an ion sheath, double sheath or an electron sheath. 
In this literature, the double sheath is often referred to as a ``space charge limited'' (SCL) sheath, and the electron sheath as an ``inverse'' sheath~\cite{sydo:09,camp:12a}. 

Consider a planar conducting electrode that is electrically floating with respect to the plasma. 
In the absence of electron emission, the sheath surrounding the electrode is expected to be an ion sheath, with the plasma potential being the floating potential higher than the electrode potential, as described in section~\ref{sec:structures}, and with the sheath structure described in section~\ref{sec:is}. 
If the electrode emits electrons, the potential profile and corresponding floating potential would be expected to change. 
Electrons emitted from the surface are accelerated by the ion sheath potential into the plasma. 
If the electron emission rate is sufficiently high, the emitted electron density can exceed the ion density near the electrode surface. 
This creates a thin region of negative space-charge near the electrode surface, forming a virtual cathode with a total sheath potential profile taking the form of a double sheath. 
Hobbs and Wesson~\cite{hobb:67} were the first to propose a fluid model for how electron emission alters the floating potential. 
Assuming that the emitted electrons originate with negligible energy, they obtained the result 
\begin{equation} 
\phi_s = - \frac{T_e}{e} \ln \biggl( \frac{1 - \Gamma}{\sqrt{2\pi m_e/m_i}} \biggr)
\end{equation}
where $\Gamma$ is the ratio of emission flux to the primary electron flux reaching the electrode. 
This predicts that the sheath potential decreases with increased emission, and the model is expected to be limited to emission values smaller than unity. 
Their theory also included a model for the double sheath potential profile. 
These basic features of electron emitting sheaths have been modeled in a variety of kinetic theories, and showed general agreement with PIC simulations, as reviewed by Schwager~\cite{schw:93}.

Usually, the emitted electrons have a distribution of energies, which may be characterized with a temperature parameter $T_\w$. 
Sheehan \emph{et al}~\cite{shee:13,shee:14} proposed a kinetic theory showing that the sheath potential depends on the ratio of the temperature of emitted electrons to the plasma temperature. 
The model predicted that the sheath potential approaches zero as this temperature ratio approaches unity. 
This prediction was found to agree with PIC simulation results~\cite{shee:13}, as well as measurements of the sheath surrounding a thermionically emitting cathode in the afterglow of an rf plasma~\cite{shee:14b}. 
This work also included a generalization of the Bohm criterion to account for secondary electron emission (see equation (5) of~\cite{shee:13}). 

A fraction of emitted electrons overcome the potential barrier between the electrode and the minimum of the virtual cathode, and are then accelerated into the plasma. 
At steady-state, this results in an electron distribution function in the plasma that consists of an electron beam moving with reference to the bulk plasma electrons. 
If this electron beam is sufficiently dense and energetic, it can be a source of electron-electron two-stream instabilities. 
Such instabilities have been observed in experiments~\cite{coak:80}, studied theoretically~\cite{moro:02} and simulated using PIC methods~\cite{sydo:07,camp:12b}. 
Secondary electron emission has been proposed as a mechanism to cause anomalous electron transport in Hall effect thrusters~\cite{kaga:07,sydo:08}.

Sydorenko \emph{et al}~\cite{sydo:09} have presented results of one-dimensional PIC simulations suggesting that under conditions of intense electron emission, a steady-state space-charge limited (SCL) sheath (what we refer to as the double sheath) may not be present. 
Instead, they observed an oscillation between a SCL state and ion sheath state. 
The SCL sheath was observed to exist for a short interval of approximately 38 ns, while the non-SCL (ion) sheath existed for 154 ns. 
The overall oscillation between these two states was regular and on the order of MHz. 
These were explained as relaxation oscillations associated with a negative differential conductivity in the sheath, and subsequent trapping of a population of cold secondary emitted electrons. 

Recent work by Campanell \emph{et al}~\cite{camp:12a} has studied boundaries with a secondary electron emission coefficient that exceeds unity, primarily using 1D PIC simulations. 
This work observed essentially no sheath when the electron emission coefficient was near unity, and an ``inverse sheath'' (what we call an electron sheath), rather than a double sheath, when the secondary electron emission coefficient significantly exceeded unity.
An electron sheath potential profile reflects some fraction of the electrons emitted by the boundary. 
Global current balance considerations, as described in section~\ref{sec:gc}, can be satisfied by either of these two solutions, and, in fact, both have been observed in the simulations~\cite{camp:17}.  
The question naturally arises that, if two possible solutions exist, why is one ``chosen'' over the other? 
Campanell \emph{et al}~\cite{camp:17} have discussed the important role of ion trapping in the potential well. 
As discussed in section~\ref{sec:it}, double sheaths in partially ionized plasmas require a mechanism for pumping ions trapped in the potential well if they are to be a steady-state solution. 
This is not possible in one-dimension, and it was suggested that for this reason the electron sheath (i.e., inverse sheath) would be observed in practice. 
However, as discussed in section~\ref{sec:it}, a variety of mechanisms for ion pumping are possible in multiple dimensions, including loss to nearby surfaces or pumping of ions back to the plasma due to sheath fluctuations and instabilities. 
It remains a matter of further research to develop a predictive capability describing which sheath solutions will form in a given experimental configuration.

\section{Fireballs\label{sec:fb}} 

\subsection{Anode glow} 

When an electrode is biased above the plasma potential electrons are accelerated toward the boundary by the electric field of the electron sheath, gaining energy $e(\phi_\E - \phi_p)$ in passing from the plasma to the electrode. 
If the electrode potential is within a few volts of the plasma potential, the electron sheath can be accurately described without regard to ionization sources, as was discussed in section~\ref{sec:es}. 
In fact, since acceleration through the electron sheath nominally causes the electron density to drop, the ionization rate in the electron sheath is much less than in the bulk plasma. 

\begin{figure*}
\begin{center}
\includegraphics[width=15cm]{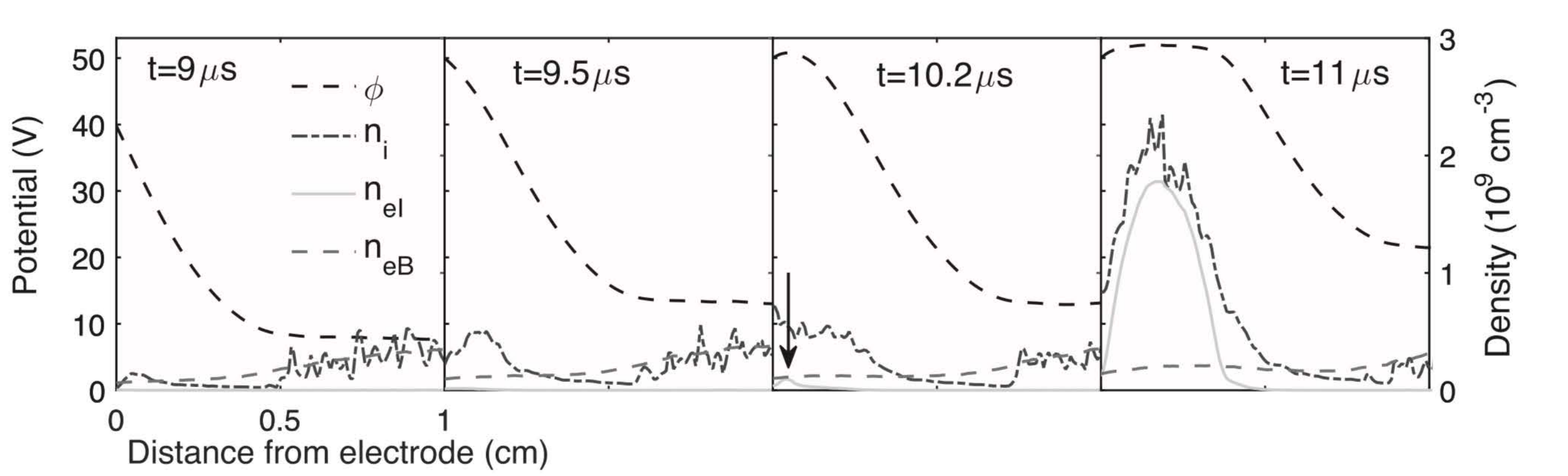}
\caption{\label{sche18fig5} Profiles of the bulk electrons $n_{eb}$, electrons from ionization $n_{eI}$, ions $n_i$, and the plasma potential $\phi$ at four different times during the fireball onset. Figure reprinted from reference~\cite{sche:17}.}
\end{center}
\end{figure*}

However, as the potential difference between the plasma and the electrode approaches the ionization potential of the neutral gas, the ionization cross section rapidly increases, and a larger fraction of the sheath-accelerated electrons ionize neutrals. 
For a large enough energy gain in the electron sheath, the ionization rate can become much higher near the electrode than in the bulk plasma. 
Since the ion generated in the ionization event is much more massive than the corresponding electron, it takes much longer for the sheath electric field to push the ion into the bulk plasma than it does for the electron to be lost to the electrode. 
The residence times are approximately $\tau_i = l_s/\sqrt{2e (\phi_{\E} - \phi_p)/m_i}$ for ions and $\tau_e = l_s/\sqrt{2e (\phi_{\E} - \phi_p)/m_e}$ for electrons, where $l_s$ is the length scale of the region where localized ionization is taking place (a subdomain of the electron sheath). 
If the ionization rate is large enough, and the ion residence time long enough, the ion density can overtake the electron density and a thin region of positive space charge forms near the electrode surface. 
This causes a flattening of the potential profile near the electrode surface, and the resulting potential profile is that of a double layer, as shown in figure~\ref{fg:type}(d). 
Because the excitation cross section increases with electron energy, similar to the electron impact ionization cross section,
this region glows brighter than the background plasma. 
An example is shown in figure~\ref{fg:spot_photo}, and is often referred to as anode glow~\cite{lang:29,emel:82}. 

Anode glow was first observed and named by Langmuir~\cite{lang:29}. 
Measurements of the potential profile using emissive probes have confirmed that the electron sheath lengthens and flattens when the anode glow is present~\cite{baal:09,song:91}. 
An analytic generalization of the Child-Langmuir law of the form of equations~(\ref{eq:cl}) or (\ref{eq:cl_e}) has not been obtained for anode glow because the local ion density is connected with the local electron density, electron velocity and the ionization cross section. 
However, Conde \emph{et al}~\cite{cond:06} have developed a semi-analytic model for the potential and density profiles (see figure 4 of \cite{cond:06}) that solves an integro-differential equation numerically. 
This model includes the effect of electron impact ionization and shows the qualitative features of lengthening and the buildup of an ion rich layer in the sheath next to the electrode that are expected from the physical arguments above. 
Anode glow was also observed in the PIC simulations of~\cite{sche:17}. 
These results are reproduced in the panel marked t = $9.5\mu$s in figure~\ref{sche18fig5}. 
Again, this figure shows the lengthening of the electron sheath and flattening of the potential profile that characterizes the anode glow double layer. 

\subsection{Fireball onset\label{sec:onset}} 

Langmuir was also the first to observe that if the electrode bias is increased further after the anode glow has formed, a critical point is eventually reached where the sheath structure bifurcates to a much larger (typically cm scale) luminous secondary discharge~\cite{lang:29}; see figure~\ref{fg:spot_photo}. 
This state is now often called a ``fireball''~\cite{song:91,song:92a,sten:11b,sten:11f,sten:11g,sten:12,levk:17}, though the same phenomenon has also been referred to as an ``anode spot''~\cite{baal:09,rube:40,park:11,park:14}, ``plasma contactor''~\cite{ahed:96,lapu:00}, ``plasma double layer''~\cite{sand:95,stra:99}, ``anode double layer''~\cite{sand:86,tang:03,gurl:05}, ``firerod''~\cite{an:94}, or other labels~\cite{seyh:88,ande:95,opre:00,muja:11}. 

\begin{figure*}
\begin{center}
\includegraphics[width=15cm]{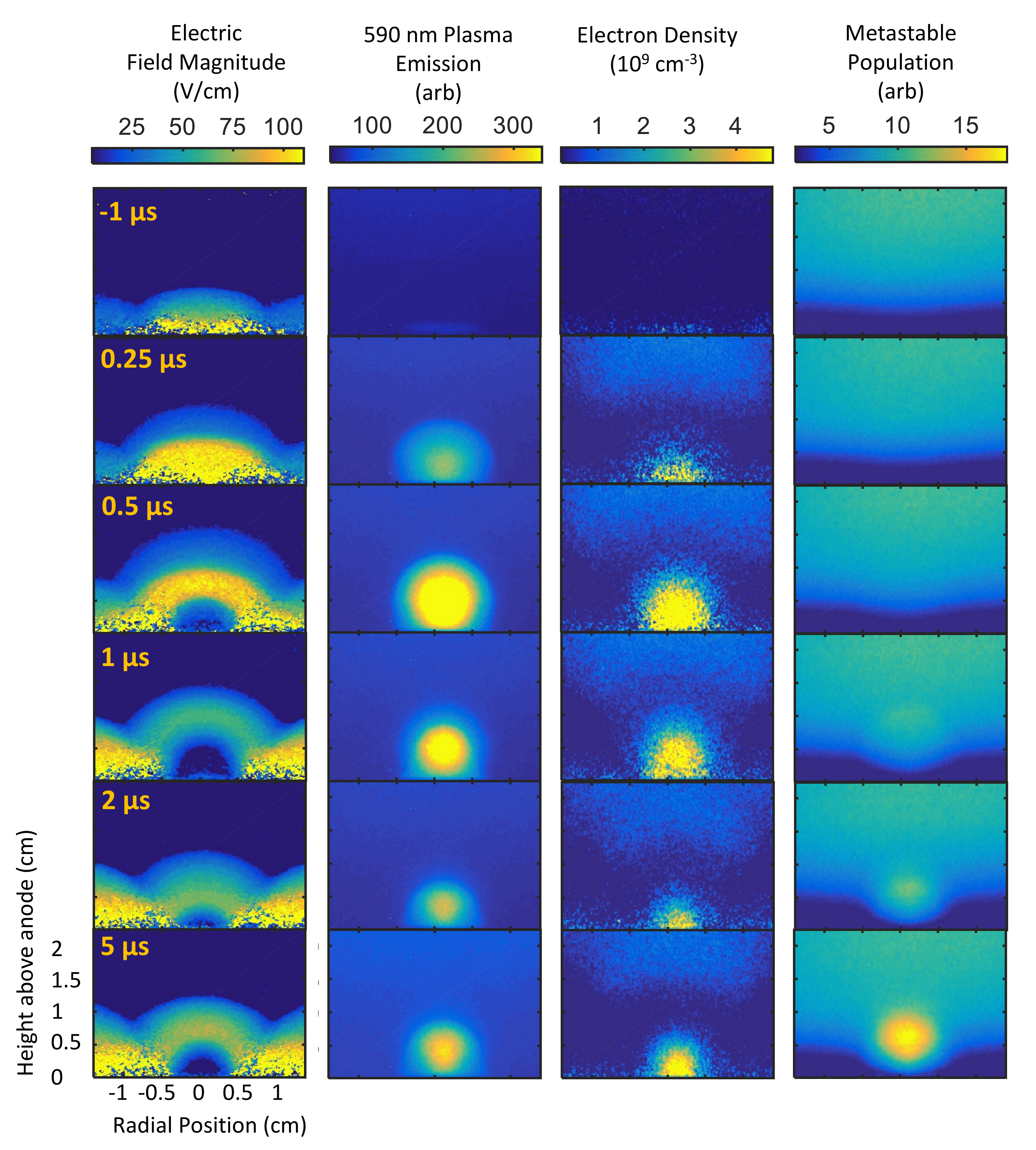}
\caption{LCIF measurements of the electric field magnitude, 590 nm plasma emission, electron density and metastable population during fireball onset. Figure reprinted from reference~\cite{sche:18}. \label{fg:LCIF} }
\end{center}
\end{figure*}

Langmuir and emissive probe measurements of the electrostatic potential profile~\cite{baal:09,song:91,trov:80} reveal that the secondary fireball discharge is a quasineutral plasma that is separated from the bulk plasma by a strong double layer electric field, as depicted schematically in figure~\ref{fg:type}. 
Recent laser collision induced fluorescence (LCIF) measurements have also revealed that the plasma density is larger inside the fireball than in the bulk plasma~\cite{weat:12,sche:18}; see figure~\ref{fg:LCIF}. 
Other basic measurements of fireball properties are that the double layer potential drop is approximately the ionization potential of the neutral gas~\cite{baal:09,song:91}, that this potential drop occurs over a few Debye lengths, and that both the diameter of the fireball and the critical bias required for onset scale approximately inversely with neutral pressure~\cite{song:91}. 
Furthermore, there is hysteresis in the onset condition, whereby the critical bias required to create a fireball from the anode glow state is higher than the bias at which the fireball will transition back to the anode glow state~\cite{baal:09,song:91,sand:95,stra:99,tang:03}.
More recently, the stability properties of fireballs have been extensively investigated~\cite{sten:11b,sten:11f,sten:11g,sten:12}.  

Fireball onset can be initiated by either increasing the electrode bias or neutral gas pressure beyond a critical value~\cite{lang:29,song:91}. 
Fireball formation is abrupt. 
Langmuir probe measurements indicate that formation occurs on a timescale of approximately $10\mu s$ in a 1 mTorr Ar plasma~\cite{sten:08,song:91} once a critical value of bias or pressure is surpassed. 
Because the fireball formation increases the effective area of the electrode for collecting electrons, formation is quickly followed by an increase of the bulk plasma potential~\cite{sche:18}. 
After a short transient period, the sheath and plasma potential adjust until the double layer potential step is near the ionization potential of the neutral gas~\cite{song:91}.

\begin{figure}
\begin{center}
\includegraphics[width=8cm]{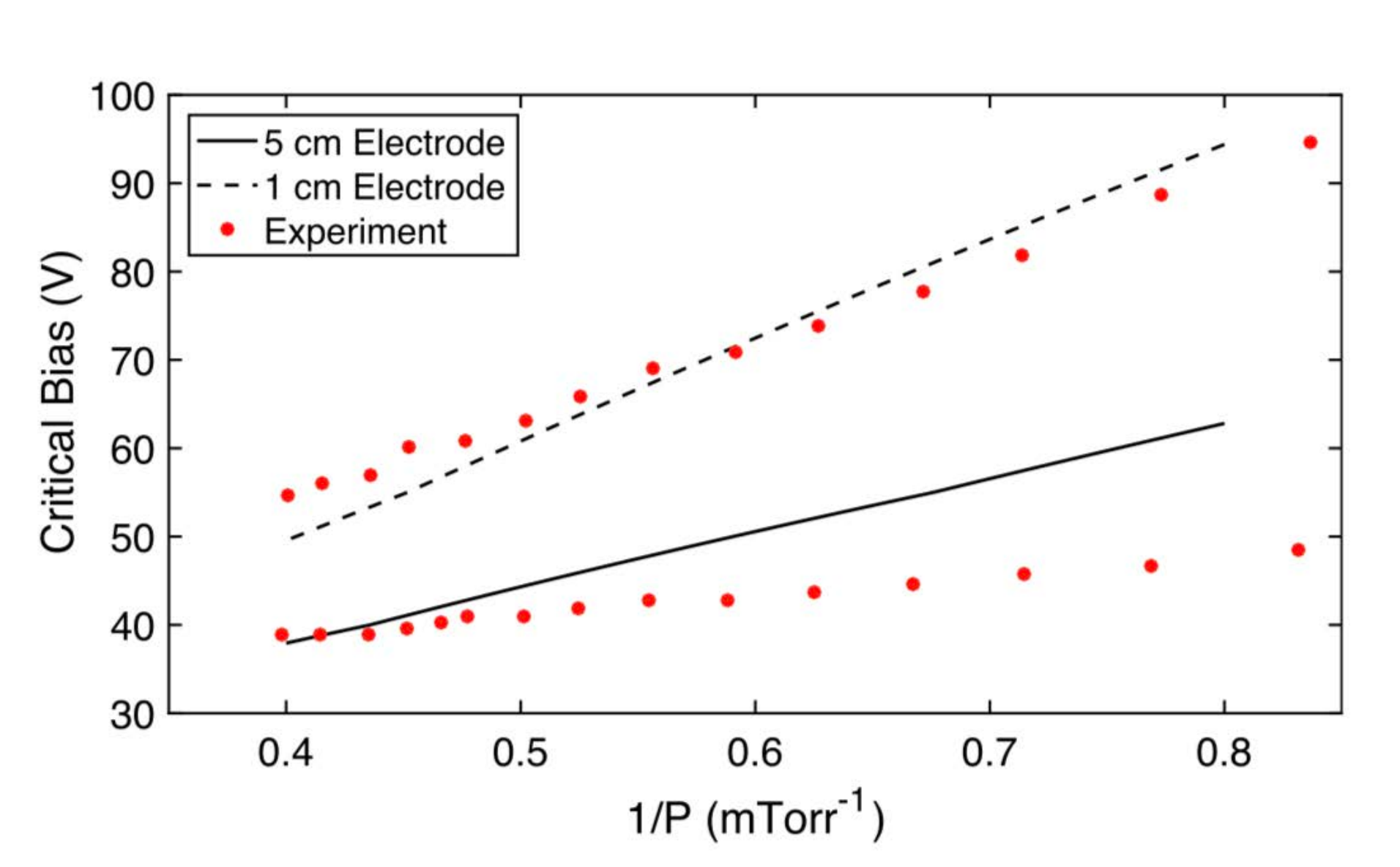}
\caption{A comparison of predictions for the critical bias and experimental measurements as a function of pressure for two different electrode sizes. Figure reprinted from reference~\cite{sche:17}. \label{sche18fig11} }
\end{center}
\end{figure}

A few models for fireball onset have been proposed. 
One of the elementary features of these models is that stationary double layers satisfy the Langmuir flux balance condition~\cite{bloc:72,bloc:78}
\begin{equation}
\label{eq:lang}
\Gamma_i = \sqrt{m_e/m_i} \Gamma_e .
\end{equation} 
Song \emph{et al}~\cite{song:91} used this condition to derive a criterion, which predicts that the fireball onsets when the ion density in the high potential side of the anode glow exceeds the bulk plasma density. 
By balancing the ionization rate with the loss rate they predicted that the critical bias scales inversely with pressure, which was found to agree well with experiments. 
Reference~\cite{sche:17} presented a slight modification of this to include the fact that sheath expansion can cause the surface area of the electron sheath associated with the anode glow to be larger than the surface area of the electrode. 
This leads to a slightly larger area for ion loss to the plasma than for electron loss to the electrode. 
The modified balance condition for the ionization and loss rates is~\cite{sche:17}
\begin{equation}
1-\frac{1}{2}\frac{A_{\s}}{A_{\E}}\sqrt{\frac{m_i}{m_e}}N\sigma z_{\s}=0,
\end{equation}
where $A_{\s}$ is the effective area of the sheath, $N$ is the neutral density, $\sigma$ is the ionization cross section, and $z_{\s}$ is the thickness of an electron sheath~\cite{sche:15}. 
A simple cylindrical shell model was proposed for the area ratio, which assumes that the electrode is a disk: $A_{\s}/A_{\E} \approx 4z_{\s}/D+1$. 
Since $z_{\s}$ and $\sigma$ are a function of the sheath potential, the critical bias relative to the plasma potential can be solved numerically. 
The results for several different neutral pressures using two different electrode sizes are compared with experimental measurements in figure~\ref{sche18fig11}, showing reasonable agreement~\cite{sche:17}. 
From the figure, it is clear that either an increase in bias or pressure for a fixed electrode size can result in onset, in agreement with previous experiments~\cite{song:91}. 

Later work~\cite{baal:09} built upon this model, suggesting that if the quasineutral region within the anode glow becomes more than a Debye length in size, a fireball will form. 
This is because ions exiting the quasineutral plasma must satisfy the Bohm criterion, which requires the formation of a presheath. 
It was suggested that the fireball is the presheath region that accelerates ions directed toward the bulk plasma. 
However, this prediction is difficult to test in the laboratory because the spatial scale is so small and probes can disrupt the anode glow properties. 

Recent progress has been made using 2D PIC simulations~\cite{sche:17} and 2D LCIF measurements~\cite{sche:18}. 
Simulations show that the ion density in the anode glow is substantially greater than the electron density just before onset; see the panel marked t=9.5~$\mu$s in figure~\ref{sche18fig5}.  
This was not a part of the model from \cite{baal:09}, which assumed a quasineutral region with a flat potential adjacent to the electrode. 
The remaining panels of figure~\ref{sche18fig5} reveal the process that results in fireball formation~\cite{sche:17}: A continued buildup of ions near the electrode surface results in the formation of a local maximum in the electrostatic potential just off of the electrode surface. 
This maximum is a potential well for low energy electrons born from electron impact ionization in its vicinity. 
As a result, some of these electrons are trapped as indicated by the arrow in the panel marked t=10.2$\mu$s. 
Once this occurs, the trapped electron density quickly increases leading to the formation of a quasineutral fireball plasma. 
The buildup of ions as a means to trap electrons born from ionization is an aspect of the fireball formation process that was revealed by PIC simulations. 

Experimental validation of these simulations was provided by recent LCIF measurements~\cite{sche:18}. 
Application of LCIF has considerably advanced understanding of fireballs because it provides a non-invasive (optical) and well-resolved measurement of electric field, electron density and electron temperature. 
A simulation of an experiment studying fireball onset and stabilization in a 100 mTorr helium discharge was found to show agreement with the measured electric field and density profiles, as well as the onset dynamics~\cite{sche:18}; see figure~\ref{fg:LCIF}. 
This provided evidence confirming the importance of the establishment of a potential well for electrons in the anode glow region prior to onset. 
PIC simulations also revealed that the dynamics of onset can depend on how rapidly the electrode bias is increased through the critical value. 
If the voltage is ramped quickly (as a step function) the initial fireball expansion quickly pushes a burst of ions into the plasma, but then subsequently collapses, before turning on again and settling to a steady state after a few microseconds. 
This ``flickering'' is not observed if the electrode bias is ramped at the same timescale as it takes for the fireball to form (a few microseconds). 
In addition to the electric field and density measurements shown in figure~\ref{fg:LCIF}, LCIF measurements have also shown that the electron temperature inside the fireball is not significantly different than in the bulk plasma~\cite{weat:12}.

\subsection{Steady-state properties}  

\subsubsection{Fireball-electrode sheath structure} 

\begin{figure}
\begin{center}
\includegraphics[width=6cm]{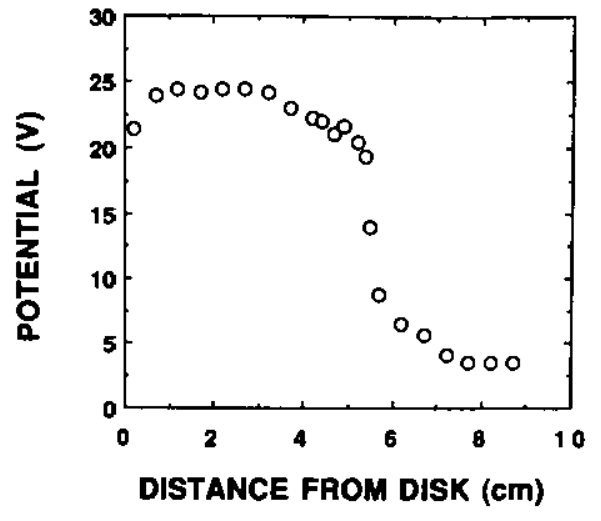}\\
\includegraphics[width=6cm]{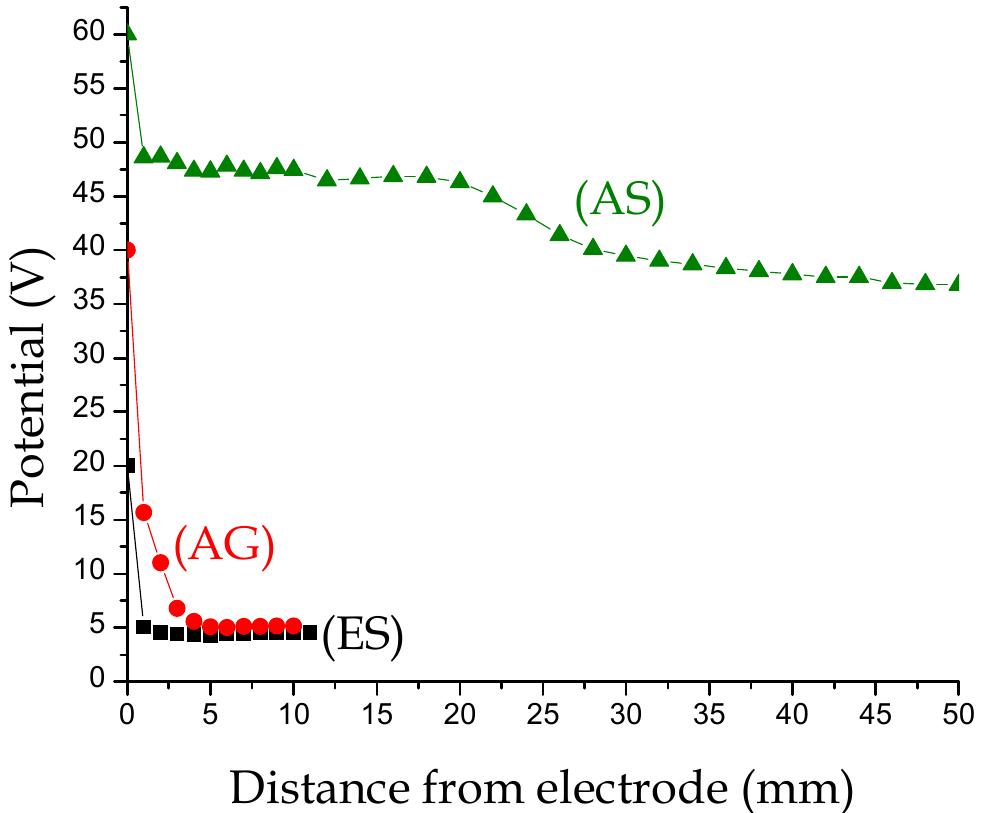}\\
\includegraphics[width=7cm]{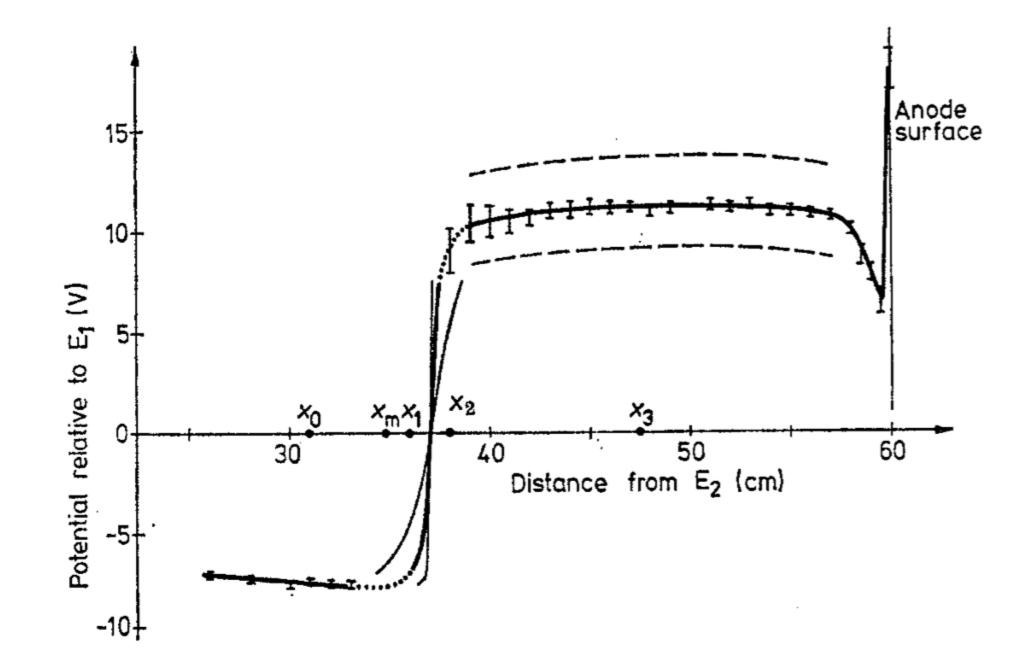}
\caption{\label{trov80fig2} Example measurements of the potential profile in a fireball showing three different possibilities for the sheath between the fireball and the electrode: (top) ion sheath, (middle) electron sheath, (bottom) double sheath.  Figures reprinted from references~\cite{song:91}, \cite{baal:09} and~\cite{trov:80}, respectively. }
\end{center}
\end{figure}

Fireballs have been observed with different sheath potential structures separating the fireball plasma from the electrode~\cite{baal:09,song:91,trov:80,sche:18}. 
Cases with an ion sheath, electron sheath, and double sheath have all been observed; see fig~\ref{trov80fig2}. 
This sheath structure can be understood using the same current balance arguments as section~\ref{sec:gc}~\cite{sche:17}, but applied to the fireball discharge. 
In this case, the wall area of the chamber $A_{\w}$ is replaced by the surface area of the fireball $A_{\F}$. 
The same principle applies because the fireball itself is a steady-state quasineutral plasma. 

First, consider the case where the sheath between the electrode and fireball plasma is an ion sheath. Ion loss in this situation occurs both through the double layer surface and to the electrode with ions traveling at their sound speed at the plasma boundaries which is determined in part by the electron temperature of the total electron distribution. 
Due to the magnitude of the double layer potential, electrons are only lost to the electrode at a rate given by the random flux of the trapped electron distribution multiplied by the Boltzmann density reduction factor associated with the sheath potential. 
The balance of currents results in an area-ratio condition, analogous to equation~(\ref{eq:A_is}), describing when an ion sheath is present~\cite{sche:17} 
\begin{equation}\label{a6}
\frac{A_\E}{A_\F} > \biggl( \frac{0.6 \sqrt{T_{eI}/T_e}}{\mu} - 1 \biggr)^{-1} \approx 1.7 \sqrt{\frac{T_{eI}}{T_e}} \mu .
\end{equation}
Here, $T_{eI}$ is the temperature of the trapped electron population born from ionization and $T_{e}$ is the total electron temperature. 
Analogous to section~\ref{sec:gc}, if $A_\E/A_\F$ is sufficiently small, the fireball potential is expected to be less than the electrode potential and a corresponding electron sheath to be present. 
In this situation, electrons are lost to the electrode at the electron thermal speed and ions are lost through the fireball surface at the sound speed. 
Assuming a constant electron density leads to a condition analogous to equation~(\ref{eq:A_es}) for an electron sheath~\cite{sche:17}
\begin{equation}\label{a3}
\frac{A_\E}{A_\F} < \frac{\mu}{\alpha_e}
\end{equation}
When the area ratio falls between the range of equations~(\ref{a6}) and (\ref{a3}), a double sheath is expected. 

Evaluating equations~(\ref{a6}) and (\ref{a3}) requires a model for the effective surface area of the fireball. 
In general, this can be complicated by different possibilities for the geometry. 
In the most common case of a sphere, accurate models have been developed that relate the diameter of the sphere to the neutral gas pressure, ionization cross section, electron-to-ion mass ratio and the double layer and sheath potential steps. 
An example will be described in the next subsection, leading to equation~(\ref{D2}). 
Calculations of equations~(\ref{a6}) and (\ref{a3}) using equation~(\ref{D2}) to model $A_\F$ were found to predict area ratio criteria that are consistent with available experimental measurements of each sheath structure, such as the electron, ion and double sheath shown in figure~\ref{trov80fig2}~\cite{sche:17}. 

\subsubsection{Size\label{sec:size}} 

The first model of the fireball diameter~\cite{song:91} combined the Langmuir condition from equation~(\ref{eq:lang}) with the requirement that the rate of ion production within the fireball must equal the rate of ion loss into the bulk plasma. 
The latter condition was modeled by treating the fireball as a complete sphere, and assuming the contribution of the electrode surface is negligible, leading to~\cite{song:91} 
\begin{equation}
\label{eq:gamma_i}
\Gamma_e D \sigma_i N = \Gamma_i
\end{equation}
where $D$ is the diameter, $N$ is the neutral gas number density, and $\sigma_i$ is the electron impact ionization cross section. 
Combining these provides a prediction for the fireball diameter 
\begin{equation}\label{D1}
D \approx \frac{1}{N \sigma_i} \sqrt{\frac{m_e}{m_i}}. 
\end{equation}
Several experiments have made note of the fact that the double layer potential is only a few volts above the ionization threshold \cite{baal:09,song:91}. 
Based on this observation, Song \emph{et al}~\cite{song:91} argued that the cross section should be evaluated at 2V above the ionization threshold and made predictions for the fireball size based on this assumption. Comparing these predictions to experiments in Ar, Kr, and Xe plasmas, they were able to verify the $D\propto 1/N$ scaling of equation~(\ref{D1}). 

\begin{figure}
\begin{center}
\includegraphics[width=8cm]{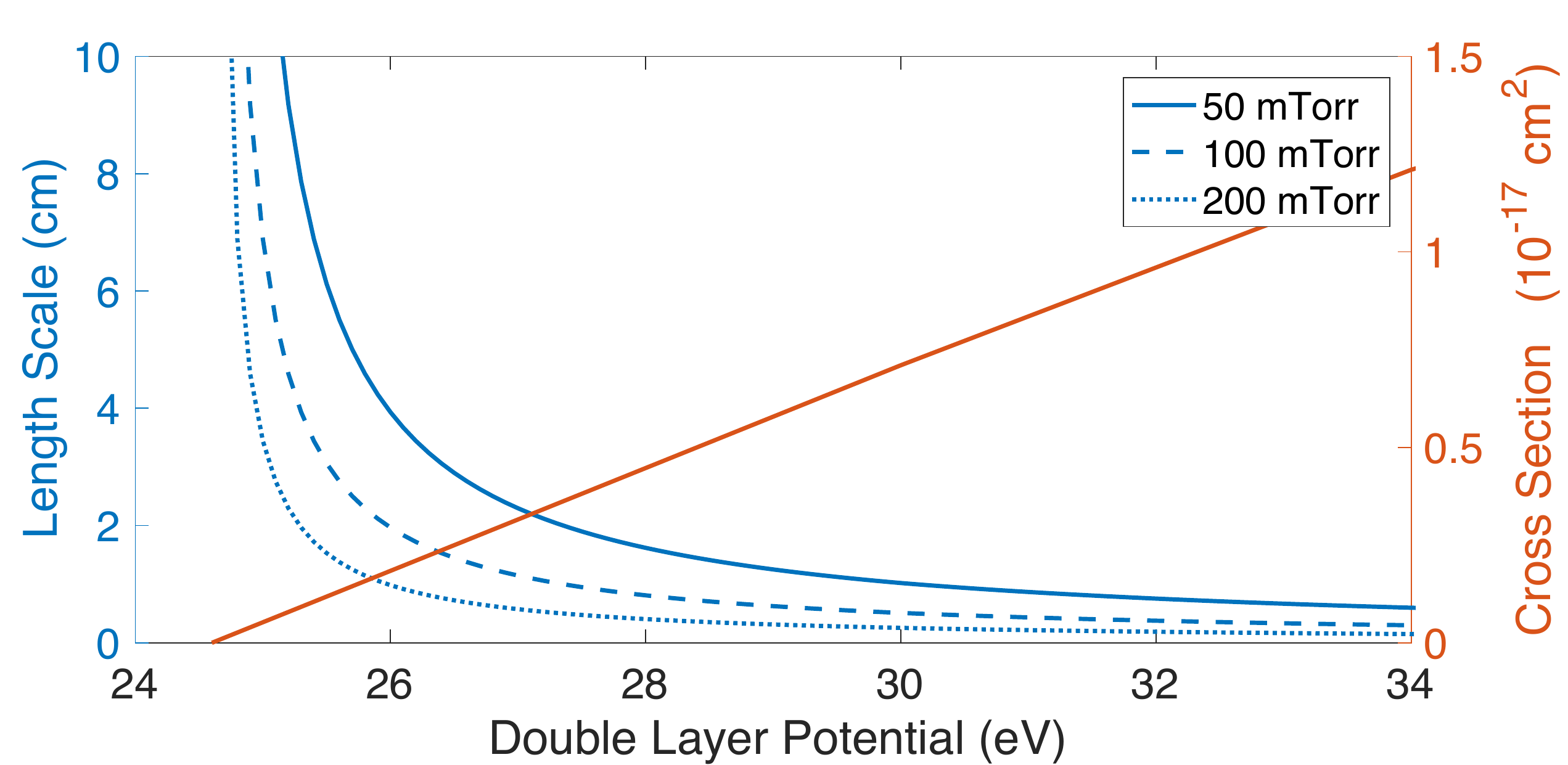}
\caption{\label{sche:17:fig13} The fireball diameter from equation~(\ref{D1}) with the cross section $\sigma_i$ evaluated at the energy an electron would gain in a double layer of potential $\Delta\phi_{\DL}$ for neutral helium at 50, 100, and 200 mTorr. 
The energy dependence of the electron impact ionization cross section of He is also shown. 
Figure reprinted from reference~\cite{sche:17}.  }
\end{center}
\end{figure}

A later estimate of the fireball size was presented in Ref.~\cite{baal:09} using flux balance arguments without invoking the Langmuir condition. 
Balancing the ionization rate with the flux of ions leaving the fireball in a 1D Cartesian model, and assuming equality between the beam electron density and fireball ion density, the length scale was estimated to be 
\begin{equation}\label{D2}
L = \frac{1}{N\sigma_i}\sqrt{\frac{m_e}{m_i}}\sqrt{\frac{\Delta\phi_\s}{2\Delta\phi_{\DL}}},
\end{equation}
where $\Delta \phi_\s$ is the potential of the ion presheath within the fireball at the high potential side of the double layer and $\Delta\phi_{\DL}$ is the double layer potential through which electrons are accelerated. 
The length estimate of equation~(\ref{D2}) reproduces the previously observed scaling of $L\propto 1/N$. 
The numerical estimates for the fireball size produced values which were within a factor of 2 of those measured in experiments~\cite{baal:09}.

Later work elaborated on the sensitivity of the fireball diameter to the energy at which the cross section is evaluated~\cite{sche:17}. 
This feature is demonstrated in figure~\ref{sche:17:fig13} which shows the predicted size in a helium plasma from equation~(\ref{D1}) as a function of electron energy gained by the double layer potential. 
If the double layer potential is taken to be infinitesimally close to the electron impact ionization threshold energy the predicted fireball size tends to infinity. To provide an additional constraint on the evaluation of the cross section, the double layer potential was determined by imposing a power balance relation within the fireball.
This predicts that the double layer potential is~\cite{sche:17}
\begin{equation}
e\Delta\phi_{DL}=\mathcal{E}_{i}+\mathcal{O}(T_e),
\end{equation}
where $\mathcal{E}_{i}$ is the ionization energy of the neutral gas and $\mathcal{O}(T_e)$ is a term on the order of the electron temperature, the form of which depends on the fireball potential structure. 
This result is consistent with the earlier assumptions of Ref.~\cite{song:91} when the electron temperature is 1-2eV, as is typical in many fireballs.

\begin{figure}
\begin{center}
\includegraphics[width=8.5cm]{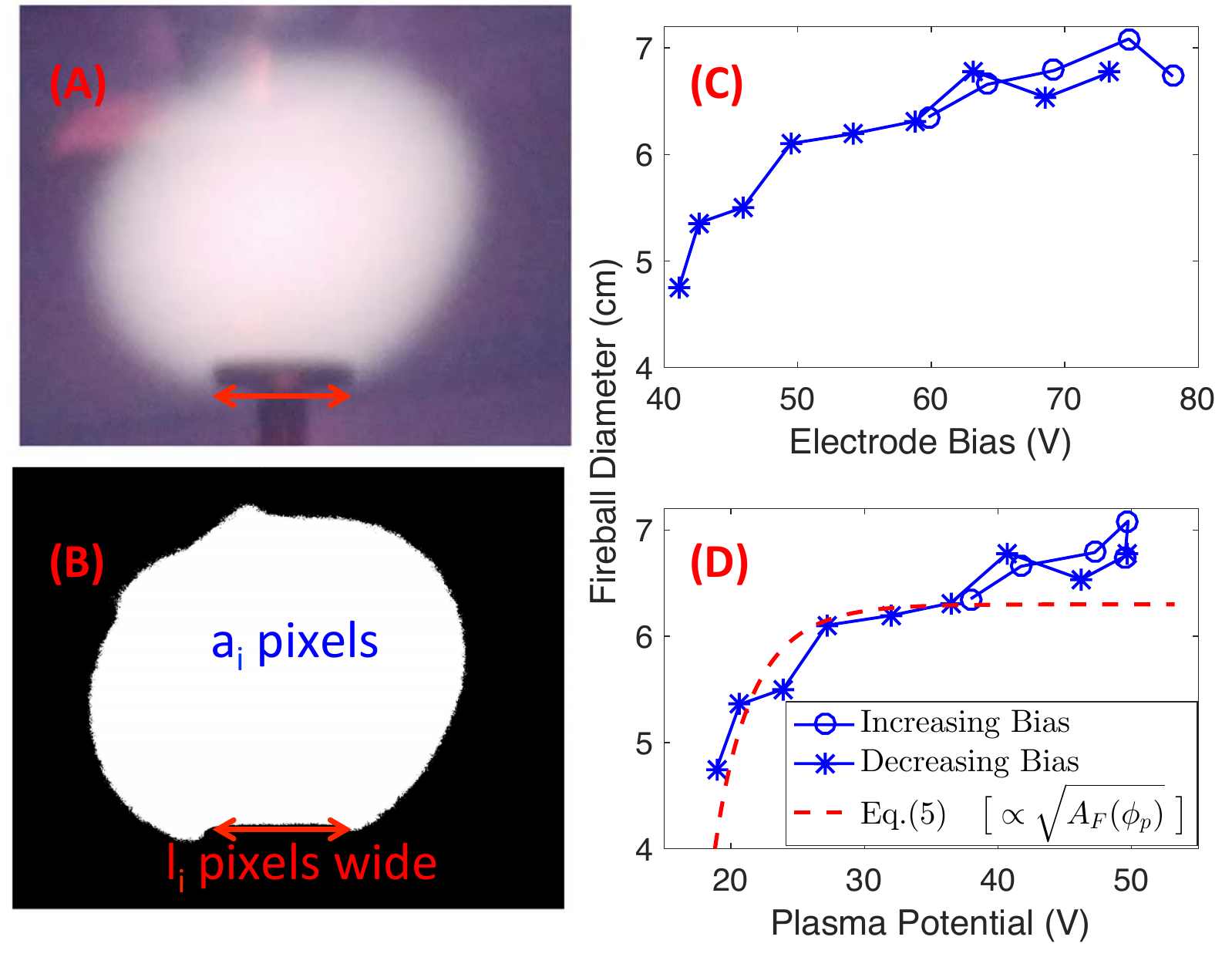}
\caption{\label{sche19fig2} (a) Photograph of a fireball and (b) the intensity threshold used to determine the fireball size. 
Fireball diameter as a function of (c) electrode bias and (d) plasma potential. The dashed line in (d) shows the scaling of $A_\F$ as a function of plasma potential. Note that ``Eq.~(5)'' in the legend refers to equation~(\ref{eq:phi_f}) of this review. Figure reprinted from reference~\cite{sche:19}. }
\end{center}
\end{figure}

\begin{figure}
\begin{center}
\includegraphics[width=7.5cm]{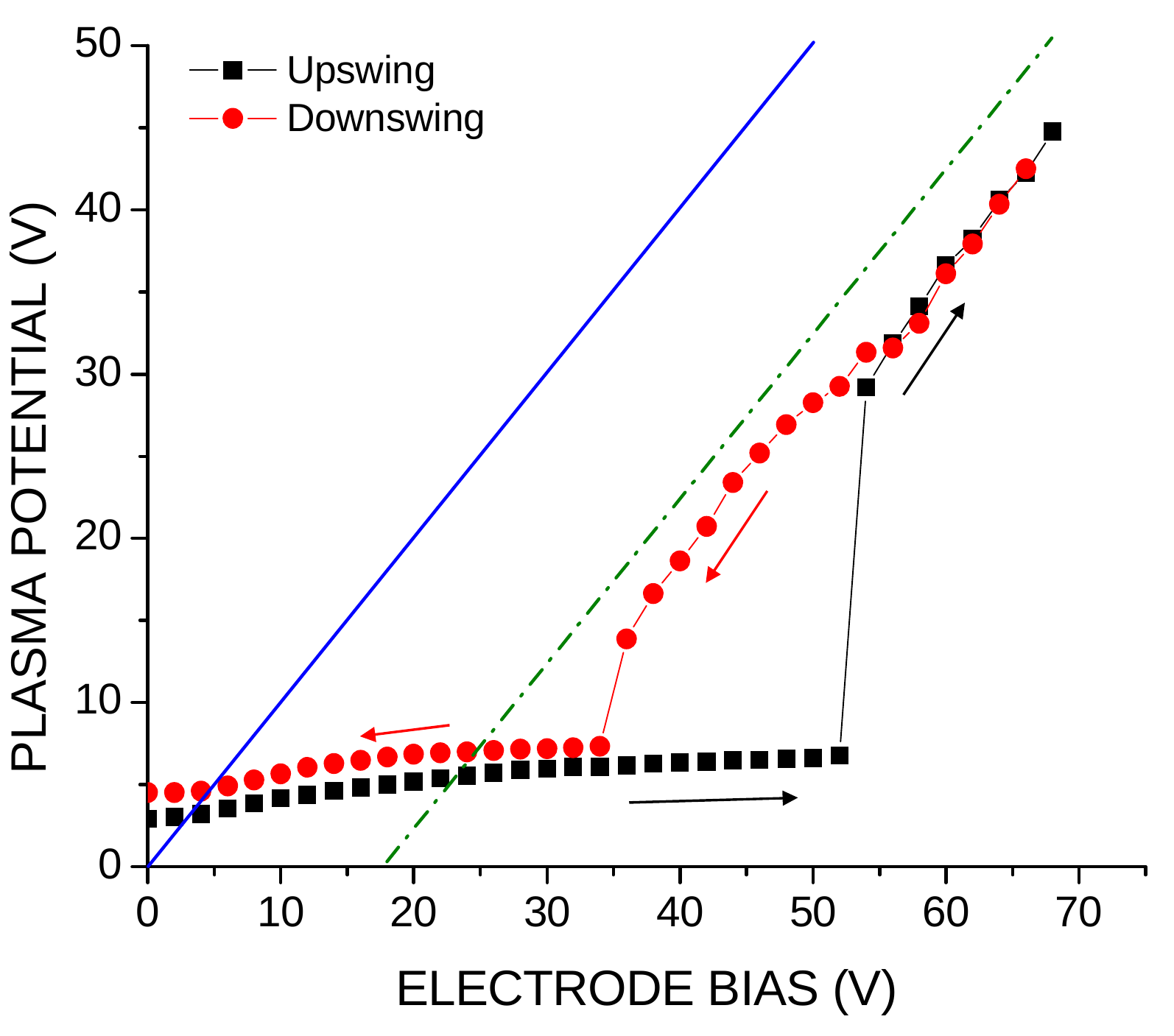}
\caption{\label{fg:F_GNF} Measured bulk plasma potential versus increasing (squares) and decreasing (circles) electrode bias. The solid line indicates an electrode bias equal to the plasma potential and the dash-dotted line shows the same line shifted by the ionization potential of argon.  Figure reprinted from reference~\cite{baal:09}. }
\end{center}
\end{figure}

Fireball size is not always tied to particle balance within the fireball itself. 
Instead, global particle balance arguments can also contribute. 
Recent experiments inferred the fireball size as a function of electrode bias by taking the intensity threshold of a photograph, as shown in figure~\ref{sche19fig2}~\cite{sche:19}. 
With a fireball present, both the plasma potential and fireball surface area were observed to increase with increasing electrode bias (figures~\ref{sche19fig2}c and d). 
Because the surface area of the fireball acts as the effective surface area for collecting electrons, it also determines the effective size of the electrode for a global current balance of the type described in section~\ref{sec:gc}. 
The plasma potential can be estimated using equation~(\ref{eq:Vp_es}), but where the electrode area $A_\E$ is replaced by the effective fireball area $A_\F$
\begin{equation}
\label{eq:phi_f}
\phi_p=-(T_e/e)\ln(\mu-A_{\F}/A_{\w}).
\end{equation} 
Figure~\ref{sche19fig2}d shows that the scaling relationship between the plasma potential and the fireball diameter $D \propto \sqrt{A_\F}$ predicted by equation~(\ref{eq:phi_f}) agrees well with the measurements.
When the fireball becomes large enough, a state of global nonambipolar flow can be established whereby electrons are only collected by the fireball surface area and ions are only collected by the chamber wall. 
This sets a maximum fireball size.  
In this global nonambipolar flow state, the plasma potential is locked to the electrode potential, as in figure~\ref{fg:Vp_NAF}, but shifted by the double layer potential drop $e\Delta\phi_{\DL}\approx\mathcal{E}_{i}$; as shown in figure~\ref{fg:F_GNF}.  

\subsubsection{Multiple fireballs} 

If the fireball size, as predicted from equation~(\ref{D2}), is sufficiently large compared to the electrode surface area, the sheath between the fireball and electrode is expected to be an electron sheath, as predicted in equation~(\ref{a3}). 
If the confinement chamber is also sufficiently large that the formation of the double layer does not raise the plasma potential enough to onset global nonambipolar flow, then the potential difference between the plasma and the confinement chamber wall is set to a fixed to a value determined by equation~(\ref{eq:phi_f}). 
Since the double layer potential drop is fixed to the ionization potential of the neutral gas, in this state further increases in the electrode bias will increase the potential difference between the electrode and the fireball; i.e., the electron sheath. 

\begin{figure}
\begin{center}
\includegraphics[width=8.5cm]{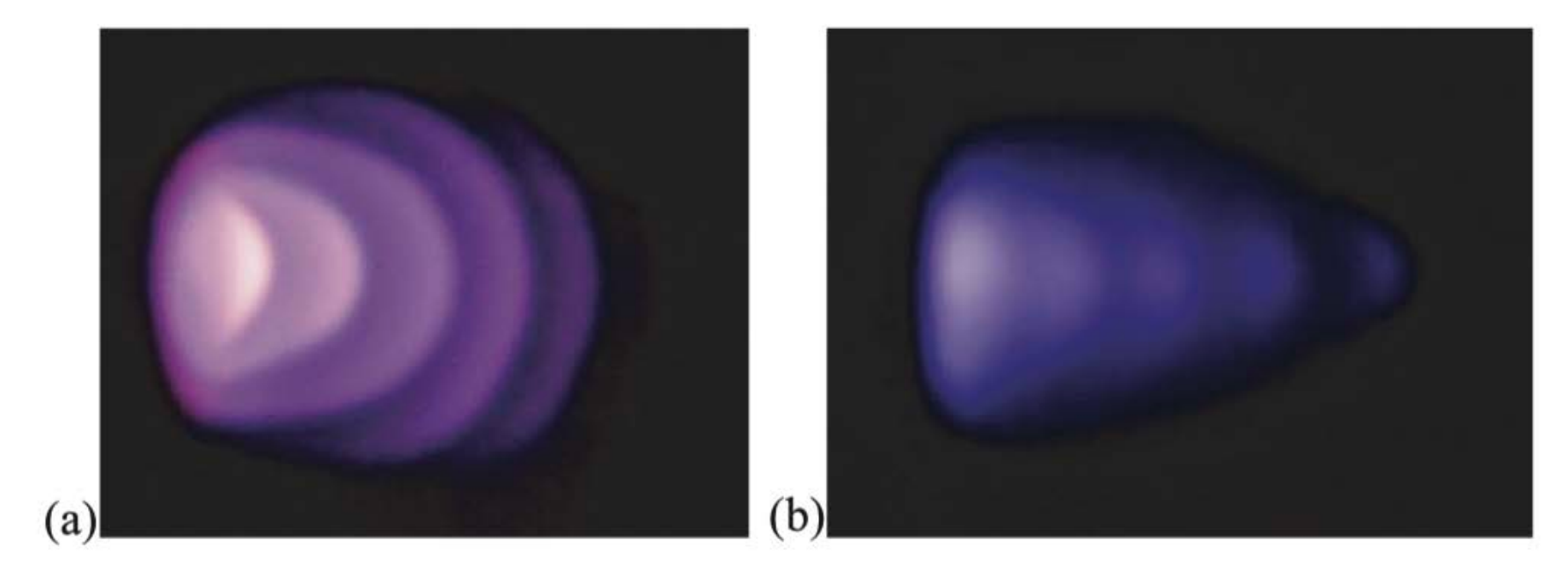}
\caption{\label{dim13fig1} Photographs of concentric fireballs in (a) stable and (b) unstable states. Figure reprinted from reference~\cite{dimi:13}.}
\end{center}
\end{figure}

\begin{figure}
\begin{center}
\includegraphics[width=8.5cm]{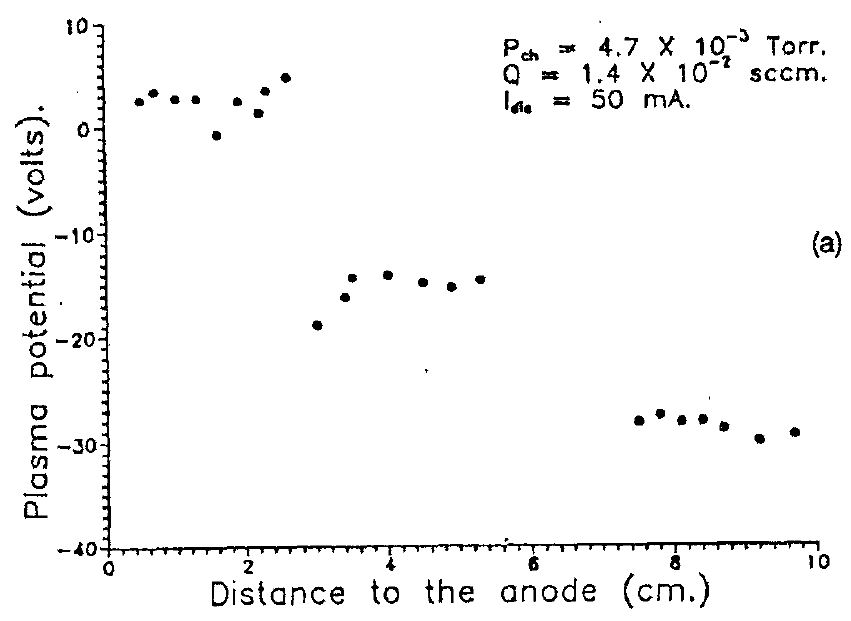}
\caption{\label{fg:mdl} Langmuir probe measurement of the plasma potential through a multiple double layer. Figure reprinted from reference~\cite{cond:94}.}
\end{center}
\end{figure}

When the electron sheath between the fireball and electrode reaches the ionization potential of the neutral gas, a second fireball would be predicted to form by the same mechanism as described in section~\ref{sec:onset}. 
Indeed, such multiple fireball formation has been observed~\cite{cond:94,neru:98,cond:99,aflo:05,ivan:05,dimi:07,dimi:13,dimi:15,gurl:14,alex:17,alex:17b,alex:18,paul:19}. 
These usually take the form of concentric spheres or hemispheres, as shown in figure~\ref{dim13fig1}~\cite{cond:94,cond:99,aflo:05}. 
In this case, the radial electrostatic potential profile is stair-cased with each step consisting of a double layer with potential drop approximately equal to the ionization potential of the neutral gas; for example see figure~\ref{fg:mdl}. 
Several concentric fireballs appear to be possible. 
Limiting factors to the maximum number of concentric fireballs might be that the fireball becomes big enough that global nonambipolar flow onsets, which locks the plasma potential to the electrode potential and prevents further fireball formation, or, perhaps, that the neutral concentration gets depleted by a plasma density so high that there is no longer enough neutrals to create another more dense fireball. 

Multiple fireballs are not always concentric. 
Side-by-side fireballs have been observed on large electrodes~\cite{dimi:07}, and even a series of several fireballs have been observed on asymmetric electrodes~\cite{ivan:05}. 
Electrodes with two conducting sides will sometimes have a fireball on each side. 
Levko~\cite{levk:17} has made 2D PIC simulations of multiple fireballs in a hollow cathode configuration, where one was observed to form inside the cathode and one outside. 
He attribute the formation of each to different mechanisms, related to the presence of electron emission from the boundaries of the hollow cathode. 
Clearly the formation mechanisms for multiple fireballs are complex. 

Fireballs can also be generated on multiple electrodes inside of a single background plasma. 
For example, Dimitriu \emph{et al}~\cite{dimi:15} studied the interaction of two fireballs generated on separate electrodes that were moved to be in proximity to one another. 
They observed a complex interplay between the fireballs, measuring current oscillations in few to tens of kHz frequency range, with the frequency observed to peak at a fixed distance of a few cm between the fireballs. 
This was modeled using an application of the scale relativity theory, and chaotic aspects of the oscillations were also observed. 

\subsubsection{Geometry} 

\begin{figure}
\begin{center}
\includegraphics[width=8cm]{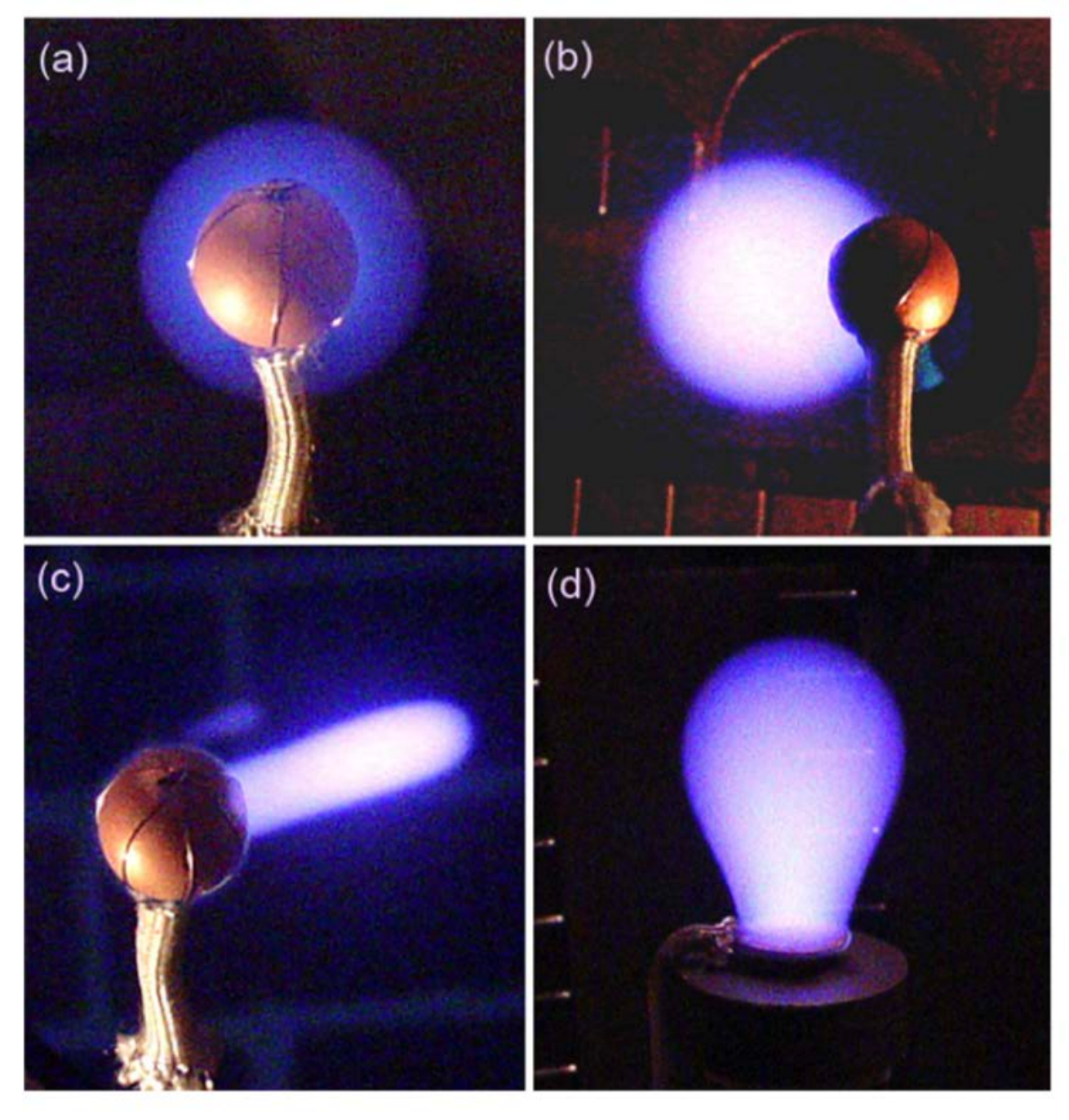}
\caption{\label{sten08fig2} Photographs of a variety of sheath and fireball geometries present at electrodes. (a) an anode glow prior to the formation of a fireball. (b) A spherical fireball. (c) A cylindrical fireball or firerod. (d) A pear shaped fireball in a dipole magnetic field. Figure reprinted from reference~\cite{sten:08}.}
\end{center} 
\end{figure}

Fireballs are not always spherical. 
While spherical fireballs are common, a variety of geometries have been observed (figure~\ref{sten08fig2}) both with and without the presence of a magnetic field. 
Near a spherical electrode in an unmagnetized plasma, Stenzel \emph{et al}~\cite{sten:08} observed fireball states to include both a sphere that uniformly surrounds the electrode, as in figure~\ref{sten08fig2}a, as well as states where the fireball is found on only one side, or portion, of the electrode, as shown in figure~\ref{sten08fig2}b~\cite{sten:08}. 
Cylindrical fireballs, commonly called ``firerods,'' such as those in figure~\ref{sten08fig2}c have been observed in magnetized plasmas~\cite{an:94}, but they have also been observed in the absence of an applied field~\cite{sten:08,baal:09}. 
Firerod formation in the absence of an applied magnetic field was observed near large electrodes and was attributed to the formation of double sheaths on regions of the electrode surface surrounding the firerod~\cite{baal:09}. 
Presence of the double sheath was suggested to be necessary to limit the effective size of the electrode for current collection, because with the fireball formation the effective size would be too large to meet the global current balance requirements described in section~\ref{sec:gc}.
The double sheaths reduce the effective collecting area, and also have the effect of constraining the fireball's radial dimension so that it has a cylindrical shape~\cite{baal:09}. 

The observations of the transition from fireball to firerod with increasing magnetic field~\cite{an:94} may be related to the dynamical motion of electrons being constrained to within a gyroradius of magnetic field lines, but it also may be influenced by changes to the global current balance of the system. 
The magnetic field reduces the ion loss rate to the chamber walls, resulting in the necessity of double sheaths on regions of the electrode surface in order to maintain global current balance. 
Pear shaped fireballs, such as those in figure~\ref{sten08fig2}d and other more complicated structures have been observed in dipole and other more complicated field geometries~\cite{sten:12,sten:08}. 
Magnetic fields can be used to shape fireballs. 

Stenzel~\emph{et al}~\cite{sten:11f,sten:11g} have also investigated an ``inverted fireball'' configuration where a spherical high density plasma is generated inside of a wire grid formed into a sphere. 
This has some similar properties to fireballs on the surface of electrodes, such as being produced by increased ionization due to energetic electrons and having a double layer electric field, but it is also particular to the transparent anode grid geometry.

\subsection{Hysteresis} 

\begin{figure}
\begin{center}
\includegraphics[width=8cm]{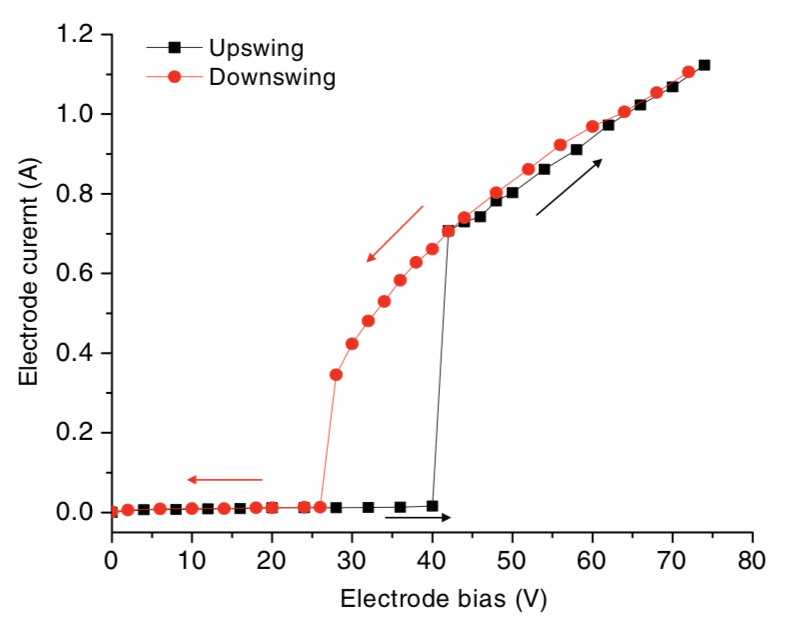}
\caption{\label{baal09fig6} Current voltage hysteresis between the upswing and downswing of the electrode bias. Figure reprinted from reference~\cite{baal:09}.}
\end{center}
\end{figure}

One of the most frequently observed features of fireballs is the hysteresis in the current voltage (I-V) characteristic of the electrode bias such as that shown in figure~\ref{baal09fig6}~\cite{baal:09,cart:87,song:91,sche:19}. 
Prior to the formation of a fireball, the electrode initially has an electron sheath present and is collecting the electron saturation current. 
As the electrode bias increases past a critical bias a fireball abruptly forms, accompanied by a large increase in electron current collected by the electrode. 

This hysteresis can be understood as follows~\cite{sche:19}: Initially, the electrode bias must exceed the plasma potential by a critical value for fireball onset as described in section~\ref{sec:onset}. 
After the fireball is present, the electron current collected by the electrode is significantly greater, owing to the greater electron collection surface area. 
Immediately after formation, the fireball surface area expands, raising the plasma potential. 
If the expanded fireball is large enough, global nonambipolar flow onsets and the plasma potential becomes locked to a fixed offset of the electrode bias, the offset being determined by the double layer potential drop $e\Delta\phi_{\DL}\approx\mathcal{E}_i$; see figure~\ref{fg:F_GNF}. 
Subsequent lowering of the electrode bias results in a lower plasma potential. 
The lower plasma potential allows more electrons to be lost to the chamber walls. 
Global current balance requires that the electron collecting surface area of the fireball must be reduced to compensate. 
This results in a contraction of the fireball. 
Eventually, at a sufficiently low bias, the fireball collapses when the plasma potential is unable to decrease beyond a minimum value set by the global current balance condition, equation~(\ref{eq:phi_f}), which depends on the area ratio $A_\F/A_\w$. 
Often times, this is simply when the plasma potential approaches a few volts above the wall potential, so the electrode critical bias for collapse is a few volts above the ionization potential of the neutral gas, as in figure~\ref{baal09fig6}. 

Hysteresis has also been observed in current as a function of other quantities including neutral pressure~\cite{song:91}, electron temperature~\cite{baal:09}, plasma potential~\cite{baal:09} and magnetic field strength~\cite{an:94}. 
In general, the threshold values of these quantities needed for onset are larger than those needed to sustain the fireball. 

\subsection{Stability}

\subsubsection{Macroscopic Instability}

\begin{figure}
\begin{center}
\includegraphics[width=8cm]{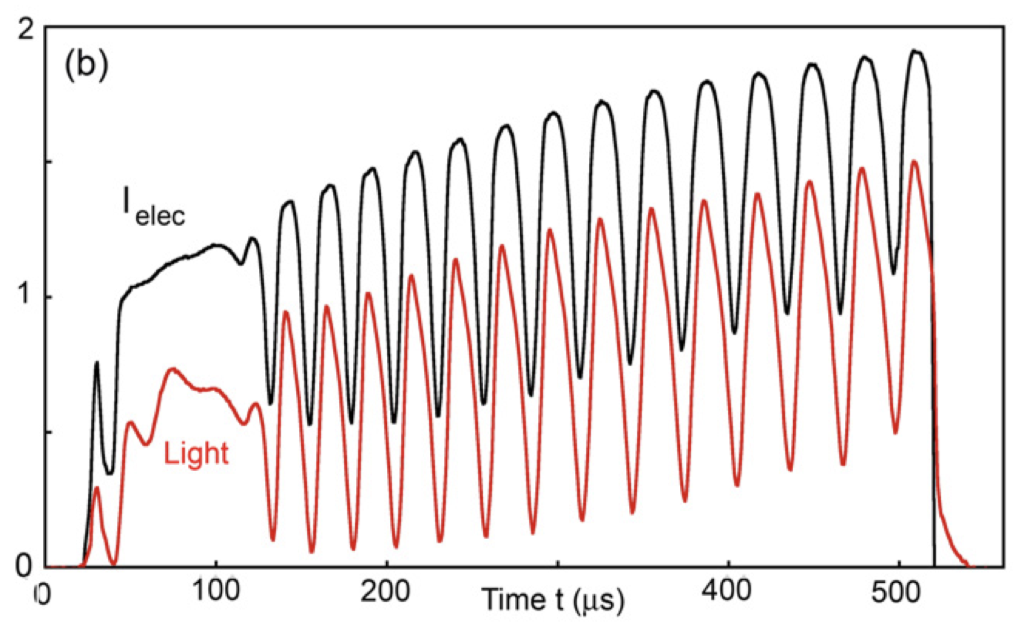}
\caption{\label{sten08fig3} The time dependence of fireball light emission and electrode current for the macroscopically unstable fireball. The coincidence of the emission and current collection suggest that the peaks and troughs are coincident with fireball formation and collapse. Figure reprinted from reference~\cite{sten:08}.}
\end{center}
\end{figure}

Stenzel \emph{et al}~\cite{sten:08,sten:11b,sten:11f,sten:11g,sten:12,sten:11c,sten:11d,sten:12c} and others~\cite{chir:07,cond:04} have carried out a series of investigations describing the complex stability properties of fireballs over a wide range of conditions and configurations. 
Fireballs have been observed in macroscopically stable (MS) and macroscopically unstable (MU) states. 
In the MS state, the fireball remains fixed to the electrode with little change in current collection or brightness. 
In the MU state the fireball forms and extinguishes as indicated by a periodic increase and decrease of electron current collection and light emission, as shown in figure~\ref{sten08fig3}. 
This work emphasizes the ways in which fireballs are not a local phenomenon, but are an integral part of the entire discharge. 
Specifically, macroscopic stability properties can be influenced not just by local ionization or plasma drifts within the fireball, but also global current and power balance requirements as well as external circuit interactions. 
Typical oscillation periods for the MU fireball are on the order of $\sim20-100$~$\mu s$~\cite{sten:08,song:91}. 
The oscillation period of the MU state has been observed to depend on the electrode voltage and current, discharge voltage and current, neutral gas species and pressure, and pulse length and repetition rate for fireballs formed in afterglow plasmas~\cite{sten:08}. 

Stenzel \emph{et al}~\cite{sten:12c} describe the process as follows: When ions are expelled from the electrode faster than they can be created by ionization, the fireball collapses. 
This leads to electron drifts exceeding the thermal velocity. 
After collapse, the plasma density decreases causing the sheath to again expand. 
Ionization in the sheath triggers the grown of a new fireball. 
Under certain plasma conditions, this processes repeats. 
This sequence is similar to the ``flickering'' phenomenon observed in experiments and simulations of pulsed electrodes~\cite{sche:18}, described in section~\ref{sec:onset}. 

Yip and Hershkowitz~\cite{yip:13} presented a model for the oscillation period based on a rate equation for electron generation and loss. 
The model was able to describe the qualitative features of their experiment, where fireballs formed near a biased wire array. 
It has not been tested in other experimental arrangements. 
Stenzel \emph{et al}~\cite{sten:08} have suggested that the instability may be due to limitations on the amount of current that can be drawn from the plasma. 
Likewise, it has been suggested that instability results from an incompatibility between the electron loss through the fireball surface area after onset and the electrode-to-wall area ratio needed to balance of global current loss with a positive electrode~\cite{sche:19}. 
The cause of instability is likely more complicated as experiments have measured disconnected regions of instability in the electrode bias-neutral pressure phase space~\cite{yip:13}. 

\begin{figure}
\begin{center}
\includegraphics[width=7.5cm]{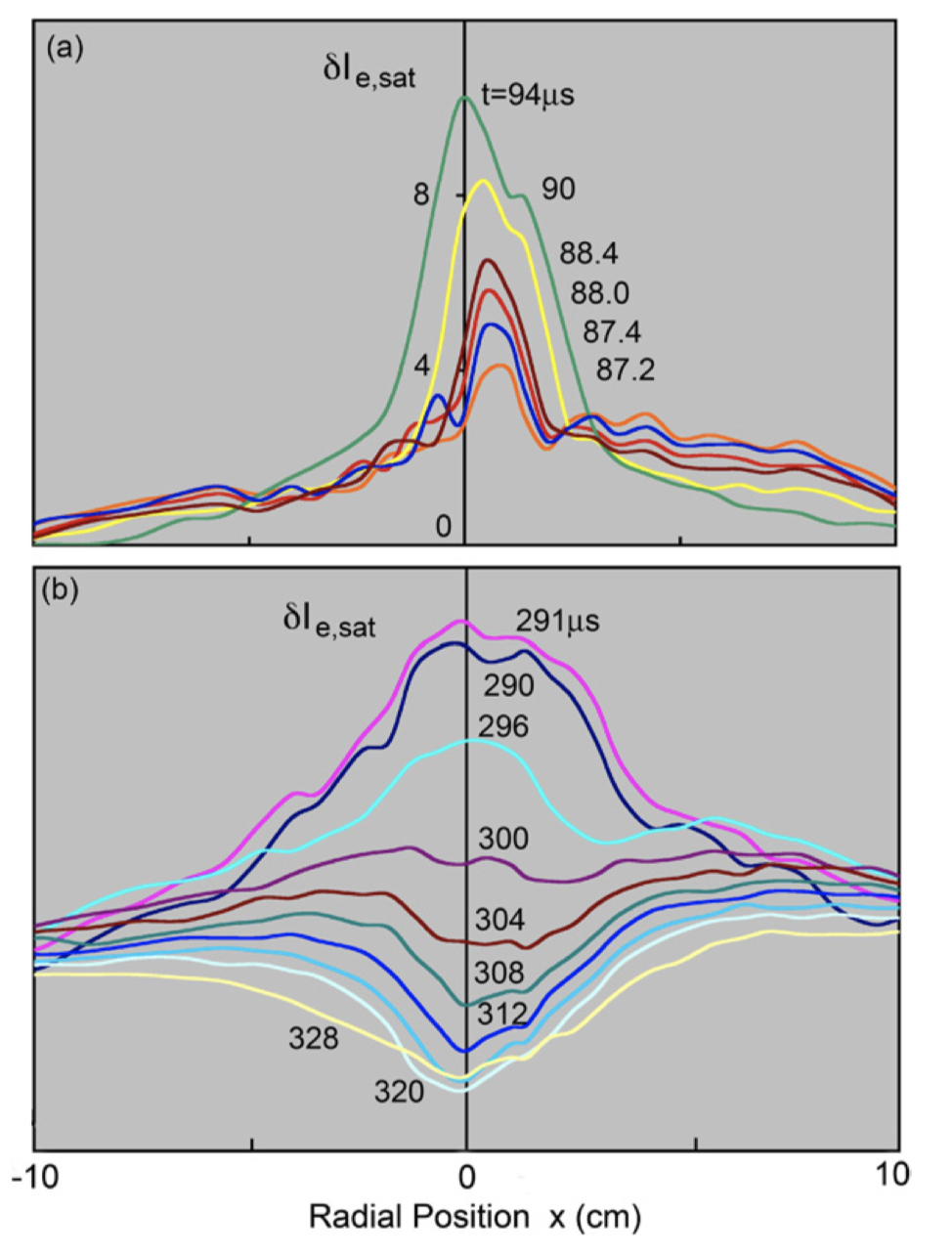}
\caption{\label{sten08fig6} Radial profiles of the measured electron saturation current ($\propto n_e\sqrt{T_e}$) at multiple times in the MU fireball during (a) onset and (b) collapse. Measurements indicate a density hole coincides with the collapse. Figure reprinted from reference~\cite{sten:08}. }
\end{center}
\end{figure}

The potential structure and density of the fireball has been studied during its formation and collapse in the MU state~\cite{sten:08,song:91,sten:12}. 
In Ref.~\cite{song:91}, a Langmuir probe was used to measure the plasma potential profile at different locations during fireball formation and collapse. In both cases, the period of formation and collapse in 1 mTorr of argon were approximately $10-20~\mu s$. This was comparable to the ionization time of 5 $\mu s$ estimated using
\begin{equation}
t_{\scriptsize \textrm{ionize}} \approx \frac{1}{N\sigma_i \sqrt{2e\Delta\phi_{\DL}/m_e}}. 
\end{equation}
Further measurements in Ref.~\cite{sten:08} found a consistent conclusion.
By measuring radial profiles of the electron saturation current 0.5 cm from the electrode surface, as shown in figure~\ref{sten08fig6}, changes in density were inferred via the relation $I_{e, \scriptsize \textrm{sat}}\propto n_e\sqrt{T_e}$. 
Figure~\ref{sten08fig6} demonstrates that the fireball formation initiates at the center of the fireball and likewise the fireball collapse initiates at the same location. 
A depletion of the measured saturation current indicates a loss of the highest energy electrons at the center of the fireball and a decrease of density due to a reduction of the ionization rate. 
PIC simulations of transient behavior after the application of a stepped electrode bias suggest that the fireball collapse may be due to a raise in the bulk plasma potential, reducing the double layer potential and thus the ionization rate~\cite{sche:18}. 
An experimental study of fireballs in a hollow cathode geometry reached a similar conclusion that the double layer potential is associated with the MU fireball collapse~\cite{park:14}. 
This study concluded that stability could be obtained by increasing ionization within the bulk plasma to compensate for electron loss through the fireball. 
Such observations are consistent with the suggestion that instability is related to limits on the amount of electron current which can be drawn from the plasma.    

\subsubsection{Electron frequency instabilities}

Microscopic instabilities have also been observed. 
High frequency instabilities near the electron plasma frequency have been associated with the presence of the fireball and have been attributed to the beam plasma instability~\cite{trov:80}, which is commonly found on the upstream side of a double layer due to the accelerated electron beam interacting with the trapped electron population on the high potential side. 
The beam-plasma instability has been observed in both MU and MS fireballs. 

The sheath-plasma resonance instability~\cite{sten:88} is another type of high-frequency instability in the range of the electron plasma frequency that has been observed after fireball collapse in the MU state~\cite{sten:11b,sten:11c,sten:11d}. 
While this instability is not intrinsic to the fireball, it results from the presence of an electron sheath after collapse. 
This instability is associated with a variety of nonlinear behaviors such as amplitude clipping of the collected electron current and bursty transient behavior as indicated by the fluctuations in electrode current. 
A high frequency transit time instability has also been extensively studied in inverted fireballs which are present on the interior of large gridded spherical electrodes~\cite{sten:11f,sten:11g,sten:12c}. 
The instability frequency in such cases is related to the electron transit time across the fireball.

\subsubsection{Ion frequency instabilities}

Instabilities in the ion-plasma frequency range have been observed at the low potential side of the double layer in PIC simulations~\cite{sche:17}. 
The low potential side of the fireball appears electrically much like an electron sheath to the electrons and ions. 
As discussed section~\ref{sec:es}, electrons flow into an electron sheath at a speed near their thermal speed. 
At the same time, ions enter the plasma from the fireball with an energy of at least the double layer potential energy $V_i \geq \sqrt{2e\Delta \phi_\DL/m_i}$. 
This creates a large differential flow between electrons and thermal ions that can excite electron-ion two-stream instabilities in the ion-acoustic instability branch~\cite{sche:15}; see section~\ref{sec:es_lf}.   
In addition, the relative drift between the ions emitted from fireball and plasma ions may excite ion-ion two-stream instabilities. 
Ion-ion two-stream instabilities are typically predicted to have a larger growth rate than ion-acoustic instabilities. 
In fact, simulations indicate that the strength of fluctuations at the low potential side is greater when a fireball is present than for an electron sheath~\cite{sche:17}.

\subsection{Fireballs in magnetized plasmas\label{sec:fb_mag}} 

It has long been observed that even relatively weak magnetic fields can influence fireballs~\cite{sten:12,tang:03,trov:80,alpo:86,song:92b,sten:11a,chau:16a,chau:16b,rane:18}. 
Magnetic fields as low as 10's of Gauss can be used to shape fireballs, as shown in figure~\ref{sten08fig2}. 
Many magnetic field configurations have been explored, including uniform~\cite{an:94}, diverging~\cite{cart:87,song:92a,alpo:86}, mirror~\cite{sten:12} and cusp~\cite{sten:12} geometries. 
Some of the early work on magnetized fireballs was motivated by creating a laboratory experiment to study double layers observed in space environments, such as the magnetosphere~\cite{kell:84}. 
Such double layers are thought to contribute to particle acceleration near magnetic cusps, which may be a source of auroral generation~\cite{shaw:78}. 

Alport~\emph{et al}~\cite{alpo:86} provided a detailed potential map of the double layer electric field of a firerod in the presence of an inhomogeneous magnetic field. 
They also showed that both ion and electron cyclotron instabilities are excited by the firerod, which likely represent the magnetized plasma analogs of the ion and electron frequency instabilities discussed in the previous two subsections.  
Stenzel~\emph{et al}~\cite{sten:10,sten:12,sten:11a} has also extended the stability analysis and experiments described in the previous subsection to magnetized plasmas. 
One distinguishing feature of magnetization is the observation of high-frequency waves excited by the $\vc{E}\times \vc{B}$ drift of the electrons, which generate toroidal eigenmodes~\cite{sten:10}. 

Song~\emph{et al}~\cite{song:92b} showed that the diverging nature of the magnetic field can stabilize the position of the firerod. 
This work also provided a modified model for fireball onset that includes the effect of ion reflection by the magnetic field gradient due to the magnetic mirror force. 
This model is similar to that discussed in section~\ref{sec:size}, but where equation~(\ref{eq:gamma_i}) is replaced by
\begin{equation}
\varepsilon N \sigma_i z_D \Gamma_e = \Gamma_i
\end{equation}
where $z_D$ is the length scale of the firerod in the magnetic field direction (which replaces the diameter $D$ of an unmagnetized fireball) and $\varepsilon$ is the fraction of ions transmitted through the magnetic mirror. 
This model was successfully able to predict the position of the double layer (i.e., size of the firerod) in the inhomogenous magnetic field. 

Later work by An~\emph{et al}~\cite{an:94} developed a model for the steady-state properties of firerods in a uniform magnetic field and tested it experimentally. 
This work considered a positively biased disk shaped electrode with a surface normal to a uniform magnetic field. 
It showed that the minimum magnetic field strength for a fireball to transition to a firerod is associated with the condition that ions be weakly magnetized. 
Defining $K \sim \omega_{ci}/\nu_{\textrm{\scriptsize in}}$ as the ratio of the ion cyclotron frequency and the ion-neutral collision frequency, and using $\nu_{\textrm{\scriptsize in}} = N\sigma_i v_{Ti}$, this relationship can be expressed as
\begin{equation}
B = \frac{K}{e} (m_i k_B T_i)^{1/2} \sigma_i N
\end{equation} 
where $K$ is a number of order 1. 
This predicts that the minimum magnetic field strength required for a firerod to onset is proportional to the neutral pressure, which was found to agree well with experiments~\cite{an:94}. 

A model was also developed for the firerod length~\cite{an:94}. 
This is similar to that presented in section~\ref{sec:size}, but is complicated by the fact that particles can escape the double layer both through end and sides of the cylinder and that motion of electrons and ions in the direction perpendicular to the magnetic field is suppressed by the magnetic field. 
Nevertheless, a relatively simple model was proposed (see equations (5)--(7) of \cite{an:94}) that reproduced at least the qualitative features of the measurements.  
Finally, this work also mapped out a stabilty diagram in terms of pressure and magnetic field strength which showed that there is also a maximum magnetic field strength beyond which the firerod can no longer be maintained. 
A stronger magnetic field will suppress radial diffusion of electrons and ions, making the geometry more and more one-dimensional with increasing field strength.  
In lieu of the global current balance conditions described above, it may be that the strong magnetization prevents the possibility of firerod formation because it effectively reduces the chamber wall area for collecting ions ($A_\w$), which may imply that the a larger effective electrode size generated by a fireball is too large to preserve the global current balance requirement. 
Further experiments would be required to test this suggestion. 

\section{Connections with Related Topics\label{sec:con}} 

This review has focused on dc biased electrodes in low-pressure discharges with simple geometric configurations, such as that depicted in figure~\ref{fg:electrode}, because it provides a simple demonstration of physical processes associated with sheaths near biased electrodes. 
However, many plasma discharges of practical interest are more complicated. 
These complexities can change some of the expectations from the above discussion in fundamental ways. 
Additionally, some of the concepts related to sheaths near biased electrodes may be applicable to understanding other phenomena, but the connection is not always obvious. 
Here we briefly mention four particularly relevant example topics: high pressure plasmas, magnetized plasmas, rf capacitively coupled discharges and electronegative plasmas.

\subsection{Anode and cathode spots in high pressure plasmas } 

Anode and cathode spots and patterns have long been a fascinating research topic in gas discharge physics~\cite{beni:88,alme:10,trel:16}. 
These are localized regions of luminous plasma that can take different shapes and form patterns near electrode surfaces. 
They have been observed in a variety of discharge types, including dc glow discharges~\cite{verr:09,shir:14}, arc discharges~\cite{trel:14}, magnetrons~\cite{yang:15}, dielectric barrier discharges~\cite{abol:11,call:14,boeu:12,bran:17,likh:07}, and microdischarges~\cite{scho:04}.
Electrode boundaries supporting such spots can be either solid or liquid~\cite{verr:09,shir:14}. 
Often times these are associated with discharges operating near atmospheric pressure conditions. 
Modeling the patterns that form is an active research topic regarding the mathematical description of bistable nonlinear dissipative systems~\cite{beni:14,bien:18}, as well as computational physics~\cite{trel:16}. 

On one hand, anode spots appear to be similar to the fireballs in low-pressure plasmas (which are sometimes also called anode spots) that were discussed in section~\ref{sec:fb}. 
Both are related to self-organized, highly-luminous and localized secondary discharges that form near biased electrodes in plasmas. 
Indeed, the experimental setups are similar enough that it may be expected that fireballs would form at low pressure and spots at high pressure in the same apparatus. 
This, perhaps, may lead one to expect that the underlying physical mechanisms are similar. 

On the other hand, the theoretical descriptions for each of these phenomena, as currently understood, are quite distinct. 
The model for fireballs described in section~\ref{sec:fb} is based on increased ionization in a localized region that bifurcates to a fireball to preserve the flux balance condition associated with a double layer potential step separating the spot discharge from the bulk plasma. 
Thus, fundamental features of these fireballs include a space-charge-limited double-layer electric field and a quasineutral interior region. 
Kinetic effects (i.e., effects associated with details of the velocity distribution functions beyond what is described in a fluid or thermodynamics model) are important to this description. 
In contrast, spots in high pressure discharges appear to be well described by fluid models, such as systems of reaction-diffusion or drift-diffusion descriptions~\cite{trel:16}. 
Here, the physical mechanism responsible for spot formation is a thermal instability that is coupled with the boundary condition of the electrode temperature (a parameter completely absent from the low-pressure fireball description). 
Because collisions between charged particles and neutrals are so frequent at high pressure, space-charge-limited electric fields, such as the double layers of low-pressure fireballs, are rarely observed. 
Furthermore, anode spots are often observed in complex patterns at distinct locations on the electrode, which interact with one another, whereas fireballs are typically of the electrode size or larger and have either a single fireball, or concentric (nested), structure. 
The linear stability of the reaction-diffusion models for pattern formation in high-pressure spots has been accurately described using the Turing parameters characteristic of an activator-inhibitor system~\cite{trel:16,rumb:19}. 

Despite these differences, interesting questions remain regarding the relationship between these two phenomena. 
The similarity of the experiments suggests that there must be a pressure range over which the low-pressure fireball transitions to a high-pressure anode spot. 
What pressure characterizes this transition? 
How does the space-charge limited double-sheath transition to the anode spot of a fluid description? 
When does the temperature of the boundary begin to be important? 
How does the large-scale fireball shrink to a pattern of anode spots? 
Does this transition occur gradually or abruptly?

\begin{figure}
\begin{center}
\includegraphics[width=8cm]{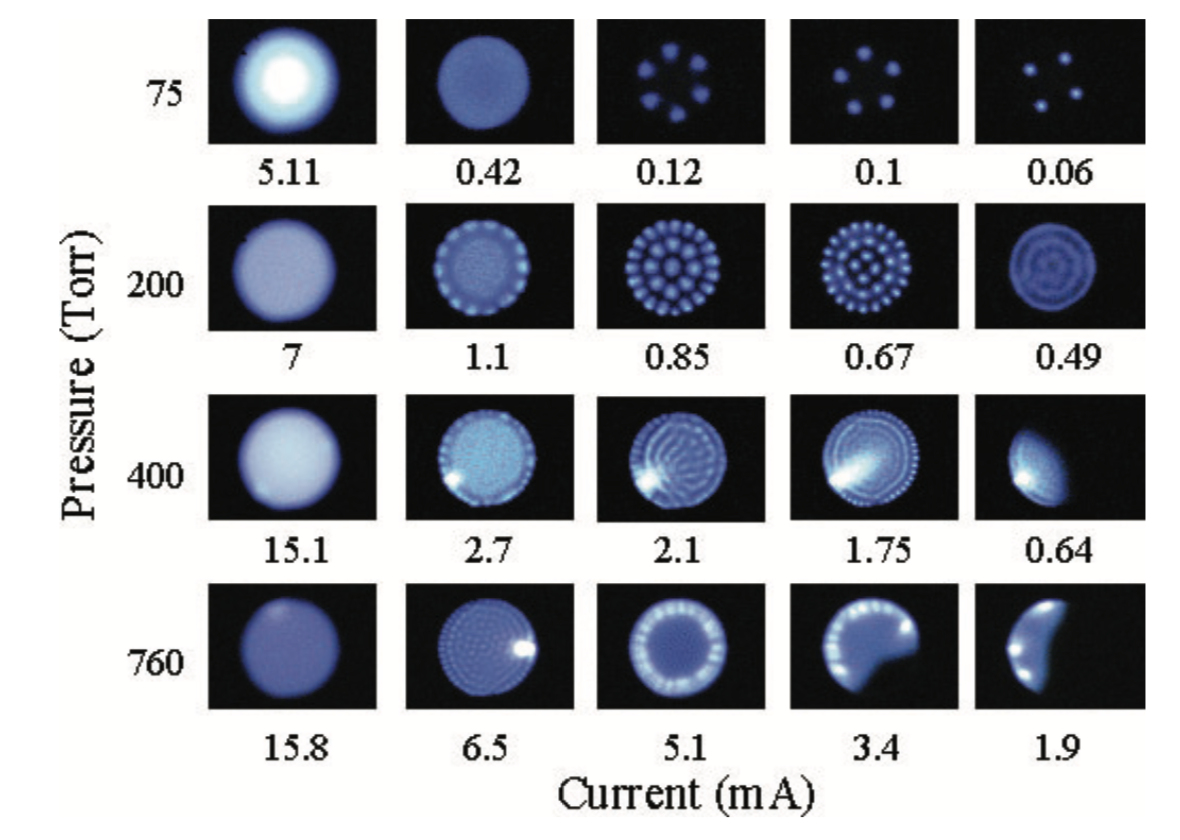}
\caption{\label{Cspot_fig} A variety of cathode spot patterns have been observed at different currents and pressures. Figure reprinted from reference~\cite{scho:04}}
\end{center}
\end{figure}

Cathode spots are a related high pressure (100s of Torr) self-organization phenomenon that have been observed in microdischarges; see figure~\ref{Cspot_fig}. 
Their formation depends on the magnitude of the current collected (or the electrode bias) and the neutral gas pressure. 
Experiments reported in \cite{scho:04} showed that varying the pressure and electrode current resulted in the formation of bright discharge-like features on the cathode, with varying degrees of azimuthal symmetry. 
Computational models of cathode spots based on two-fluid diffusion equations have reproduced the patterns observed in experiments, including the azimuthal symmetry~\cite{bien:16}. 
An analogous set of questions to that in the previous paragraph may be asked regarding how a cathode sheath that is well understood by the models described in section~\ref{sec:is} transitions to a complex spot structure at high pressure. 

\subsection{Magnetized Plasmas} 

This review focuses on unmagnetized plasmas, but many plasmas of interest are magnetized. 
Magnetization is known to fundamentally change the plasma-boundary interaction region. 
For instance, the canonical model of ion sheaths proposed by Chodura~\cite{chod:82} includes a ``magnetic presheath'' region characterized by the length scale 
\begin{equation}
\label{eq:dm}
d_m = \sqrt{6} (c_s/\omega_{ci}) \sin \psi
\end{equation}
where $\omega_{ci} = q_iB/m_i$ is the ion cyclotron frequency~\cite{chod:82}, and $\psi$ is the angle between the magnetic field and the normal direction to the boundary. 
In this model, the plasma boundary transition can be considered to consist of three layers: collisional presheath, magnetic presheath, and Debye sheath. 
Here, the ion velocity parallel to the magnetic field reaches the sound speed ($v_\parallel = c_s$) at the boundary between collisional presheath and magnetic presheath, while the ion velocity normal to the boundary surface reaches the sound speed ($v_n = c_s$) at the boundary between the magnetic presheath and the Debye sheath~\cite{riem:94,stan:95}. 
Several extensions and modifications of the Chodura model have been proposed~\cite{riem:94,stan:95,holl:93,hutc:96,ahed:97,beil:98,gera:17,gera:18,gera:19}. 
Much of the motivation, and experimental work, on magnetized sheaths stems from its importance to magnetic fusion energy experiments~\cite{stan:84,stan:00,chun:89}. 
There have also been a number of computational studies of the magnetic plasma-boundary transition, including particle trajectory simulations~\cite{dewa:87,cohe:98}, PIC simulations~\cite{chod:82,tskh:04,khaz:15,daub:98}, fluid-Monte Carlo simulations~\cite{khaz:15}, gyrokinetic simulations~\cite{gera:18} and Vlasov simulations~\cite{deva:06,coul:16}. 

Kim \emph{et al}~\cite{kim:95} measured equipotential contours in a magnetized sheath and presheath in a low temperature plasma; see also~\cite{sing:02}. 
These results agreed with qualitative features of Chodura's model, including the presence of both a collisional presheath and a magnetic presheath. 
Furthermore, the magnetic presheath thickness was measured to be approximately that predicted in equation~(\ref{eq:dm}), but where the $\sqrt{6}$ factor is replaced by unity. 
The shorter presheath may be due to the influence of ion-neutral collisions, as explained by Riemann's model~\cite{riem:94}.  
The collisional magnetic presheath was found to have an approximately $T_e/(2e)$ potential drop when $\psi \leq 40^\circ$. 
Equipotential contours were found to be parallel to the boundary in the collisional presheath, but not in the magnetic presheath. 

Recent experiments have employed LIF to measure the IVDF and associated ion flow profiles in 3D~\cite{sidd:14,sidd:16}. 
Ion-neutral collisions were found to be an important aspect of the presheath in these low temperature discharges, showing a breakdown of Chodura's model and the importance of effects described by Riemann~\cite{riem:94} and Ahedo~\cite{ahed:97}. 
Specifically, this work revealed that significant $\vc{E}\times \vc{B}$ drifts are generated, in which the ion flow has a component parallel to the boundary surface that is a significant fraction of the ion-acoustic speed. 
Models incorporating such drifts have been developed by Riemann~\cite{riem:94} and Stangeby~\cite{stan:95}. 
The experimental work also demonstrated that the $\vc{E}\times \vc{B}$ drifts are generated far from the boundary, due to the presheath electric field, and that charge exchange in the intervening region generates a population of energetic neutrals that must also be accounted for in models of wall sputtering and erosion. 
Furthermore, this work emphasizes the importance of kinetic effects in the magnetized presheath when ion flux tubes intersect the wall~\cite{daub:98}, which are not adequately included in the present theoretical models. 

This recent work shows that assumptions underlying common models can be too restrictive to apply to many real experimental circumstances. 
Since magnetization is important in so many applications, this presents an impetus for both further theoretical development and more rigorous experimental tests to extend magnetized sheath models. 
In this regard, some of the kinetic effects regarding ion sheaths presented in section~\ref{sec:is} may be applied or extended to the magnetized plasmas. 
For example, it was shown that ion-acoustic and ion-ion two-stream instabilities can be important in the unmagnetized presheath. 
Naturally, in a magnetized plasma one would expect that it may be the ion cyclotron instability that is excited. 
This may influence scattering, and other fluid properties of the plasma-boundary transition, in similar, but also predictably distinct, ways in comparison to the unmagnetized case. 

Furthermore, beyond ion sheaths, there has been very little work on the influence of a magnetic field on sheaths near biased electrodes. 
One of the few areas that have been investigated are fireballs in a magnetic field~\cite{torv:79,ande:81}, as summarized in section~\ref{sec:fb_mag}. 
Many of the new results pertaining to the other sheath types (electron sheaths, double sheaths, anode glow) would also be expected to change in response to a magnetic field. 
There is a significant opportunity for novel research directions here. 
For example, section~\ref{sec:es} described the recent discovery of an electron presheath and associated electron drift near a positively biased electrode. 
Is there a magnetic electron presheath? 
If so, does it have analogous properties to the magnetic ion presheath? 
The electron drift in the electron presheath was observed to excite strong ion-acoustic instabilities. 
Would a magnetic presheath drive strong ion cyclotron waves? 
High frequency (order $\omega_{pe}$) fluctuations are routinely observed near positively biased electrodes. 
Does the nature of these fluctuations change if the plasma is magnetized? 
Are upper hybrid modes excited in this case? 
The influence of magnetization on sheaths near biased electrodes is an area that is ripe for investigation.

\subsection{rf Capacitively Coupled Plasmas (CCP)} 

Although this review focuses on dc biased electrodes, plasmas in many experiments and industrial plasma reactors are generated by rf biasing.
A detailed and validated picture of rf ion sheaths has been developed over the years, and is well described in textbooks~\cite{lieb:05,chab:11}. 
Here, we simply comment on a couple of ways in which the recent work on dc biased electrodes may contribute to a better understanding of rf discharges. 

One aspect is the possible role of instabilities in the presheath. 
For example, many of the industrial plasmas in which rf discharges are used contain multiple species of positively charged ions. 
Does the instability-enhanced friction effect described in section~\ref{sec:is_mis} influence the ion energy at the sheath edge of an rf discharge as well? 
Bogdanova \emph{et al}~\cite{bogd:19} have recently investigated the ion composition of rf Ar/H$_2$ mixtures in an rf CCP, but the broader question of the potential role of instabilities is a complicated one that depends on the bias frequency, particle ion species and concentrations, pressure, and other plasma parameters. 
The bias frequency of many CCP discharges can be close to the expected unstable wave frequency. 
Does this stop, or otherwise change the nature of, the instabilities and related influence on transport? 
If present, can the bias frequency and plasma parameters be tailored in a way to take advantage of the instability-enhanced friction? 
A similar set of questions can also be asked in relation to ion-acoustic instabilities in a single species plasma. 

A second aspect is the role of electron sheaths near rf biased electrodes. 
Typically the plasma potential is above the electrode potential over the entire phase of the rf cycle~\cite{chab:11}. 
However, electric field reversals have also been measured~\cite{sato:90,czar:99,maho:97,gans:04,toch:92,turn:92} and simulated~\cite{turn:92,bele:90,vend:92,lero:95}. 
Here, electric field reversal means that the electrode is biased above the plasma potential for a short phase of the rf cycle. 
Electric field reversals have been attributed to the role of electron inertia near the electrode, as well as collisional drag on electrons in higher pressure regimes. 
Schulze \emph{et al}~\cite{schu:08} provide a summary of these results, and also show using both experiments and PIC simulations that electric field reversals can occur in dual frequency discharges as well. 

The recent advances in understanding dc electron sheaths presented in section~\ref{sec:es} may be applicable to electric field reversals in CCPs. 
In particular, the discovery that an electron presheath extends far into the plasma and that it causes a flow-shift of the electron distribution would be expected to influence the electron energy distribution function in CCPs.  
Electron beams have been observed in CCPs~\cite{schu:08b}. 
As in the dc electron sheath, electron drifts may excite ion-plasma frequency instabilities as electrons drift relative to the ions. 
The excited waves may influence both ion and electron transport through wave-particle collisions. 
Furthermore, high frequency (electron plasma frequency) waves that have been observed in the dc case may also occur with rf biasing. 
In fact, evidence for high frequency instabilities have been observed in PIC simulations~\cite{diom:14}. 

Finally, the geometric considerations of section~\ref{sec:gc} might be applied to CCP design. 
In particular, that section discusses the size requirements for the electrode and other boundaries that are required for a dc electrode to be biased above the plasma potential. 
How do these conditions change for rf biasing? 
If electron beams produce desirable effects in the plasma, could applying rf to electrodes that are sufficiently small to be biased above the plasma potential provide a measure of control, or other advantages? 
Could a combination of electrodes of different sizes, and different dc bias potential, be used to tailor plasma properties in a desirable way?

\subsection{Electronegative Plasmas} 

Many plasmas, including most material processing reactors, contain negatively charged ions. 
Negatively charged ions are known to significantly alter ion sheath properties~\cite{boyd:59,brai:88,fran:92,lich:94,lich:97,sher:99b,kouz:99,chab:00,chab:01,sobo:17}. 
For instance, the plasma-boundary transition of an electronegative plasma is thought to consist of three regions: an electronegative core, electropositive halo, and a positive ion rich sheath. 
Under certain conditions the core and halo are predicted to be separated by a double layer~\cite{brai:88,fran:92}. 
There is generally thought to be a good description of the plasma-boundary transition region near floating or grounded boundaries in electronegative discharges. 
Several aspects of these models have been validated experimentally, including the negative ion density profile~\cite{bere:00} and the assumption that negative ions are in Boltzmann equilibrium~\cite{ghim:09}. 

Despite the progress in understanding the ion sheath, there is relatively little understanding of the behavior of biased electrodes in electronegative discharges. 
Potentially interesting opportunities exist to apply some of the concepts discussed in this review to plasmas containing negative ions.
For instance, section~\ref{sec:gc} considered the geometric criteria required to bias an electrode above the plasma potential. 
Reconsidering this condition in the presence of a negative ion species, one may expect that a much larger electrode could be biased positive with respect to the plasma since the area ratio criterion for doing so is characterized by $\sqrt{m_-/m_+}$ where $m_-$ is the mass of the negative charge carrier and $m_+$ is the mass of the positive charge carrier. 
Such biased electrodes might be used to control the concentration of negative ions in the plasma. 
Furthermore, the basics of the analog of electron sheaths in a plasma containing negative ions is relatively unexplored. 
How does a negatively charged sheath transition from a thin electron sheath at low (or no) electronegativity to a negative ion sheath at high electronegativity? 
What are the basic properties of such a sheath? 
A variety of instabilities exist in electronegative discharges~\cite{chab:01} because positive and negative ions flow in opposite directions in response to a boundary electric field. 
Does the instability-enhanced friction mechanism influence the relative drifts of positive and negative ions, as it was seen to do amongst positive ion species in section~\ref{sec:is_mis}? 
Does this change the predicted flow profiles, or electric field characteristics, of the plasma-boundary transition region?

\section{Summary\label{sec:conc}} 

Positively biased electrodes are a common feature in many plasma diagnostics and applications. 
How these electrodes influence the plasma, and how the plasma influences the electrodes, are both governed by the properties of the sheath structure near the electrode surface. 
This sheath can take a wide variety of forms, including ion sheath, electron sheath, double sheath, anode glow double layer, or fireball double layer. 
Each is different and influences plasma-boundary interactions in different ways. 
Knowing which will form can be critical to design of applications and diagnostics, but this can be a complicated question to answer because it depends not only on local properties of the electrode or plasma, but also on the global confinement properties. 
An estimate of the plasma potential can be obtained from considerations of global current balance, but a precise determination depends on factors such as the effective area for collecting species at each boundary, sticking coefficients, emission properties, plasma gradients, and material composition. 

This review summarized a number of recent advances in our understanding of each of these types of sheaths. 
One theme of the recent results is kinetic effects. 
Since the sheath and presheath is often shorter, or comparable to, relevant collision mean free paths, the boundary selectively removes certain classes of particles from the velocity phase-space distributions. 
For ion sheaths, these effects were shown to be especially significant when an electrode is biased near the plasmas potential; particularly when it is within an electron temperature of the plasma potential. 
For example, a subsonic Bohm criterion for the ion speed at the sheath edge was observed to arise due to absorption of electrons at the electrode. 
Kinetic effects were also shown to be important to the EVDF near an electron sheath. 
Here, the EVDF was observed to consist of a flow-shift on the order of the electron thermal speed in addition to the traditionally predicted depletion of the EVDF associated with electrons lost to the electrode. 
This is a substantial change from the conventional picture, which did not include a flow shift, and it led to the prediction and validation of an electron-sheath equivalent of the Bohm criterion that is satisfied primary by the electron flow shift. 
Kinetic effects were also observed in both double sheaths and fireballs. 
In fireballs, a new category of kinetic effect was observed to arise from ionization. 
One example was the formation mechanism, which was found to be associated with a local potential well that forms due to increased ionization near the electrode surface. 

A second major theme of recent results is the importance of flow-induced instabilities. 
Traditional theories of dc sheaths are typically based on steady-state fluid or kinetic descriptions in which plasma parameters smoothly transition from the plasma to the boundary in a laminar manner. 
The new research has revealed that as ions or electrons flow toward the boundary they often spontaneously excite flow-driven instabilities (sometimes called kinetic or Vlasov instabilities). 
These instabilities can feed back to influence plasma properties. 
In fact, rather than being a rare occurrence, they were found to lead to observable, and sometimes important, effects in each of the types of sheaths discussed. 
In ion sheaths, ion-acoustic instabilities were observed to increase the ion-ion collision rate, creating a thermalizing effect on the IVDF. 
These were also predicted to influence the EVDF. 
In a two-ion-species plasma, ion-ion two-stream instabilities were found to lead to an enhanced ion-ion friction force that significantly influences the speed of each species as they traverse the presheath and enter the sheath. 
In electron sheaths, the fast electron flow was observed to excite large-amplitude ion-acoustic instabilities that cause the sheath to fluctuate. 
High-frequency instabilities, near the electron plasma frequency, have also been observed near electron sheaths. 
In double sheaths, counter streaming ion populations were observed to induce ion-ion two-stream instabilities, which caused a significant reduction in their flow speeds due to the associated enhanced friction force. 
Electron-electron two-stream instabilities were observed in the presence of electron emission. 
Finally, ion-streaming inside of a fireball was observed to lead to similar ion-acoustic type instabilities as in an ion sheath. 
Other global relaxation-type instabilities were also observed to be a prominent feature of fireballs.  

In addition to recent advances, a number of unresolved questions were identified. 
These include the need for more direct measurements of the presence of instabilities and their influence on ion and electron distribution functions. 
A number of issues remain unresolved with respect to the conditions at which double sheaths form, including the role of ion pumping, determining when double sheaths are steady-state or transitory solutions, as well as the role of electron emission from boundaries. 
Similarly, although basic mechanisms of formation and steady-state properties of fireballs have been identified, many questions remain. 
There is not a good explanation for the observed oblong shape of fireballs in some circumstances, particularly in describing the role of a magnetic field, nor for the observations where the shape can be non-spherical in seemingly low-magnetic field experiments. 
A wide range of possible applications may be possible by utilizing fireballs, or any of the other observed sheath types. 

A number of open questions remain in topics that overlap with what was discussed in this review, but which were not discussed directly. 
One very important area is the influence of magnetic fields on sheaths~\cite{sidd:16}, including the variety of sheath types discussed here. 
Another is high pressure discharges. 
This review concentrated on low-pressure (mTorr range) discharges, but recent trends in the field have focused on pressures near atmospheric pressure~\cite{adam:17}. 
It is certain that much, if not most, of the basic properties of sheaths that were discussed in this review will need to be substantially modified in these highly collisional situations. 
Likewise, in many of these applications boundary material plays a more active role in the local plasma physics, such as evaporation from liquid boundaries. 
The transfer of matter and chemical reactivity from the plasma through the boundary is also a key issue in this field, which is related to sheaths, but is not often addressed in low-pressure discharges. 

Advances in diagnostics, particularly non-invasive diagnostics such as optical-based methods, are likely to help advance this field. 
Similarly, the ability to perform more sophisticated computer simulations of sheaths in two and three dimensions, using kinetic methods such as PIC are helping to accelerate the rate of advances. 
The importance of sheaths to the wide-variety of plasma-based applications, and the slew of unanswered questions that remain, ensure that sheaths will continue to remain a vibrant research topic.

\section*{Acknowledgments}

The first author thanks Noah Hershkowitz for introducing him to all of the topics discussed in this review, and for many educational conversations over the past fifteen years.  
We also thank Ryan Hood, Lucas Beving, Fred Skiff, Robert Merlino, Greg Severn, and Chi-Sung Yip for helpful conversations related to this material. 
This work was supported by the Office of Fusion Energy Science at the U.S. Department of Energy under contracts DE-AC04-94SL85000 and DE-SC0016473.


\end{document}